\documentstyle[11pt,epsfig]{article}
\def\appendix{\par
 \setcounter{section}{0}
 \setcounter{subsection}{0}
 \def\thesection{Appendix \Alph{section}}
 \def\theequation{\Alph{section}.\arabic{equation}}
 \setcounter{equation}{0}}

\def\dis{\displaystyle}
\def\le{\left(}
\def\ri{\right)}

\def\no{\nonumber}

\def\rar{\rightarrow}

\def\e{\epsilon}
\def\f12{\frac{1}{2}}

\def\pd{\partial}

\def\L{\lambda}

\begin{document}
\begin{titlepage}
\flushright{USM-TH-220}
\vskip 2cm
\begin{center}
{\Large \bf Gluon self-interaction in the position space in Landau gauge}\\
\vskip 1cm  
Gorazd Cveti\v{c}$^{a}$ and  Igor Kondrashuk$^{a,b}$\\
\vskip 5mm  
{\it  (a) Centro de Estudios Subat\'omicos y Departamento de F\'\i sica, \\
Universidad T\'ecnica Federico Santa Mar\'\i a, Casilla 110-V, Valpara\'\i so, Chile} \\
{\it  (b) Departamento de Ciencias B\'asicas, \\
Universidad del B\'\i o-B\'\i o, Campus Fernando May, Casilla 447, Chill\'an, Chile} \\
\end{center}
\vskip 20mm
\begin{abstract}
We propose a method to treat the three-gluon self-interaction vertex in the position space
in $D=4 - 2 \e$ dimensions. 
As an example, we calculate a two-loop contribution to
auxiliary $Lcc$ vertex in the Landau gauge which contains the three-gluon vertex 
for $SU(N)$ Yang-Mills theory. 
We represent the integral expression as a sum of separate contributions so that
each of the contributions is a double finite integral or single integral
(singular or finite) in the position space. In each double finite integral 
we use the freedom to shift exponents in powers in the denominator of integrands 
by some multiples of $\e$, in order to perform at least one of the integrations 
by the uniqueness technique without corrupting the first term of 
the decomposition in $\e.$
\end{abstract}
\end{titlepage}

\section{Introduction}

Transversality of the vector propagator in the Landau gauge causes the existence 
of the finite scalar auxiliary three-point vertex $Lcc$ in the 
Wess-Zumino gauge of ${\cal N} = 4$ supersymmetric Yang-Mills theory. 
As it has been shown in Refs.~\cite{Cvetic:2004kx}-\cite{Cvetic:2007fp},
all the poles in $\e$ disappear in all number of loops for that vertex. 
This result has been derived from the results of 
Refs.~\cite{Cvetic:2002dx}-\cite{Kondrashuk:2003tw}.
This vertex does not depend on any scale, ultraviolet (UV) or infrared (IR), 
to all orders of the perturbation theory for maximally supersymmetric 
Yang-Mills theory.
Parameter $\e$ is the parameter of dimensional regularization, 
$ D = 4-2\e $ is the dimension of the space-time. The first two vertex-like 
two-loop contributions that correpond to diagrams $(a)$ and $(b)$ 
were calculated in the previous papers, Refs.~\cite{Cvetic:2006iu,Cvetic:2007fp}, 
respectively. In this work we calculate the contribution that corresponds 
to diagram $(c).$  The notation that is used here is the notation 
of Ref.~\cite{Cvetic:2006iu}.

The $Lcc$ vertex is the vertex in which the auxiliary field 
$L$ couples to two (self-adjoint) Faddeev-Popov ghost fields $c.$  
It is superficially convergent in the Landau gauge. 
This fact can be checked by index counting and by noting that 
two derivatives from the ghost propagators 
can always be integrated out of the diagram due to the transversality of the 
gauge propagator. It means that the field $c$ does not have renormalization 
in the Landau gauge.
Formally, this result holds to all orders of perturbation theory due to the 
so called antighost equation \cite{Blasi:1990xz}. In the nonsupersymmetric 
theories this vertex is
not finite and a calculation of the anomalous dimension of operator $cc$ 
has been performed in \cite{Dudal:2003pe,Dudal:2003np}.  

In Refs.~\cite{Cvetic:2004kx,Cvetic:2006kk,Kondrashuk:2004pu} it has been shown that if the vertex $Lcc$ is known to all loop orders of the maximally supersymmetric
Yang-Mills theory, one can obtain 
all the other correlators of dressed mean fields by solving the Slavnov-Taylor identity. However, despite all-order UV and IR
finiteness of this scalar vertex, its all-order calculation presents 
in itself an outstanding task for an analytic programmer. At this stage, we can present only full planar two-loop result in the Landau gauge. The difficult part is the algebra of Lorentz indices in the gluon self-interaction part.  We have solved this problem
at two-loop level by doing algebra of convolution of the derivatives and integrating by parts. This trick allows us to keep the extensive Lorentz algebra under control 
and to use, at the subsequent stage, {\it Mathematica} software \cite{Mathematica}. 
For convenience, we present 
in this manuscript details of calculation up to a certain stage.

Dressed mean fields are the usual effective fields redefined by integral convolutions with 
parts of propagators \cite{Cvetic:2004kx,Cvetic:2006kk,Kondrashuk:2004pu}. 
The difference with the usual BPHZ renormalization procedure \cite{BoSh} is that these parts of propagators contain not only poles in $\e$ but also space-dependent logarithmic
part (or momentum-dependent logarithmic part if one works in the momentum space) constructed
typically in the $\overline{MS}$ scheme procedure \cite{Vasil}. This is why integral convolution 
was used instead of traditional simple multiplication. All poles in $\e$ in Wess-Zumino gauge are contained in these dressing functions of the effective 
fields.
Physical meaning have the kernels of those dressed effective fields in the effective action. On-shell values in the momentum space correspond to the amplitudes of 
physical particles, for example gluons.

As it has been shown in Refs.~\cite{Cvetic:2004kx,Cvetic:2006kk,Kondrashuk:2004pu}, scale-independence of the correlators of dressed mean fields corresponding to physical particles 
is a direct consequence of the Slavnov-Taylor identity \cite{Slavnov:1972fg,Taylor:1971ff,Slavnov:1974dg,Faddeev:1980be,Lee:1973hb,Zinn-Justin:1974mc}.
The latter is a consequence of the BRST symmetry \cite{Becchi:1974md,Tyutin:1975qk}. The scale independence of the kernels of dressed mean fields and the vanishing of the beta function of the coupling 
in the maximally supersymmetric Yang-Mills theory at any number of loops suggest that the correlators of the dressed mean fields can be analyzed by the methods of conformal 
field theory in the position space, since the vanishing of the beta function is a consequence of the conformal symmetry of ${\cal N} = 4$ supersymmetric  Yang--Mills theory.

In the momentum space, by using unitarity methods, 
it has been demonstrated that only the diagrams created due to the ``rung rule'' contribute in the four-point gluon amplitude 
up to three-loop level \cite{Bern:2005iz}.  This ``rung rule,'' combined with conformal invariance of the ladder diagrams of Refs.~\cite{Usyukina:1992jd,Usyukina:1993ch,Broadhurst:1993ib} in the momentum space, allowed to classify all conformally invariant contributions in the momentum space up to four-loop level \cite{Drummond:2006rz,Bern:2006ew,Nguyen:2007ya}. 
The same conformal symmetry appears in Alday-Maldacena approach \cite{Alday:2007hr} to calculate gluon
scattering amplitude on the string side at strong coupling. Conformal invariant contributions to four-point gluon amplitude  reproduce the known results for the anomalous 
dimensions of twist-two operators in maximally supersymmetric Yang-Mills theory \cite{Kotikov:2000pm,Kotikov:2002ab,Kotikov:2003fb,Kotikov:2004er,Kotikov:2006ts}. 
Even if structure of amplitudes at the weak coupling can be found by using iterative formula of Ref.~\cite{Bern:2005iz}, the structure of off-shell correlators is still unknown. 
For some sub-class of diagrams it can be found exactly \cite{Isaev:2003tk}. The Slavnov-Taylor identity can be used as an alternative method for study the full structure 
of the off-shell correlators.

This manuscript is organized as follows. In Sec.~2 the gluon self-interaction vertex of diagram $(c)$ is considered. In Sec.~3 the integral structure and most important  
details of calculation are presented. In Subsec.~3.1 the basic ideas of the method are presented in more detail.
The result consists of three parts. Each part corresponds to a particular position of the derivative of the gluon vertex. In Sec.~4 the total result 
for the diagram is written. Appendices A, B, C and D contain further technical details.

\section{Diagram $(c)$}

\begin{figure}[ht]
\setlength{\unitlength}{2mm}
\begin{center}
\epsfig{file=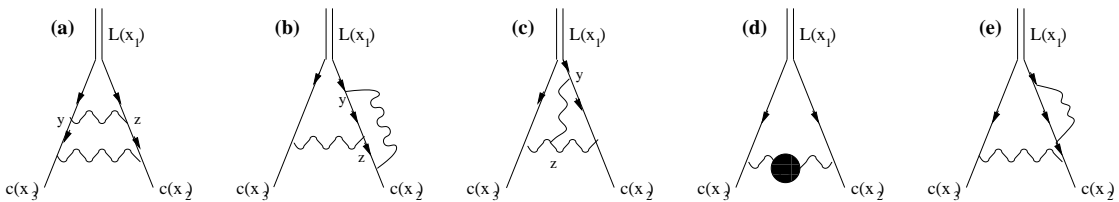, width=15.cm}

\end{center}
\vspace{0.0cm}
\caption{\footnotesize
The two-loop diagrams for the $Lcc$ vertex. The wavy lines represent the gluons, the straight lines the ghosts. Black disk stands for the total one-loop correction to the gluon propagator.}
\label{Lccfig}
\end{figure}

The planar two-loop correction  to $Lcc$ vertex can be represented as 
a combination of five diagrams depicted in Fig.~\ref{Lccfig}.
We introduce, for brevity, the notation $$[yz] = (y-z)^2, ~~~~ [y1] = (y-x_1)^2,....$$ 
and so on. We use the notation $\Pi_{\mu\nu}^{ab}(xy)$  for the gluon propagator
from point $x$ to point $y.$  In the exact four-dimesional  $D =4$ space it is
\begin{eqnarray*}
\Pi_{\mu\nu}^{ab}(xy) = \le\frac{g_{\mu\nu}}{[xy]} + 2\frac{(xy)_\mu (xy)_\nu}{[xy]^2}\ri\delta^{ab} \ .
\end{eqnarray*}
It satisfies the condition of transversality in $D = 4.$ From the Yang-Mills Lagrangian we have the following three-gluon vertex which is in the center of the diagram $c.$
\begin{eqnarray*}
\int~Dz~\le\pd_\mu A^A_\rho(z)\ri A^B_\mu(z) A^C_\rho(z) f^{ABC} \ .
\end{eqnarray*}
We work with $SU(N)$ group where the structure constants 
are completely antisymmetric in indices. This vertex creates the interaction for the diagram. In the rest of the paper we use three
triples of indices $(y,\lambda,a),$ $(2,\sigma,b),$ and $(3,\nu,c).$  Thus, for 
three-point gluon function we have the following contribution, which is in the center of
the diagram $c$:
\begin{eqnarray*}
\dis{\int~Dz \le\pd^{(z)}_\mu\Pi_{\rho\lambda}^{Aa}(zy)\ri\Pi_{\mu\sigma}^{Bb}(z2)\Pi_{\rho\nu}^{Cc}(z3)f^{ABC}
+ \int~Dz \le\pd^{(z)}_\mu\Pi_{\rho\lambda}^{Aa}(zy)\ri\Pi_{\mu\nu}^{Bc}(z3)\Pi_{\rho\sigma}^{Cb}(z2)f^{ABC}} \\
\dis{+ \int~Dz \le\pd^{(z)}_\mu\Pi_{\rho\sigma}^{Ab}(z2)\ri\Pi_{\mu\lambda}^{Ba}(zy)\Pi_{\rho\nu}^{Cc}(z3)f^{ABC}
+ \int~Dz \le\pd^{(z)}_\mu\Pi_{\rho\sigma}^{Ab}(z2)\ri\Pi_{\mu\nu}^{Bc}(z3)\Pi_{\rho\lambda}^{Ca}(zy)f^{ABC}} \\
\dis{+ \int~Dz \le\pd^{(z)}_\mu\Pi_{\rho\nu}^{Ac}(z3)\ri\Pi_{\mu\lambda}^{Ba}(zy)\Pi_{\rho\sigma}^{Cb}(z2)f^{ABC}
+ \int~Dz \le\pd^{(z)}_\mu\Pi_{\rho\nu}^{Ac}(z3)\ri\Pi_{\mu\sigma}^{Bb}(z2)\Pi_{\rho\lambda}^{Ca}(zy)f^{ABC}} \ .
 \end{eqnarray*}
Taking into account the antisymmetry of the indices, we obtain
\begin{eqnarray*}
\int~Dz \left[\le\pd^{(z)}_\mu\Pi_{\rho\lambda}(zy)\ri\Pi_{\mu\sigma}(z2)\Pi_{\rho\nu}(z3) - \le\pd^{(z)}_\mu\Pi_{\rho\lambda}(zy)\ri\Pi_{\mu\nu}(z3)\Pi_{\rho\sigma}(z2) \right] f^{abc} \\
+ \int~Dz \left[ - \le\pd^{(z)}_\mu\Pi_{\rho\sigma}(z2)\ri\Pi_{\mu\lambda}(zy)\Pi_{\rho\nu}(z3) + \le\pd^{(z)}_\mu\Pi_{\rho\sigma}(z2)\ri\Pi_{\mu\nu}(z3)\Pi_{\rho\lambda}(zy) \right]f^{abc} \\
+ \int~Dz \left[ \le\pd^{(z)}_\mu\Pi_{\rho\nu}(z3)\ri\Pi_{\mu\lambda}(zy)\Pi_{\rho\sigma}(z2) - \le\pd^{(z)}_\mu\Pi_{\rho\nu}(z3)\ri\Pi_{\mu\sigma}(z2)\Pi_{\rho\lambda}(zy) \right]f^{abc} \ .
 \end{eqnarray*}
We can re-organize the expression to the following form:
\begin{eqnarray}
\int~Dz \left[\le\pd^{(z)}_\mu\Pi_{\rho\lambda}(zy)\ri\Pi_{\mu\sigma}(z2)\Pi_{\rho\nu}(z3) - \le\pd^{(z)}_\mu\Pi_{\rho\nu}(z3)\ri\Pi_{\mu\sigma}(z2)\Pi_{\rho\lambda}(zy) \right]f^{abc} \no\\
+ \int~Dz \left[ - \le\pd^{(z)}_\mu\Pi_{\rho\sigma}(z2)\ri\Pi_{\mu\lambda}(zy)\Pi_{\rho\nu}(z3) + \le\pd^{(z)}_\mu\Pi_{\rho\nu}(z3)\ri\Pi_{\mu\lambda}(zy)\Pi_{\rho\sigma}(z2) \right]f^{abc}
\no\\
+ \int~Dz \left[\le\pd^{(z)}_\mu\Pi_{\rho\sigma}(z2)\ri\Pi_{\mu\nu}(z3)\Pi_{\rho\lambda}(zy) - \le\pd^{(z)}_\mu\Pi_{\rho\lambda}(zy)\ri\Pi_{\mu\nu}(z3)\Pi_{\rho\sigma}(z2) \right] f^{abc} =
\no\\
\no\\
\no\\
= \int~Dz \left[\le\pd^{(z)}_\mu\Pi_{\rho\lambda}(zy)\ri\Pi_{\rho\nu}(z3) - \le\pd^{(z)}_\mu\Pi_{\rho\nu}(z3)\ri\Pi_{\rho\lambda}(zy) \right]\Pi_{\mu\sigma}(z2)f^{abc} \no\\
+ \int~Dz \left[ - \le\pd^{(z)}_\mu\Pi_{\rho\sigma}(z2)\ri\Pi_{\rho\nu}(z3) + \le\pd^{(z)}_\mu\Pi_{\rho\nu}(z3)\ri\Pi_{\rho\sigma}(z2) \right]\Pi_{\mu\lambda}(zy)f^{abc} \no\\
+ \int~Dz \left[\le\pd^{(z)}_\mu\Pi_{\rho\sigma}(z2)\ri\Pi_{\rho\lambda}(zy) - \le\pd^{(z)}_\mu\Pi_{\rho\lambda}(zy)\ri\Pi_{\rho\sigma}(z2) \right] \Pi_{\mu\nu}(z3)f^{abc} 
\ .
\label{start} 
\end{eqnarray}

\section{Integral structure}

The idea of the calculation is simple. The ghost propagator differs slightly  from Dirac $\delta$-function. Actually, one derivative convoluted 
with the ghost propagator
immediately produces 
the $\delta$-function in the position space. Each vector propagator at least once has such a convolution with the ghost propagator. 
We can apply the integration by parts (IBP) for such a  convolution of two derivatives. The convolution can be represented as a difference of d'Alambertians. 
After this, the $\delta$-functions  appear. They will remove one of the 
integrations.

Our purpose is to reduce Lorentz algebra in 
the integrands in order to obtain {\em finite} double integrals with less number of indices, or single integrals that can be finite or singular. 
We use the following terminology. ``Finite integrals'' means that they have smooth limit in $\e,$ ``singular integrals'' means that they have poles in $\e.$
In paper \cite{Cvetic:2007fp} devoted to diagram $(b)$ we 
performed one of the integrations first by the uniqueness technique and 
then carried out the Lorentz algebra to scalarize the second 
single integration. Such an approach was justified there, since the number of Lorentz indices for the diagram with two vector propagators was relatively small.
In the diagram $(c)$ there is one three-gluon vertex.
Therefore, the number of Lorentz indices and the number of resulting double integrals 
are huge, and to keep 
the indices under control we apply IBP to reduce their number. 

We note by index counting that all three lines of Eq. (\ref{start}) are finite separately. No IR or UV poles should arise. 
UV behavior can be checked in the momentum space while IR behavior can be checked in the position space. In the UV region all subgraphs 
corresponding to each separate line of Eq. (\ref{start}) are finite 
(it does not matter where the derivative of the three-gluon vertex stands precisely). 

\subsection{The first line of Eq. (\ref{start})}

In this Subsection we analyze the first line of Eq. (\ref{start}). We have for the first line the following expression:
\begin{eqnarray}
T_1 \equiv \frac{(31)_\nu}{[31]^2}\int~Dy~\frac{(2y)_\sigma}{[2y]^2}\frac{(1y)_\L}{[1y]^{2}}
\int~Dz \le\pd^{(z)}_\mu\Pi_{\rho\nu}(z3)\ri\Pi_{\rho\lambda}(zy) \Pi_{\mu\sigma}(z2) \label{FL} 
\ .
\label{line1}
\end{eqnarray}
The integration here is performed in $D=4-2\e$ dimensions. 
All the notations are taken from Ref.~\cite{Cvetic:2006iu}.  We use simple algebra to represent integral $T_1$ as a 
combination of terms with simpler Lorentz structure,
\begin{eqnarray}
\Pi_{\rho\L}(zy) \frac{(1y)_\L}{[1y]^{2-\e}} =  \le \frac{g_{\rho\L}}{[yz]^{1-\e}} + 2(1-\e)\frac{(yz)_\rho (yz)_\L}{[yz]^{2-\e}}\ri \frac{(1y)_\L}{[1y]^{2-\e}} = \no\\
\le \frac{2g_{\rho\L}}{[yz]^{1-\e}} - \pd_\L^{(y)} \frac{(yz)_{\rho}}{[yz]^{1-\e}} \ri \frac{(1y)_\L}{[1y]^{2-\e}} =
\frac{2(1y)_\rho}{[yz]^{1-\e}[1y]^{2-\e}}  - \le\pd_\L^{(y)} \frac{(yz)_{\rho}}{[yz]^{1-\e}} \ri\frac{(1y)_\L}{[1y]^{2-\e}}  = \no\\
\frac{2(1y)_\rho}{[yz]^{1-\e}[1y]^{2-\e}}  - \frac{1}{2(1-\e)}\le\pd_\L^{(y)} \frac{(yz)_{\rho}}{[yz]^{1-\e}}\ri\le \pd_\L^{(y)} \frac{1}{[1y]^{1-\e}}\ri  = \no\\
\frac{2(1y)_\rho}{[yz]^{1-\e}[1y]^{2-\e}}  - \frac{1}{4(1-\e)}\left[ \pd^2_{(y)}\le \frac{(yz)_{\rho}}{[yz]^{1-\e}[1y]^{1-\e}}\ri 
- \le \pd^2_{(y)}\frac{(yz)_{\rho}}{[yz]^{1-\e}} \ri\frac{1}{[1y]^{1-\e}} \right. \no\\
\left. - \frac{(yz)_{\rho}}{[yz]^{1-\e}} \le \pd^2_{(y)} \frac{1}{[1y]^{1-\e}} \ri\right] 
\ .
\label{IBP1}
\end{eqnarray}
Now we use the formula of Ref.~\cite{Drummond:2006rz}\footnote{We have taken this formula from the first version of Ref. \cite{Drummond:2006rz}.}
\begin{eqnarray*}
\pd^2_{(y)} \frac{1}{[1y]^{1-\e}} = k(\e)\delta^{(4-2\e)}(1y)
\ ,
\end{eqnarray*}
where $k$ is some coefficient that we do not specify at this moment. Then, integral $T_1$ takes the form
\begin{eqnarray}
T_1 = \frac{(31)_\nu}{[31]^2}\int~Dy~\frac{(2y)_\sigma}{[2y]^{2-\e}}\frac{(1y)_\L}{[1y]^{2-\e}}\int~Dz \le\pd^{(z)}_\mu\Pi_{\rho\nu}(z3)\ri\Pi_{\rho\lambda}(zy) \Pi_{\mu\sigma}(z2) = \no\\
\frac{(31)_\nu}{[31]^2}\int~Dy~\frac{(2y)_\sigma}{[2y]^{2-\e}}\int~Dz \le\pd^{(z)}_\mu\Pi_{\rho\nu}(z3)\ri \Pi_{\mu\sigma}(z2) \times \no\\
\times\left\{\frac{2(1y)_\rho}{[yz]^{1-\e}[1y]^{2-\e}}  - \right. \no\\
- \left. \frac{1}{4(1-\e)}\left[ \pd^2_{(y)}\le \frac{(yz)_{\rho}}{[yz]^{1-\e}[1y]^{1-\e}} \ri  - \le \pd^2_{(y)}\frac{(yz)_{\rho}}{[yz]^{1-\e}} \ri
\frac{1}{[1y]^{1-\e}} - \frac{(yz)_{\rho}}{[yz]^{1-\e}} \le \pd^2_{(y)} \frac{1}{[1y]^{1-\e}} \ri\right]\right\} = \no\\
\frac{(31)_\nu}{[31]^2}\int~Dy~\frac{(2y)_\sigma}{[2y]^{2-\e}}\int~Dz \le\pd^{(z)}_\mu\Pi_{\rho\nu}(z3)\ri \Pi_{\mu\sigma}(z2)\times \no\\
\times
\left\{\frac{2(1y)_\rho}{[yz]^{1-\e}[1y]^{2-\e}}  + \frac{1}{4(1-\e)}\le \pd^2_{(y)}\frac{(yz)_{\rho}}{[yz]^{1-\e}} \ri \frac{1}{[1y]^{1-\e}} \right\} - \no\\
- \frac{1}{4(1-\e)}\frac{(31)_\nu}{[31]^2}\int~Dy~\frac{(2y)_\sigma}{[2y]^{2-\e}}\int~Dz \le\pd^{(z)}_\mu\Pi_{\rho\nu}(z3)\ri \Pi_{\mu\sigma}(z2) \pd^2_{(y)}
\frac{(yz)_{\rho}}{[yz]^{1-\e}[1y]^{1-\e}} +   \no\\
+ \frac{k}{4(1-\e)}\frac{(31)_\nu}{[31]^2}\int~Dy~\frac{(2y)_\sigma}{[2y]^{2-\e}}\int~Dz \le\pd^{(z)}_\mu\Pi_{\rho\nu}(z3)\ri \Pi_{\mu\sigma}(z2)
\frac{(yz)_{\rho}}{[yz]^{1-\e}}\delta^{(4-2\e)}(1y)  = \no\\
\frac{(31)_\nu}{[31]^2}\int~Dy~\frac{(2y)_\sigma}{[2y]^{2-\e}}\int~Dz \le\pd^{(z)}_\mu\Pi_{\rho\nu}(z3)\ri \Pi_{\mu\sigma}(z2)
\left\{\frac{2(1y)_\rho}{[yz]^{1-\e}[1y]^{2-\e}}  - \frac{(yz)_{\rho}}{[yz]^{2-\e}[1y]^{1-\e}} \right\} + \no\\
+ \frac{1}{8(1-\e)^2}\frac{(31)_\nu}{[31]^2}\pd_\sigma^{(2)}\int~Dy~\frac{1}{[2y]^{1-\e}}\int~Dz \le\pd^{(z)}_\mu\Pi_{\rho\nu}(z3)\ri \Pi_{\mu\sigma}(z2)
\pd^2_{(y)} \frac{(yz)_{\rho}}{[yz]^{1-\e}[1y]^{1-\e}}+   \no\\
+ \frac{k}{4(1-\e)}\frac{(31)_\nu}{[31]^2}\frac{(21)_\sigma}{[12]^{2-\e}}\int~Dz \le\pd^{(z)}_\mu\Pi_{\rho\nu}(z3)\ri \Pi_{\mu\sigma}(z2) \frac{(1z)_{\rho}}{[1z]^{1-\e}}  = \no\\
\frac{(31)_\nu}{[31]^2}\int~Dy~\frac{(2y)_\sigma}{[2y]^{2-\e}}\int~Dz \le\pd^{(z)}_\mu\Pi_{\rho\nu}(z3)\ri \Pi_{\mu\sigma}(z2)
\left\{\frac{2(1y)_\rho}{[yz]^{1-\e}[1y]^{2-\e}}  - \frac{(yz)_{\rho}}{[yz]^{2-\e}[1y]^{1-\e}} \right\} + \no\\
+ \frac{k}{8(1-\e)^2}\frac{(31)_\nu}{[31]^2}\pd_\sigma^{(2)}\int~Dy~\delta(2y)\int~Dz \le\pd^{(z)}_\mu\Pi_{\rho\nu}(z3)\ri \Pi_{\mu\sigma}(z2) \frac{(yz)_{\rho}}{[yz]^{1-\e}[1y]^{1-\e}}+
\no  \\
+ \frac{k}{4(1-\e)}\frac{(31)_\nu}{[31]^2}\frac{(21)_\sigma}{[12]^{2-\e}}\int~Dz \le\pd^{(z)}_\mu\Pi_{\rho\nu}(z3)\ri \Pi_{\mu\sigma}(z2) \frac{(1z)_{\rho}}{[1z]^{1-\e}}  = \no\\
\frac{(31)_\nu}{[31]^2}\int~Dy~\frac{(2y)_\sigma}{[2y]^{2-\e}}\int~Dz \le\pd^{(z)}_\mu\Pi_{\rho\nu}(z3)\ri \Pi_{\mu\sigma}(z2)
\left\{\frac{2(1y)_\rho}{[yz]^{1-\e}[1y]^{2-\e}}  - \frac{(yz)_{\rho}}{[yz]^{2-\e}[1y]^{1-\e}} \right\} + \no\\
+ \frac{k}{8(1-\e)^2}\frac{(31)_\nu}{[31]^2}\pd_\sigma^{(2)}\frac{1}{[12]^{1-\e}}\int~Dz \le\pd^{(z)}_\mu\Pi_{\rho\nu}(z3)\ri \Pi_{\mu\sigma}(z2) \frac{(2z)_{\rho}}{[2z]^{1-\e}} + \no\\
+ \frac{k}{4(1-\e)}\frac{(31)_\nu}{[31]^{2-\e}}\frac{(21)_\sigma}{[12]^{2-\e}}\int~Dz \le\pd^{(z)}_\mu\Pi_{\rho\nu}(z3)\ri \Pi_{\mu\sigma}(z2) \frac{(1z)_{\rho}}{[1z]^{1-\e}}  \no\\
\equiv  T_{11} + T_{12} + T_{13} 
\ .
\label{FLT}
\end{eqnarray}
It is simpler  to investigate singularities of single integrals than singularities of double integrals. Integrals $T_{12}$ and $T_{13}$ are finite. 
Since $T_1$ is finite, it means that $T_{11}$ is also finite. Moreover, the finiteness (UV and IR)  of $T_{11}$ can be checked by index counting directly. 
UV-finiteness in the momentum space 
can be checked for the Fourier transform 
of integral (\ref{FLT}) for each subgraph and for both momentum integrations together, 
according to BPHZ $R$-operation. 
IR-finiteness of the double integrals can be checked directly for integral (\ref{FLT}), for each integration separately and for 
both integrations together in the area
$|y| \rar \infty$, $|z| \rar \infty $, 
in complete analogy with $R$-operation of the momentum space.        
\begin{eqnarray*}
T_{11} = - \frac{(31)_\nu}{[31]^2}\int~Dy~\frac{(2y)_\sigma}{[2y]^{2-\e}}\int~Dz \Pi_{\rho\nu}(z3) \Pi_{\mu\sigma}(z2)\times \\
\times\left\{\frac{2(1y)_\rho}{[1y]^{2-\e}}\le \pd^{(z)}_\mu \frac{1}{[yz]^{1-\e}} \ri  - \frac{1}{[1y]^{1-\e}}\le\pd^{(z)}_\mu \frac{(yz)_{\rho}}{[yz]^{2-\e}} \ri \right\} \equiv 
T_{111} + T_{112}
\ .
\end{eqnarray*}
We calculate $T_{111}$  in detail in the main body of the paper, and 
the calculation of $T_{112}$ can be found in 
Appendix A.  The technique of calculation used in this paper is described 
in Ref.~\cite{Cvetic:2006iu}. It is based on the uniqueness method \cite{Vasiliev:1981dg,Vasil, Kazakov:1984bw,Kazakov:1984km}, 
and Gegenbauer polynomial technique introduced in 
Refs.~\cite{Tkachov:1981wb,Chetyrkin:1981qh,Chetyrkin:1980pr,Celmaster:1980ji,Terrano:1980af,Lampe:1982av} and further developed in Ref.~\cite{Kotikov:1995cw}.  
All the 
results are obtained in terms of the Davydychev integral $J(1,1,1)$ explicitly 
found in Ref.~\cite{Davydychev:1992xr}, 
and logarithms of ratios of the space-time differences 
of the coordinates of the effective fields in the position space. 
New integral representation for the Davydychev integral has been found in Ref.~\cite{Cvetic:2006iu}.

\subsubsection{Calculation of $T_{111}$ }

\begin{eqnarray*}
- T_{111} \equiv \frac{(31)_\nu}{[31]^2}\int~Dy~\frac{(2y)_\sigma}{[2y]^{2-\e}}\int~Dz \Pi_{\rho\nu}(z3) \Pi_{\mu\sigma}(z2)\frac{2(1y)_\rho}{[1y]^{2-\e}}
\le \pd^{(z)}_\mu \frac{1}{[yz]^{1-\e}} \ri 
\ .
\end{eqnarray*}
Again, we apply simple algebra to reduce the number of indices further
\begin{eqnarray}
\Pi_{\mu\sigma}(2z) \le\pd^{(z)}_\mu \frac{1}{[yz]^{1-\e}} \ri  =  \le \frac{g_{\mu\sigma}}{[2z]^{1-\e}}
+ 2(1-\e)\frac{(2z)_\mu (2z)_\sigma}{[2z]^{2-\e}}\ri \le\pd^{(z)}_\mu \frac{1}{[yz]^{1-\e}} \ri = \no\\
\le \frac{2g_{\mu\sigma}}{[2z]^{1-\e}} - \pd_\mu^{(2)} \frac{(2z)_{\sigma}}{[2z]^{1-\e}} \ri \le\pd^{(z)}_\mu \frac{1}{[yz]^{1-\e}} \ri = \no\\
\frac{4(1-\e)(yz)_\sigma}{[2z]^{1-\e}[yz]^{2-\e}}  + \le\pd_\mu^{(z)} \frac{(2z)_{\sigma}}{[2z]^{1-\e}}\ri\le\pd^{(z)}_\mu \frac{1}{[yz]^{1-\e}} \ri = \no\\
\frac{4(1-\e)(yz)_\sigma}{[2z]^{1-\e}[yz]^{2-\e}}   +  \frac{1}{2}\left[ \pd^2_{(z)}\le \frac{(2z)_{\sigma}}{[2z]^{1-\e}[yz]^{1-\e}} \ri
- \le \pd^2_{(z)}\frac{(2z)_{\sigma}}{[2z]^{1-\e}} \ri \frac{1}{[yz]^{1-\e}} \right. \no\\  
\left. - \frac{(2z)_{\sigma}}{[2z]^{1-\e}} \le \pd^2_{(z)} \frac{1}{[yz]^{1-\e}} \ri\right] 
\ .
\label{trick2}
\end{eqnarray}
The first contribution to $-T_{111}$  (with factor $2(1-\e)$) is
\begin{eqnarray*}
J_1 \equiv \frac{(31)_\nu}{[31]^2}\int~Dy~\frac{(2y)_\sigma}{[2y]^{2-\e}}\int~Dz \Pi_{\rho\nu}(z3)\frac{2(yz)_\sigma}{[2z]^{1-\e}[yz]^{2-\e}} \frac{2(1y)_\rho}{[1y]^{2-\e}}  = \\
-\frac{1}{2(1-\e)^2}\frac{(31)_\nu}{[31]^2}\int~Dy~Dz \Pi_{\rho\nu}(z3)\le\pd^{(y)}_\sigma \frac{1}{[2y]^{1-\e}} \ri
\le\pd^{(y)}_\sigma \frac{1}{[yz]^{1-\e}} \ri\frac{1}{[2z]^{1-\e}} \frac{2(1y)_\rho}{[1y]^{2-\e}}  = \\
\frac{1}{4(1-\e)^3}\frac{(31)_\nu}{[31]^2} \pd_\rho^{(1)}\int~Dy~Dz \Pi_{\rho\nu}(z3)\frac{1}{[2y]^{1-\e}[yz]^{1-\e}[2z]^{1-\e}} k \delta(1y)    \\
+ \frac{1}{4(1-\e)^2}\frac{(31)_\nu}{[31]^2}\int~Dy~Dz \Pi_{\rho\nu}(z3)\frac{1}{[2y]^{1-\e}} k\delta(yz)\frac{1}{[2z]^{1-\e}} \frac{2(1y)_\rho}{[1y]^{2-\e}}  \\
+ \frac{1}{4(1-\e)^2} \frac{(31)_\nu}{[31]^2}\int~Dy~Dz \Pi_{\rho\nu}(z3) k\delta(2y)\frac{1}{[yz]^{1-\e}}\frac{1}{[2z]^{1-\e}} \frac{2(1y)_\rho}{[1y]^{2-\e}}  = \\
\frac{k}{4(1-\e)^3}\frac{(31)_\nu}{[31]^2} \pd_\rho^{(1)}\int~Dz \Pi_{\rho\nu}(z3) \frac{1}{[21]^{1-\e}[1z]^{1-\e}[2z]^{1-\e}}     \\
+ \frac{k}{4(1-\e)^2}\frac{(31)_\nu}{[31]^2}\int~Dz \Pi_{\rho\nu}(z3)\frac{1}{[2z]^{2-2\e}} \frac{2(1z)_\rho}{[1z]^{2-\e}}  \\
+ \frac{k}{4(1-\e)^2} \frac{(31)_\nu}{[31]^2}\frac{2(12)_\rho}{[12]^{2-\e}}\int~Dz \Pi_{\rho\nu}(z3) \frac{1}{[2z]^{2-2\e}} = \\
\frac{k}{4}\frac{(31)_\nu}{[31]^2} \pd_\rho^{(1)}\frac{1}{[21]}\int~Dz \Pi_{\rho\nu}(z3) \frac{1}{[1z][2z]} \\
+ \frac{k}{4(1-\e)^2}\frac{(31)_\nu}{[31]^2}\int~Dz \Pi_{\rho\nu}(z3)\frac{1}{[2z]^{2-2\e}} \frac{2(1z)_\rho}{[1z]^{2-\e}}  \\
+ \frac{k}{4(1-\e)^2} \frac{(31)_\nu}{[31]^2}\frac{2(12)_\rho}{[12]^{2-\e}}\int~Dz \Pi_{\rho\nu}(z3) \frac{1}{[2z]^{2-2\e}} \equiv \\ 
\equiv \frac{k}{4}J_{12} + \frac{k}{4(1-\e)^2}M_{11}  + \frac{k}{4(1-\e)^2}M_{10} 
\ .
\end{eqnarray*}
Integral $J_1$ is singular in UV.  Integral $J_{12}$ is finite in UV and IR. Integrals $M_{11}$ and $M_{10}$ are singular. The second contribution (with factor is $1/2$) to $-T_{111}$ is 
\begin{eqnarray*}
J_2 \equiv \frac{(31)_\nu}{[31]^2}\int~Dy~\frac{(2y)_\sigma}{[2y]^{2-\e}}\int~Dz \Pi_{\rho\nu}(z3)
\pd^2_{(z)}\le \frac{(2z)_{\sigma}}{[2z]^{1-\e}[yz]^{1-\e}} \ri \frac{2(1y)_\rho}{[1y]^{2-\e}}  = \\
\frac{(31)_\nu}{[31]^2}\int~Dy~\frac{(2y)_\sigma}{[2y]^{2-\e}}\int~Dz~
\le \frac{2g_{\rho\nu}}{[3z]^{1-\e}} - \pd_\nu^{(3)} \frac{(3z)_{\rho}}{[3z]^{1-\e}} \ri
\pd^2_{(z)}\le \frac{(2z)_{\sigma}}{[2z]^{1-\e}[yz]^{1-\e}} \ri \frac{2(1y)_\rho}{[1y]^{2-\e}} = \\
\frac{2}{1-\e}\frac{(31)_\nu}{[31]^2}\pd_\nu^{(3)}\pd_\rho^{(1)}\pd_\rho^{(3)} \int~Dy~\frac{(2y)_\sigma}{[2y]^{2-\e}}
\int~Dz~\frac{(2z)_{\sigma}}{[2z]^{1-\e}[yz]^{1-\e}[1y]^{1-\e}[3z]^{1-\e}}\\
- \frac{(31)_\nu}{[31]^2}\pd_\nu^{(1)}\int~Dy~\frac{(2y)_\sigma}{[2y]^{2-\e}}\int~Dz~2k\delta(3z)
\frac{(2z)_{\sigma}}{[2z]^{1-\e}[yz]^{1-\e}[1y]^{1-\e}} = \\
2\frac{(31)_\nu}{[31]^2}\pd_\nu^{(3)}\pd_\rho^{(1)}\pd_\rho^{(3)} \int~Dy~\frac{(2y)_\sigma}{[2y]^{2-2\e}}
\int~Dz~\frac{(2z)_{\sigma}}{[2z]^{1-2\e}[yz][1y][3z]}\\
- 2k\frac{(31)_\nu}{[31]^2}\frac{(23)_{\sigma}}{[23]^{1-\e}}\pd_\nu^{(1)}\int~Dy~\frac{(2y)_\sigma}{[2y]^{2}[y3][1y]} 
\equiv 2J_{21} -2kJ_{22}
\ .
\end{eqnarray*}
We shifted exponents in powers by $\sim \e$ since $J_2,$ $J_{21}$ and $J_{22}$ are finite in IR and UV. We always apply this trick to finite integrals when we need to use 
the uniqueness relation
at least for one of the integrations (this is the trick of Ref.~\cite{Cvetic:2006iu}). The third contribution (with  factor $-1/2$) to $-T_{111}$ is 
\begin{eqnarray*}
J_3 \equiv \frac{(31)_\nu}{[31]^2}\int~Dy~\frac{(2y)_\sigma}{[2y]^{2-\e}}\int~Dz \Pi_{\rho\nu}(z3)
\le \pd^2_{(z)}\frac{(2z)_{\sigma}}{[2z]^{1-\e}} \ri \frac{1}{[yz]^{1-\e}}\frac{2(1y)_\rho}{[1y]^{2-\e}}  = \\
2\frac{(31)_\nu}{[31]^2}\int~Dy~\frac{(2y)_\sigma}{[2y]^{2-\e}}\int~Dz \Pi_{\rho\nu}(z3)
\le \pd_\sigma^{(2)} \frac{1}{[2z]^{1-\e}} \ri \frac{1}{[yz]^{1-\e}}\frac{2(1y)_\rho}{[1y]^{2-\e}}  = \\
-\frac{1}{1-\e}\frac{(31)_\nu}{[31]^2}\int~Dy~ \le \pd_\sigma^{(2)} \frac{1}{[2y]^{1-\e}} \ri \int~Dz \Pi_{\rho\nu}(z3)
\le \pd_\sigma^{(2)} \frac{1}{[2z]^{1-\e}} \ri \frac{1}{[yz]^{1-\e}}\frac{2(1y)_\rho}{[1y]^{2-\e}}  = \\
-\frac{1}{2(1-\e)}\frac{(31)_\nu}{[31]^2}  \pd_{(2)}^{2} \int~DyDz~\Pi_{\rho\nu}(z3)\frac{2(1y)_\rho}{[2y]^{1-\e}[2z]^{1-\e}[yz]^{1-\e}[1y]^{2-\e}}   \\
+ \frac{1}{2(1-\e)}\frac{(31)_\nu}{[31]^2} \int~DyDz~\Pi_{\rho\nu}(z3)k\delta(2y)\frac{2(1y)_\rho}{[2z]^{1-\e}[yz]^{1-\e}[1y]^{2-\e}}   \\
+ \frac{1}{2(1-\e)}\frac{(31)_\nu}{[31]^2} \int~DyDz~\Pi_{\rho\nu}(z3)k\delta(2z)\frac{2(1y)_\rho}{[2y]^{1-\e}[yz]^{1-\e}[1y]^{2-\e}}  = \\
-\frac{1}{2}\frac{(31)_\nu}{[31]^2}  \pd_{(2)}^{2} \int~DyDz~\Pi_{\rho\nu}(z3)\frac{2(1y)_\rho}{[2y][2z][yz][1y]^{2}}   \\
+ \frac{k}{2(1-\e)}\frac{(31)_\nu}{[31]^2} \frac{2(12)_\rho}{[12]^{2-\e}}\int~Dz~\Pi_{\rho\nu}(z3)\frac{1}{[2z]^{2-2\e}} \\
+ \frac{k}{2(1-\e)}\frac{(31)_\nu}{[31]^2} \Pi_{\rho\nu}(23)\int~Dy~\frac{2(1y)_\rho}{[2y]^{2-2\e}[1y]^{2-\e}} \equiv \\
 \equiv -\frac{1}{2}J_{31}   + \frac{k}{2(1-\e)}M_{10} + \frac{k}{2(1-\e)}M_{12}
\ .
\end{eqnarray*}
Integral $J_{3}$ is divergent in UV. Integral $J_{31}$ is finite, integrals $M_{10}$ and $M_{12}$  are singular. The fourth contribution (with factor $-1/2$) to $-T_{111}$ \\
\begin{eqnarray*}
J_4 \equiv \frac{(31)_\nu}{[31]^2}\int~Dy~\frac{(2y)_\sigma}{[2y]^{2-\e}}\int~Dz \Pi_{\rho\nu}(z3)
\frac{(2z)_{\rho}}{[2z]^{1-\e}} \le \pd^2_{(z)} \frac{1}{[yz]^{1-\e}} \ri \frac{2(1y)_\rho}{[1y]^{2-\e}}  = \\
\frac{(31)_\nu}{[31]^2}\int~Dy~\frac{(2y)_\sigma}{[2y]^{2-\e}}\int~Dz \Pi_{\rho\nu}(z3)\frac{(2z)_{\sigma}}{[2z]^{1-\e}} k\delta(yz)
\frac{2(1y)_\rho}{[1y]^{2-\e}}  = \\
k\frac{(31)_\nu}{[31]^2}\int~Dz~\frac{(2z)_\sigma}{[2z]^{2-\e}}\Pi_{\rho\nu}(z3)\frac{(2z)_{\sigma}}{[2z]^{1-\e}}
\frac{2(1z)_\rho}{[1z]^{2-\e}}  = \\
k\frac{(31)_\nu}{[31]^2}\int~Dz~\frac{1}{[2z]^{2-2\e}}\Pi_{\rho\nu}(z3)\frac{2(1z)_\rho}{[1z]^{2-\e}} = k M_{11}
\ .
\end{eqnarray*}
This single integral is singular. We collect and calculate all the singular integrals from $J_1,$ $J_3$ and $J_4$ together in order to demonstrate that poles 
finally disappear in the sum of all the singular integrals. The sum of all those singular integrals is
\begin{eqnarray*}
M_1 \equiv \frac{k}{2(1-\e)} M_{11} + \frac{k}{2(1-\e)} M_{10} - \frac{k}{4(1-\e)} M_{10} - \frac{k}{4(1-\e)} M_{12} - \frac{k}{2} M_{11} 
\ .
\end{eqnarray*}
For the beginning, we do the calculation of the first integral
\begin{eqnarray*}
\frac{(31)_\nu}{[31]^2}\int~Dz \Pi_{\rho\nu}(z3)\frac{1}{[2z]^{2-2\e}} \frac{2(1z)_\rho}{[1z]^{2-\e}}
\ .
\end{eqnarray*}
The shift of exponents in powers $1/[1z]^{2-\e} \mapsto 1/[1z]^{2+\e}$ 
will be performed in the last denominator of the integrand in order to use the uniqueness 
method for the case of three factors in the denominators.
Moreover, this factor $1/[1z]^{2-\e}$ is present in all other integrands. 
In other singular integrands we will shift that exponent too. 
The sum of all singular integrals must be finite. 
The shift of exponents in powers must be the same in all divergent integrands.
\begin{eqnarray*}
M_{11} = \frac{(31)_\nu}{[31]^2}\int~Dz \Pi_{\rho\nu}(z3)\frac{1}{[2z]^{2-2\e}} \frac{2(1z)_\rho}{[1z]^{2-\e}} \rar
\frac{(31)_\nu}{[31]^2}\int~Dz \Pi_{\rho\nu}(z3)\frac{1}{[2z]^{2-2\e}} \frac{2(1z)_\rho}{[1z]^{2+\e}} = \no\\
\frac{(31)_\nu}{[31]^2}\int~Dz \le \frac{g_{\rho\nu}}{[3z]^{1-\e}} + 2(1-\e)\frac{(3z)_\rho (3z)_\nu}{[3z]^{2-\e}}\ri \frac{1}{[2z]^{2-2\e}} \frac{2(1z)_\rho}{[1z]^{2+\e}} = \no\\
\frac{(31)_\nu}{[31]^2}\int~Dz \le \frac{2g_{\rho\nu}}{[3z]^{1-\e}} - \pd_\nu^{(3)} \frac{(3z)_{\rho}}{[3z]^{1-\e}} \ri \frac{1}{[2z]^{2-2\e}} \frac{2(1z)_\rho}{[1z]^{2+\e}} = \no\\
\frac{(31)_\nu}{[31]^2}\int~Dz \frac{4(1z)_\nu }{[3z]^{1-\e} [2z]^{2-2\e} [1z]^{2+\e}}
- \frac{(31)_\nu}{[31]^2}\pd_\nu^{(3)}  \int~Dz \frac{2(3z)_{\rho}(1z)_\rho} {[3z]^{1-\e}[2z]^{2-2\e} [1z]^{2+\e}} = \no\\
- \frac{2}{1+\e}\frac{(31)_\nu}{[31]^2}\pd_\nu^{(1)} \int~Dz \frac{1}{[3z]^{1-\e} [2z]^{2-2\e} [1z]^{1+\e}}
- \frac{(31)_\nu}{[31]^2}\pd_\nu^{(3)}  \int~Dz \frac{2(3z)_{\rho}(1z)_\rho} {[3z]^{1-\e}[2z]^{2-2\e} [1z]^{2+\e}} = \no\\
- \frac{2}{1+\e}\frac{(31)_\nu}{[31]^2}\pd_\nu^{(1)} J(1+\e,2-2\e,1-\e)  - \frac{(31)_\nu}{[31]^2}\pd_\nu^{(3)}  \int~Dz \frac{2(31)_{\rho}(1z)_\rho
+ 2(1z)_{\rho}(1z)_\rho} {[3z]^{1-\e}[2z]^{2-2\e} [1z]^{2+\e}} = \no\\
- \frac{2}{1+\e}\frac{(31)_\nu}{[31]^2}\pd_\nu^{(1)} J(1+\e,2-2\e,1-\e)
+ \frac{1}{1+\e}\frac{(31)_\nu}{[31]^2}\pd_\nu^{(3)}(31)_{\rho} \pd_\rho^{(1)}  \int~Dz \frac{1} {[3z]^{1-\e}[2z]^{2-2\e} [1z]^{1 +\e}}  \no \\
-  2\frac{(31)_\nu}{[31]^2}\pd_\nu^{(3)}  J(1+\e,2-2\e,1-\e)  = \no\\
- \frac{2}{1+\e}\frac{(31)_\nu}{[31]^2}\pd_\nu^{(1)} J(1+\e,2-2\e,1-\e) + \frac{1}{1+\e}\frac{(31)_\nu}{[31]^2}\pd_\nu^{(3)}(31)_{\rho} \pd_\rho^{(1)} J(1+\e,2-2\e,1-\e)   \no   \\
-  2\frac{(31)_\nu}{[31]^2}\pd_\nu^{(3)}  J(1+\e,2-2\e,1-\e)  = \no\\
\left[- \frac{2}{1+\e}\frac{(31)_\nu}{[31]^2}\pd_\nu^{(1)} + \frac{1}{1+\e}\frac{(31)_\nu}{[31]^2}\pd_\nu^{(3)}(31)_{\rho} \pd_\rho^{(1)}
-  2\frac{(31)_\nu}{[31]^2}\pd_\nu^{(3)}\right] \frac{A(1+\e,2-2\e,1-\e)}{[12][23]^{1-2\e}[31]^{\e}}, \no\\  
\\
\\
M_{10} \equiv \frac{(31)_\nu}{[31]^2}\frac{2(12)_\rho}{[12]^{2-\e}}\int~Dz \Pi_{\rho\nu}(z3) \frac{1}{[2z]^{2-2\e}} \rar
\frac{(31)_\nu}{[31]^2}\frac{2(12)_\rho}{[12]^{2+\e}}\int~Dz \Pi_{\rho\nu}(z3) \frac{1}{[2z]^{2-2\e}} =  \no\\
\frac{(31)_\nu}{[31]^2}\frac{2(12)_\rho}{[12]^{2+\e}}\int~Dz \le \frac{g_{\rho\nu}}{[3z]^{1-\e}} + 2(1-\e)\frac{(3z)_\nu (3z)_\rho}{[3z]^{2-\e}}\ri \frac{1}{[2z]^{2-2\e}} = \no\\
\frac{(31)_\nu}{[31]^2}\frac{2(12)_\rho}{[12]^{2+\e}}\int~Dz \le \frac{2g_{\rho\nu}}{[3z]^{1-\e}} - \pd_\nu^{(3)} \frac{(3z)_{\rho}}{[3z]^{1-\e}} \ri \frac{1}{[2z]^{2-2\e}} = \no\\
\frac{(31)_\nu}{[31]^2}\frac{4(12)_\nu}{[12]^{2+\e}}\int~Dz \frac{1}{[3z]^{1-\e}[2z]^{2-2\e}} -
\frac{(31)_\nu}{[31]^2}\frac{2(12)_\rho}{[12]^{2+\e}}\pd_\nu^{(3)} \int~Dz \frac{(3z)_{\rho}}{[3z]^{1-\e}}\frac{1}{[2z]^{2-2\e}} = \no\\
\frac{(31)_\nu}{[31]^2}\frac{4(12)_\nu}{[12]^{2+\e}}\frac{A(1-\e,2-2\e,1+\e)}{[23]^{1-2\e}} -
\frac{1}{2\e}\frac{(31)_\nu}{[31]^2}\frac{2(12)_\rho}{[12]^{2-\e}}\pd_\nu^{(3)}\pd_\rho^{(3)}\int~Dz \frac{1}{[3z]^{-\e}}\frac{1}{[2z]^{2-2\e}} = \no\\
\frac{(31)_\nu}{[31]^2}\frac{4(12)_\nu}{[12]^{2+\e}}\frac{A(1-\e,2-2\e,1+\e)}{[23]^{1-2\e}} -
\frac{1}{2\e}\frac{(31)_\nu}{[31]^2}\frac{2(12)_\rho}{[12]^{2+\e}}\pd_\nu^{(3)}\pd_\rho^{(3)}\frac{A(-\e,2-2\e,2+\e)}{[23]^{-2\e}} = \no\\
\frac{A(1-\e,1-2\e,1+\e)}{\e(1-2\e)}\frac{4(31)_\nu(12)_\nu}{[12]^{2+\e}[23]^{1-2\e}[31]^2}              \label{mumu4} \\
- \frac{A(1-\e,1-2\e,1+\e)}{4\e^2(1-2\e)(1+\e)}\frac{(31)_\nu}{[31]^2}\frac{2(12)_\rho}{[12]^{2+\e}}\pd_\nu^{(3)}\pd_\rho^{(3)}\frac{1}{[23]^{-2\e}}, \no\\
\\
\\
M_{12} \equiv \frac{(31)_\nu}{[31]^2} \Pi_{\rho\nu}(23)\int~Dy~\frac{2(1y)_\rho}{[2y]^{2-2\e}[1y]^{2-\e}} \rar
\frac{(31)_\nu}{[31]^2}\Pi_{\rho\nu}(23)\int~Dy~\frac{2(1y)_\rho}{[2y]^{2-2\e}[1y]^{2+\e}} = \\
- \frac{1}{1+\e}\frac{(31)_\nu}{[31]^2} \Pi_{\rho\nu}(23) \pd_\rho^{(1)}\int~Dy~\frac{1}{[2y]^{2-2\e}[1y]^{1+\e}} = \no\\
- \frac{1}{1+\e}\frac{(31)_\nu}{[31]^2} \Pi_{\rho\nu}(23) \pd_\rho^{(1)} \frac{A(1+\e,2-2\e,1-\e)}{[12]} =       \no\\
- \frac{A(1+\e,1-2\e,1-\e)}{\e(1+\e)(1-2\e)}\frac{(31)_\nu}{[31]^2} \Pi_{\rho\nu}(23) \pd_\rho^{(1)}\frac{1}{[12]}  \ . \no
\end{eqnarray*}
Working with all these expresions in a program written 
in {\it Mathematica}, 
it results that $M_1$ is not singular and in the limit $\e \rar 0$ 
\begin{eqnarray*}
M_1 \equiv k\left[\frac{-1/4}{[23]^2[31]^2} + \frac{-1}{[12][23][31]^2} + \frac{5/4}{[12]^2[31]^2} + \frac{1/2}{[12][23]^2[31]} + \frac{- 1}{[12]^2[23][31]} 
+ \frac{- 1/4}{[12]^2[23]^2}\right] \no\\
+ k\left[\frac{1/4}{[23]^2[31]^2} + \frac{1/4}{[12][23][31]^2} + \frac{-1/2}{[12]^2[31]^2}  + \frac{-1/2}{[12][23]^2[31]}  + \frac{1/4}{[12]^2[23][31]}  
+ \frac{1/4}{[12]^2[23]^2}\right] \ln[12]  \no\\  
+ k\left[\frac{-1/4}{[23]^2[31]^2} + \frac{-1/4}{[12][23][31]^2} + \frac{1/2}{[12]^2[31]^2}  + \frac{1/2}{[12][23]^2[31]}  + \frac{-1/4}{[12]^2[23][31]}  
+ \frac{-1/4}{[12]^2[23]^2}\right] \ln[23]
\ .
\end{eqnarray*}
Now we calculate double finite integrals. All together, the finite integrals from $J_1,$ $J_2,$ $J_3,$ and $J_4$  can be organized as
\begin{eqnarray*}
K_1 \equiv  \frac{k}{2}J_{12}  + \frac{1}{2}J_2 + \frac{1}{4}J_{31}
\ .
\end{eqnarray*}
The first double integral to calculate is
\begin{eqnarray*}
J_{31} \equiv \frac{(31)_\nu}{[31]^2}  \pd_{(2)}^{2} \int~DyDz~\Pi_{\rho\nu}(z3)\frac{2(1y)_\rho}{[2y][2z][yz][1y]^{2}} \rightarrow \no\\
\frac{(31)_\nu}{[31]^2} \pd^2_{(2)}\int~Dy~Dz \Pi_{\rho\nu}(z3) \frac{2(1y)_\rho}{[1y]^{2}[2z]^{1-2\e}[2y]^{1-2\e}[yz]} = \no\\
\frac{(31)_\nu}{[31]^2} \int~Dy~Dz \Pi_{\rho\nu}(z3) \frac{-4\e(1-2\e)2(1y)_\rho}{[1y]^{2}[2z]^{2-2\e}[2y]^{1-2\e}[yz]}  \no\\
+ \frac{(31)_\nu}{[31]^2} \int~Dy~Dz \Pi_{\rho\nu}(z3) \frac{-4\e(1-2\e)2(1y)_\rho}{[1y]^{2}[2z]^{1-2\e}[2y]^{2-2\e}[yz]}  \no\\
+ \frac{(31)_\nu}{[31]^2} \int~Dy~Dz \Pi_{\rho\nu}(z3) \frac{8 (1-2\e)^2 2(1y)_\rho (2z)_\L(2y)_\L}{[1y]^{2}[2z]^{2-2\e}[2y]^{2-2\e}[yz]} = \no\\
\frac{(31)_\nu}{[31]^2}\int~Dy~Dz \Pi_{\rho\nu}(z3) \frac{(4- 20\e + 24\e^2) 2(1y)_\rho}{[1y]^{2}[2z]^{1-2\e}[2y]^{2-2\e}[yz]}  \no\\
+ \frac{(31)_\nu}{[31]^2}\int~Dy~Dz \Pi_{\rho\nu}(z3) \frac{(4- 20\e + 24\e^2) 2(1y)_\rho}{[1y]^{2}[2z]^{2-2\e}[2y]^{1-2\e}[yz]}  \no\\
- \frac{(31)_\nu}{[31]^2}\int~Dy~Dz \Pi_{\rho\nu}(z3) \frac{4 (1-2\e)^2 2(1y)_\rho}{[1y]^{2}[2z]^{2-2\e}[2y]^{2-2\e}} \no \\
\equiv (4- 20\e + 24\e^2)J_{311} + (4- 20\e + 24\e^2)J_{312} - 4 (1-2\e)^2 J_{313}
\ .
\end{eqnarray*}
We present a set of integrals by using formulas of Ref.~\cite{Cvetic:2007fp},
\begin{eqnarray*}
J_{312} \equiv \frac{(31)_\nu}{[31]^2}\int~Dy~Dz \Pi_{\rho\nu}(z3) \frac{2(1y)_\rho}{[1y]^{2}[2z]^{2-2\e}[2y]^{1-2\e}[yz]}  \\
= \frac{(31)_\nu}{[31]^2}\int~Dy~Dz \le \frac{g_{\rho\nu}}{[3z]} + \frac{2}{1-2\e}\frac{(3z)_\rho (3z)_\nu}{[3z]^{2}}\ri \frac{2(1y)_\rho}{[1y]^{2}[2z]^{2-2\e}[2y]^{1-2\e}[yz]}  \\
= \frac{(31)_\nu}{[31]^2}\int~Dy~Dz \le \frac{2-2\e}{1-2\e}\frac{g_{\rho\nu}}{[3z]} - \frac{1}{1-2\e}\pd_\nu^{(3)}\frac{(3z)_\rho}{[3z]}\ri
\frac{2(1y)_\rho}{[1y]^{2}[2z]^{2-2\e}[2y]^{1-2\e}[yz]}  \\
= \frac{2-2\e}{1-2\e}\frac{(31)_\nu}{[31]^2}\int~Dy~Dz \frac{2(1y)_\nu}{[3z][1y]^{2}[2z]^{2-2\e}[2y]^{1-2\e}[yz]} \\
- \frac{1}{1-2\e} \frac{(31)_\nu}{[31]^2}\pd_\nu^{(3)}\int~Dy~Dz \frac{2(3z)_\rho(1y)_\rho}{[1y]^{2}[2z]^{2-2\e}[2y]^{1-2\e}[yz][3z]} = \\
= - \frac{(2-2\e)A(2-2\e,1,1)}{1-2\e}\left[\frac{1}{[31][23]^{1-\e}}J(2,2-3\e,\e)  + \frac{1}{[31]^2[23]^{1-\e}}J(1,2-3\e,\e) \right.\\
- \left. \frac{1}{[31]^2[23]^{1-\e}}  J(2,2-3\e,\e-1) \right] \\
- \frac{1}{1-2\e} \frac{(31)_\nu}{[31]^2}\pd_\nu^{(3)}\left[\int~Dy~Dz \frac{[3z]+[3y]-[yz]}{[1y]^{2}[2z]^{2-2\e}[2y]^{1-2\e}[yz][3z]} \right.\\
+ \left. \int~Dy~Dz \frac{2(3z)_\rho(13)_\rho}{[1y]^{2}[2z]^{2-2\e}[2y]^{1-2\e}[yz][3z]} \right]  \\
= - \frac{(2-2\e)A(2-2\e,1,1)}{1-2\e}\left[\frac{1}{[31][23]^{1-\e}}J(2,2-3\e,\e)  + \frac{1}{[31]^2[23]^{1-\e}}J(1,2-3\e,\e) \right.\\
- \left. \frac{1}{[31]^2[23]^{1-\e}}  J(2,2-3\e,\e-1) \right] \\
- \frac{A(1,1,2-2\e)}{1-2\e} \frac{(31)_\nu}{[31]^2}\pd_\nu^{(3)}\left[\frac{1}{[23]^{1-\e}}J(2,2-3\e,\e-1) - \frac{1}{[23]^{1-\e}}J(2,1-2\e,0)\right] \\
+ \frac{A(1,2,1-2\e)}{1-2\e} \frac{(31)_\nu}{[31]^2}\pd_\nu^{(3)}\left[\frac{[13]}{[12]^{1-\e}}J(1+\e,2-3\e,1) - \frac{1}{[12]^{1-\e}}J(\e,2-3\e,1)\right],
\\
\\
J_{311} \equiv \frac{(31)_\nu}{[31]^2}\int~Dy~Dz \Pi_{\rho\nu}(z3) \frac{2(1y)_\rho}{[1y]^{2}[2z]^{1-2\e}[2y]^{2-2\e}[yz]} = \\
\frac{(31)_\nu}{[31]^2}\int~Dy~Dz \le \frac{2-2\e}{1-2\e}\frac{g_{\rho\nu}}{[3z]} - \frac{1}{1-2\e}\pd_\nu^{(3)}\frac{(3z)_\rho}{[3z]}\ri
\frac{2(1y)_\rho}{[1y]^{2}[2z]^{1-2\e}[2y]^{2-2\e}[yz]} = \\
\frac{2-2\e}{1-2\e} \frac{(31)_\nu}{[31]^2}\int~Dy~Dz \frac{2(1y)_\nu}{[1y]^{2}[2z]^{1-2\e}[2y]^{2-2\e}[yz][3z]} \\
- \frac{1}{1-2\e}\frac{(31)_\nu}{[31]^2}\pd_\nu^{(3)} \int~Dy~Dz \frac{2(3z)_\rho(1y)_\rho}{[1y]^{2}[2z]^{1-2\e}[2y]^{2-2\e}[yz][3z]} = \\
\frac{2-2\e}{1-2\e} \frac{(31)_\nu}{[31]^2}\int~Dy~Dz \frac{2(1y)_\nu}{[1y]^{2}[2z]^{1-2\e}[2y]^{2-2\e}[yz][3z]} \\
- \frac{1}{1-2\e}\frac{(31)_\nu}{[31]^2}\pd_\nu^{(3)} \int~Dy~Dz \frac{2(3z)_\rho(3y)_\rho}{[1y]^{2}[2z]^{1-2\e}[2y]^{2-2\e}[yz][3z]}  \\
- \frac{1}{1-2\e}\frac{(31)_\nu}{[31]^2}\pd_\nu^{(3)} \int~Dy~Dz \frac{2(3z)_\rho(13)_\rho}{[1y]^{2}[2z]^{1-2\e}[2y]^{2-2\e}[yz][3z]} = \\
\\
- \frac{2-2\e}{1-2\e} \frac{(31)_\nu}{[31]^2} \pd_\nu^{(1)}\frac{A(1,1,2-2\e)}{[12]^{1-\e}}J(\e,2-3\e,1) \\
- \frac{A(1,1,2-2\e)}{1-2\e}\frac{(31)_\nu}{[31]^2}\pd_\nu^{(3)} \Bigr[\le 1+\e-(31)_\nu\pd_\nu^{(1)}) \ri \frac{1}{[12]^{1-\e}}J(\e,2-3\e,1) \Bigl.\\
+ \Bigl.  \frac{1-\e}{[12]^{2-\e}}J(\e-1,2-3\e,1)  - \frac{1-\e}{[12]^{2-\e}}J(\e,1-3\e,1) \Bigr]
\ .
\end{eqnarray*}
The following formulas derived by IBP are necessary to obtain the above representation,
\begin{eqnarray}
J(\e-1,2-3\e,1) = (2-3\e)\Bigl[J(\e-1,3-3\e,0)-[23]J(\e-1,3-3\e,1)\Bigr] \no\\
+ (\e-1)\Bigl[J(\e,2-3\e,0)- [13]J(\e,2-3\e,1)\Bigr] \no\\
J(\e,1-3\e,1) = (1-3\e)\Bigl[J(\e,2-3\e,0)-[23]J(\e,2-3\e,1)\Bigr] \no\\
+ \e\Bigl[J(1+\e,1-3\e,0)- [13]J(1+\e,1-3\e,1)\Bigr] 
\ .
\label{mumu9}
\end{eqnarray}
The third integral in $J_{31}$ is simple,
\begin{eqnarray*}
J_{313} \equiv \frac{(31)_\nu}{[31]^2}\int~Dy~Dz \Pi_{\rho\nu}(z3) \frac{2(1y)_\rho}{[1y]^{2}[2z]^{2-2\e}[2y]^{2-2\e}} = \\
- \frac{(31)_\nu}{[31]^2} \pd_\rho^{(1)}  \frac{A(1,2-2\e,1)}{[12]^{1-\e}}  \int~Dz \Pi_{\rho\nu}(z3) \frac{1}{[2z]^{2-2\e}} = \\
-\frac{(31)_\nu}{[31]^2}\pd_\rho^{(1)}\frac{A(1,2-2\e,1)}{[12]^{1-\e}}\left[ \frac{2-2\e}{1-2\e} g_{\rho\nu} J(0,2-2\e,1) + \frac{1}{1-2\e}  \pd_\nu^{(3)} (23)_\rho J(0,2-2\e,1) \right.\\
+ \left. \frac{1}{2(1-2\e)^2}\pd_\nu^{(3)}\pd_\rho^{(2)}J(0,1-2\e,1) \right]
\ .
\end{eqnarray*}
All these formulas can be programmed in {\it Mathematica} and the result for $J_{31}$ is
\begin{eqnarray*}
J_{31} = \left[\frac{-8}{[12][23]^2} + \frac{-4}{[12]^2[31]} + \frac{4}{[23]^2[31]} + \frac{4[31]}{[12]^2[23]^2}\right]J[1,1,1] \\
+ \left[\frac{-4}{[12]^2[23]^2} + \frac{4}{[12][23]^2[31]} + \frac{4}{[12]^2[23][31]} \right]\ln{[12]} + \frac{-8}{[12]^2[23][31]} \ln{[23]} \\
+ \left[\frac{4}{[12]^2[23]^2} + \frac{-4}{[12][23]^2[31]} + \frac{4}{[12]^2[23][31]} \right]\ln{[31]}
\ .
\end{eqnarray*}
The second finite double integral to calculate is
\begin{eqnarray*}
2J_{21} \equiv 2\frac{(31)_\nu}{[31]^2}\pd_\nu^{(3)}\pd_\rho^{(1)}\pd_\rho^{(3)} \int~Dy~Dz\frac{(2y)_\sigma (2z)_{\sigma} }{[2y]^{2-2\e}[2z]^{1-2\e}[yz][1y][3z]} 
\ .
\end{eqnarray*}
This is seen as a differential operator applied to the integral
\begin{eqnarray*}
2 \int~Dy~Dz\frac{(2y)_\sigma(2z)_\sigma}{[2y]^{2-2\e}[2z]^{1-2\e}[yz][1y][3z]} = \no \\
 \int~Dy~Dz\frac{1}{[2y]^{1-2\e}[2z]^{1-2\e}[yz][1y][3z]} +  \int~Dy~Dz\frac{1}{[2y]^{2-2\e}[2z]^{-2\e}[yz][1y][3z]} \no \\
-  \int~Dy~Dz\frac{1}{[2y]^{2-2\e}[2z]^{1-2\e}[1y][3z]} \equiv  I_1 + I_2 - I_3 \no
\ .
\end{eqnarray*}
$I_1$ is finite, and the difference $I_2 - I_3$ must give a finite result:
\begin{eqnarray*}
I_2 - I_3 = \frac{A(1,1,1-2\e)}{\e(1-2\e)}\frac{1}{[12]^{1-\e}}J(\e,1-3\e,1)  + \frac{A^2(1,1,1-2\e)}{\e^2(1-2\e)}\frac{1}{[12]^{1-\e}[23]^{-\e}}
\ .
\end{eqnarray*}
Taking into account this relation, Eq. (\ref{mumu9}), and the relation 
\begin{eqnarray}
\pd_\rho^{(1)}\pd_\rho^{(3)} I_1 = \int~Dy~Dz\frac{4(1y)_\rho(3z)_\rho}{[2y]^{1-2\e}[2z]^{1-2\e}[yz][1y]^2[3z]^2} = \no\\
2 A(1,1-2\e,2)\left[- \frac{[31]}{[12]^{1-\e}}J(1+\e,1-3\e,2) + \frac{1}{[12]^{1-\e}}J(\e,1-3\e,2) \right.\no\\
\left. + \frac{1}{[23]^{1-\e}}J(2,1-3\e,\e) - \frac{1}{[12]^{1-\e}}J(0,1-3\e,2) \right] \equiv I_4 \no
\ ,
\end{eqnarray}
we obtain by {\it Mathematica}
\begin{eqnarray*}
2J_{21} = \frac{(31)_\nu}{[31]^2}\pd_\nu^{(3)}\left[ \pd_\rho^{(1)}\pd_\rho^{(3)} \le I_2 - I_3 \ri    + I_4 \right] =  \\
\left[\frac{-2}{[12]^2[23]^2} + \frac{-2}{[12]^2[31]^2}  + \frac{- 2}{[23]^2[31]^2} + \frac{-4}{[12][23][31]^2} + \frac{4}{[12][23]^2[31]}  + \frac{4}{[12]^2[23][31]} \right] \\
+ \left[\frac{2}{[12][23]^2} + \frac{-2}{[12][31]^2} + \frac{2[12]}{[23]^2[31]^2} + \frac{-4}{[23][31]^2}  + \frac{-4}{[23]^2[31]} \right]J(1,1,1) \\
+ \left[ \frac{2}{[23]^2[31]^2} + \frac{2}{[12][23][31]^2} + \frac{-2}{[12][23]^2[31]}  \right]\ln{[12]} \\
+ \left[\frac{2}{[12]^2[23]^2} + \frac{2}{[23]^2[31]^2} + \frac{-2}{[12][23][31]^2} + \frac{-4}{[12][23]^2[31]}   + \frac{-2}{[12]^2[23][31]}   \right]\ln{[23]} \\
+ \left[\frac{-2}{[12]^2[23]^2} + \frac{-4}{[23]^2[31]^2} + \frac{6}{[12][23]^2[31]} + \frac{2}{[12]^2[23][31]}  \right]\ln{[31]}
\ .
\end{eqnarray*}
The first finite single integral is
\begin{eqnarray*}
J_{12} = \frac{(31)_\nu}{[31]^2} \pd_\rho^{(1)}\frac{1}{[21]}\int~Dz \Pi_{\rho\nu}(z3) \frac{1}{[1z][2z]} = \\ 
\left[\frac{2}{[12][31]^2} + \frac{-2[23]}{[12]^2[31]^2} + \frac{2}{[12]^2[31]} \right] J(1,1,1) \\
+ \left[\frac{1}{[12]^2[31]^2} + \frac{1}{[12][23][31]^2} +    \frac{-1}{[12]^2[23][31]}\right] \ln{[12]} + \frac{-2}{[12]^2[31]^2} \ln{[23]} \\ 
+ \left[\frac{1}{[12]^2[31]^2} + \frac{-1}{[12][23][31]^2} +    \frac{1}{[12]^2[23][31]}\right] \ln{[31]}
\ .
 \end{eqnarray*}
The second single finite integral is
\begin{eqnarray*}
2J_{22} = 2\frac{(31)_\nu}{[31]^2}\frac{(23)_{\sigma}}{[23]^{1-\e}}\pd_\nu^{(1)}\int~Dy~\frac{(2y)_\sigma}{[2y]^{2}[y3][1y]} =  \frac{2}{[12][23][31]^2}
+ \frac{1}{[23][31]^2} J(1,1,1)  \\
+ \left[\frac{-1}{[12]^2[31]^2} + \frac{1}{[12]^2[23][31]}\right] \ln{[23]} + \left[\frac{1}{[12]^2[31]^2} + \frac{-1}{[12]^2[23][31]}\right] \ln{[31]}
\ .
\end{eqnarray*}
Taking into account the contribution of singular integrals, $M_1$, we obtain
\begin{eqnarray}
- T_{111} = K_1 + M_1 = \no\\
k\left[\frac{-1/4}{[12]^2[23]^2} +  \frac{5/4}{[12]^2[31]^2} +
\frac{-1/4}{[23]^2[31]^2} + \frac{-2}{[12][23][31]^2}  + \frac{1/2}{[12][23]^2[31]} +  \frac{-1}{[12]^2[23][31]}     \right] \no\\
+ k\left[\frac{1}{[12][31]^2} +  \frac{-1/2}{[23][31]^2}  + \frac{-[23]}{[12]^2[31]^2} +\frac{1}{[12]^2[31]}   \right]J(1,1,1) \no\\
+ k\left[ \frac{1/4}{[12]^2[23]^2} + \frac{1/4}{[23]^2[31]^2} + \frac{3/4}{[12][23][31]^2}  + \frac{-1/2}{[12][23]^2[31]}  + \frac{-1/4}{[12]^2[23][31]} \right]\ln{[12]} \no\\
+ k\left[\frac{-1/4}{[12]^2[23]^2} + \frac{-1/4}{[23]^2[31]^2}   + \frac{-1/4}{[12][23][31]^2} + \frac{1/2}{[12][23]^2[31]}  + \frac{-3/4}{[12]^2[23][31]} \right]\ln{[23]} \no\\
+ k\left[\frac{-1/2}{[12][23][31]^2} + \frac{1}{[12]^2[23][31]}\right]\ln{[31]} \no\\
\no\\
+ \left[\frac{-1}{[12]^2[23]^2} + \frac{-1}{[12]^2[31]^2}  + \frac{-1}{[23]^2[31]^2} +  \frac{-2}{[12][23][31]^2} + \frac{2}{[12][23]^2[31]} + \frac{2}{[12]^2[23][31]} \right] \no\\
+ \left[\frac{-1}{[12][23]^2} + \frac{-1}{[12][31]^2} + \frac{[12]}{[23]^2[31]^2} + \frac{-2}{[23][31]^2}  \right.\no\\
\left. +  \frac{-1}{[12]^2[31]} + \frac{-1}{[23]^2[31]}  + \frac{[31]}{[12]^2[23]^2} \right]J(1,1,1) \no\\
+ \left[ \frac{-1}{[12]^2[23]^2} + \frac{1}{[23]^2[31]^2} + \frac{1}{[12][23][31]^2}  + \frac{1}{[12]^2[23][31]} \right]\ln{[12]} \no\\
+ \left[\frac{1}{[12]^2[23]^2} + \frac{1}{[23]^2[31]^2} + \frac{-1}{[12][23][31]^2} +  \frac{-2}{[12][23]^2[31]}  + \frac{-3}{[12]^2[23][31]}\right]\ln{[23]} \no\\
+ \left[\frac{-2}{[23]^2[31]^2} + \frac{2}{[12][23]^2[31]} + \frac{2}{[12]^2[23][31]}  \right]\ln{[31]} 
\ .
\label{FLT2.1}
\end{eqnarray}

\subsubsection{Calculation of $T_{112}$}

\begin{eqnarray} 
T_{112} = - \frac{(31)_\nu}{[31]^2}\int~Dy~\frac{(2y)_\sigma}{[2y]^{2-\e}}\int~Dz \le\pd^{(z)}_\mu\Pi_{\rho\nu}(z3)\ri \Pi_{\mu\sigma}(z2)\frac{(yz)_{\rho}}{[yz]^{2-\e}[1y]^{1-\e}} = \no\\
k\left[\frac{-1/16}{[12]^2[31]^2}  + \frac{1/16}{[12]^2[31]^2}  + \frac{3/16}{[23]^2[31]^2} + \frac{-1/2}{[12][23][31]^2}   +  \frac{-1/8}{[12][23]^2[31]} \right] \no \\
+ k\left[\frac{1/4}{[12][31]^2}  +  \frac{-1/8[12]}{[23]^2[31]^2}   +  \frac{1/8}{[23]^2[31]} \right]J(1,1,1) \no\\
+ k\left[\frac{1/4}{[23]^2[31]^2} +  \frac{3/8}{[12][23][31]^2} +  \frac{1/8}{[12][23]^2[31]}   \right]\ln[12] \no\\
+ k \left[\frac{-1/16}{[12]^2[23]^2} + \frac{-3/8}{[12]^2[31]^2} + \frac{-5/16}{[23]^2[31]^2}   + \frac{-1/16}{[12][23][31]^2}  \right. \no\\
+ \left. \frac{3/8}{[12][23]^2[31]}  + \frac{-1/16}{[12]^2[23][31]} \right] \ln[23] \no \\
+ k \left[\frac{1/16}{[12]^2[23]^2} + \frac{3/8}{[12]^2[31]^2} + \frac{1/16}{[23]^2[31]^2}   + \frac{-5/16}{[12][23][31]^2}   \right.\no\\
+ \left.  \frac{-1/2}{[12][23]^2[31]} + \frac{1/16}{[12]^2[23][31]} \right] \ln[31] \no \\
\no \\
+ \left[\frac{1/2}{[12]^2[23]^2} + \frac{-1/2}{[12][23]^2[31]}  + \frac{-1/2}{[12]^2[23][31]}   \right]  + \left[\frac{-1}{[12][23]^2} + \frac{1}{[23]^2[31]} \right]J(1,1,1) \no\\
+ \left[\frac{-1/2}{[12]^2[23]^2} + \frac{1/2}{[12][23]^2[31]} +  \frac{-1/2}{[12]^2[23][31]} \right]\ln[23] \no\\
+ \left[\frac{1/2}{[12]^2[23]^2} + \frac{-1/2}{[12][23]^2[31]} +  \frac{1/2}{[12]^2[23][31]} \right] \ln[31]   
\ .
\label{FLT2.2}
\end{eqnarray}
 Details of this calculation are explained in Appendix A.

\subsubsection{Calculation of $T_{12}$  and $T_{13}$}
Eq. (\ref{FLT}) for $T_1$ contains two finite single integrals, $T_{12}$ and $T_{13}$:
\begin{eqnarray}
T_{13} \equiv \frac{k}{4}\frac{(31)_\nu}{[31]^{2}}\frac{(21)_\sigma}{[12]^{2}}\int~Dz \le\pd^{(z)}_\mu\Pi_{\rho\nu}(z3)\ri \Pi_{\mu\sigma}(z2) \frac{(1z)_{\rho}}{[1z]^{1-\e}} = \no\\
k\left[\frac{1/4}{[12]^2[23]^2} + \frac{1/4}{[12]^2[31]^2}+ \frac{1/4}{[23]^2[31]^2} + \frac{-1/2}{[12][23]^2[31]} \right] \no\\
+ k\left[ \frac{1/4}{[12][31]^2}  +   \frac{-1/2[23]}{[12]^2[31]^2} + \frac{1/4}{[12]^2[31]}   \right]J(1,1,1) \no\\
+ k\left[ \frac{1/8}{[12]^2[23]^2} + \frac{-1/2}{[12]^2[31]^2} + \frac{-1/8}{[23]^2[31]^2} + \frac{-1/8}{[12][23][31]^2} + \frac{1/8}{[12]^2[23][31]} \right]\ln{[12]} \no\\
+ k\left[\frac{1}{[12]^2[31]^2} \right]\ln{[23]} \no\\
+ k\left[ \frac{-1/8}{[12]^2[23]^2} + \frac{-1/2}{[12]^2[31]^2} + \frac{1/8}{[23]^2[31]^2} + \frac{1/8}{[12][23][31]^2}+ \frac{-1/8}{[12]^2[23][31]} \right]\ln{[31]} 
\ .
\label{FLT3}
\end{eqnarray}
The second finite integral in Eq. (\ref{FLT}) is
\begin{eqnarray}
T_{12} \equiv \frac{k}{8(1-\e)^2}\frac{(31)_\nu}{[31]^2}\pd_\sigma^{(2)}\frac{1}{[12]^{1-\e}}\int~Dz \le\pd^{(z)}_\mu\Pi_{\rho\nu}(z3)\ri \Pi_{\mu\sigma}(z2) 
\frac{(2z)_{\rho}}{[2z]^{1-\e}} = \no\\
k\left[\frac{-1/8}{[12]^2[23]^2} + \frac{1/4}{[12]^2[31]^2}  +\frac{-1/8}{[23]^2[31]^2}  + \frac{-1/8}{[12][23][31]^2}   +  \frac{1/4}{[12][23]^2[31]} +  \frac{-1/8}{[12]^2[23][31]}  \right]
\ .
\label{FLT4}
\end{eqnarray}

\subsubsection{Total result for $T_1$}

Combining Eqs. (\ref{FLT2.1}), (\ref{FLT2.2}), (\ref{FLT3}) and (\ref{FLT4}), 
we obtain for $T_1$
\begin{eqnarray*}
T_1 = T_{111} + T_{112} + T_{12} + T_{13} = \\
 k\left[\frac{5/16}{[12]^2[23]^2} +  \frac{-11/16}{[12]^2[31]^2} + \frac{9/16}{[23]^2[31]^2} + \frac{11/8}{[12][23][31]^2}  
+ \frac{-7/8}{[12][23]^2[31]} +  \frac{7/8}{[12]^2[23][31]}  \right] \no\\
+ k\left[\frac{-1/2}{[12][31]^2} +  \frac{-1/8[12]}{[23]^2[31]^2}  +  \frac{1/2}{[23][31]^2}    + \frac{1/2[23]}{[12]^2[31]^2} + \frac{-3/4}{[12]^2[31]}  + \frac{1/8}{[23]^2[31]}  
\right]J(1,1,1) \no\\
+ k\left[ \frac{-1/8}{[12]^2[23]^2} + \frac{-1/2}{[12]^2[23]^2} + \frac{-1/8}{[23]^2[31]^2} + \frac{-1/2}{[12][23][31]^2}  + \frac{5/8}{[12][23]^2[31]}  
+ \frac{3/8}{[12]^2[23][31]} \right]\ln{[12]} \no\\
+ k\left[\frac{3/16}{[12]^2[23]^2}  + \frac{5/8}{[12]^2[31]^2}  + \frac{-1/16}{[23]^2[31]^2}   + \frac{3/16}{[12][23][31]^2} + \frac{-1/8}{[12][23]^2[31]}  
+ \frac{11/16}{[12]^2[23][31]} \right]\ln{[23]} \no\\
+ k\left[\frac{-1/16}{[12]^2[23]^2}  + \frac{-1/8}{[12]^2[31]^2}  + \frac{3/16}{[23]^2[31]^2}   + \frac{5/16}{[12][23][31]^2} + \frac{-1/2}{[12][23]^2[31]}  
+ \frac{-17/16}{[12]^2[23][31]} \right]\ln{[31]} \no\\
\no\\
+ \left[\frac{3/2}{[12]^2[23]^2} + \frac{1}{[12]^2[31]^2}  + \frac{1}{[23]^2[31]^2} +  \frac{2}{[12][23][31]^2} + \frac{-5/2}{[12][23]^2[31]} + \frac{-5/2}{[12]^2[23][31]} \right] \no\\
+ \left[\frac{1}{[12][31]^2} + \frac{-[12]}{[23]^2[31]^2} + \frac{2}{[23][31]^2}  +  \frac{1}{[12]^2[31]} + \frac{2}{[23]^2[31]}  + \frac{-[31]}{[12]^2[23]^2} \right]J(1,1,1) \no\\
+ \left[ \frac{1}{[12]^2[23]^2} + \frac{-1}{[23]^2[31]^2} + \frac{-1}{[12][23][31]^2}  + \frac{-1}{[12]^2[23][31]} \right]\ln{[12]} \no\\
+ \left[\frac{-3/2}{[12]^2[23]^2} + \frac{-1}{[23]^2[31]^2} + \frac{1}{[12][23][31]^2} +  \frac{5/2}{[12][23]^2[31]}  + \frac{5/2}{[12]^2[23][31]}\right]\ln{[23]} \no\\
+ \left[\frac{1/2}{[12]^2[23]^2} +     \frac{2}{[23]^2[31]^2} + \frac{-5/2}{[12][23]^2[31]} + \frac{-3/2}{[12]^2[23][31]}  \right]\ln{[31]}
\ .
\end{eqnarray*}
As one can see from Eq. (\ref{start}), $T_1$ contributes to the full result $T$ of diagram $(c)$ with factor (-2).

\subsection{The second line of Eq. (\ref{start})}

\begin{eqnarray}
T_2 \equiv \frac{(31)_\nu}{[31]^2}\int~Dy~\frac{(2y)_\sigma}{[2y]^{2-\e}}\frac{(1y)_\L}{[1y]^{2-\e}}\int~Dz \le\pd^{(z)}_\mu\Pi_{\rho\nu}(z3)\ri 
\Pi_{\rho\sigma}(z2) \Pi_{\mu\lambda}(zy) 
\ .
\label{SL}
\end{eqnarray}
As one can see from Eq. (\ref{start}),  integral $T_2$ contributes to 
the total result for diagram $c$ with factor $2.$ 
We use simple algebra to represent $T_2$ as combination of terms with simpler Lorentz structure. 
\begin{eqnarray}
\Pi_{\mu\L}(zy) \frac{(1y)_\L}{[1y]^{2-\e}} = \frac{2(1y)_\mu}{[yz]^{1-\e}[1y]^{2-\e}}  - \frac{1}{4(1-\e)}\left[ \pd^2_{(y)}\le \frac{(yz)_{\mu}}{[yz]^{1-\e}[1y]^{1-\e}} \ri  - \right.\no\\
\left. \le \pd^2_{(y)}\frac{(yz)_{\mu}}{[yz]^{1-\e}} \ri\frac{1}{[1y]^{1-\e}} - \frac{(yz)_{\mu}}{[yz]^{1-\e}} \le \pd^2_{(y)} \frac{1}{[1y]^{1-\e}} \ri\right] 
\ .
\label{IBP2}
\end{eqnarray}
Eq. (\ref{SL}) can be transformed to a form
\begin{eqnarray}
- T_2  = \frac{(31)_\nu}{[31]^2}\int~Dy~\frac{(2y)_\sigma}{[2y]^{2-\e}}\frac{(1y)_\L}{[1y]^{2-\e}}\int~Dz \Pi_{\rho\nu}(z3) \le \pd^{(z)}_\mu 
 \Pi_{\rho\sigma}(z2)\ri \Pi_{\mu\lambda}(zy) = \no\\
\frac{(31)_\nu}{[31]^2}\int~Dy~\frac{(2y)_\sigma}{[2y]^{2-\e}}\int~Dz \Pi_{\rho\nu}(z3) \le \pd^{(z)}_\mu  \Pi_{\rho\sigma}(z2)\ri \left\{\frac{2(1y)_\mu}{[yz]^{1-\e}[1y]^{2-\e}}  
- \right. \no\\
- \left. \frac{1}{4(1-\e)}\left[ \pd^2_{(y)}\le \frac{(yz)_{\mu}}{[yz]^{1-\e}[1y]^{1-\e}} \ri  - \le \pd^2_{(y)}\frac{(yz)_{\mu}}{[yz]^{1-\e}} \ri
\frac{1}{[1y]^{1-\e}} - \frac{(yz)_{\mu}}{[yz]^{1-\e}} \le \pd^2_{(y)} \frac{1}{[1y]^{1-\e}} \ri\right]\right\} = \no\\
\frac{(31)_\nu}{[31]^2}\int~Dy~\frac{(2y)_\sigma}{[2y]^{2-\e}}\int~Dz \Pi_{\rho\nu}(z3) \le \pd^{(z)}_\mu  \Pi_{\rho\sigma}(z2)\ri\times \no\\
\times \left\{\frac{2(1y)_\mu}{[yz]^{1-\e}[1y]^{2-\e}}  + \frac{1}{4(1-\e)}\le \pd^2_{(y)}\frac{(yz)_{\mu}}{[yz]^{1-\e}} \ri \frac{1}{[1y]^{1-\e}} \right\} - \no\\
- \frac{1}{4(1-\e)}\frac{(31)_\nu}{[31]^2}\int~Dy~\frac{(2y)_\sigma}{[2y]^{2-\e}}\int~Dz \Pi_{\rho\nu}(z3) \le \pd^{(z)}_\mu  \Pi_{\rho\sigma}(z2)\ri
\pd^2_{(y)} \frac{(yz)_{\mu}}{[yz]^{1-\e}[1y]^{1-\e}} +   \no\\
+ \frac{k}{4(1-\e)}\frac{(31)_\nu}{[31]^2}\int~Dy~\frac{(2y)_\sigma}{[2y]^{2-\e}}\int~Dz \Pi_{\rho\nu}(z3) \le \pd^{(z)}_\mu  \Pi_{\rho\sigma}(z2)\ri
\frac{(yz)_{\mu}}{[yz]^{1-\e}}\delta^{(4-2\e)}(1y)  = \no\\
 \frac{(31)_\nu}{[31]^2}\int~Dy~\frac{(2y)_\sigma}{[2y]^{2-\e}}\int~Dz \Pi_{\rho\nu}(z3) \le \pd^{(z)}_\mu  \Pi_{\rho\sigma}(z2)\ri
\left\{\frac{2(1y)_\mu}{[yz]^{1-\e}[1y]^{2-\e}}  - \frac{(yz)_{\mu}}{[yz]^{2-\e}[1y]^{1-\e}} \right\} + \no\\
+ \frac{1}{8(1-\e)^2}\frac{(31)_\nu}{[31]^2}\pd_\sigma^{(2)}\int~Dy~\frac{1}{[2y]^{1-\e}}\int~Dz \Pi_{\rho\nu}(z3) \le \pd^{(z)}_\mu  \Pi_{\rho\sigma}(z2)\ri
\pd^2_{(y)} \frac{(yz)_{\mu}}{[yz]^{1-\e}[1y]^{1-\e}}+   \no\\
+ \frac{k}{4(1-\e)}\frac{(31)_\nu}{[31]^2}\frac{(21)_\sigma}{[12]^{2-\e}}\int~Dz \Pi_{\rho\nu}(z3) \le \pd^{(z)}_\mu  \Pi_{\rho\sigma}(z2)\ri \frac{(1z)_{\mu}}{[1z]^{1-\e}}  = \no\\
 \frac{(31)_\nu}{[31]^2}\int~Dy~\frac{(2y)_\sigma}{[2y]^{2-\e}}\int~Dz \Pi_{\rho\nu}(z3) \le \pd^{(z)}_\mu  \Pi_{\rho\sigma}(z2)\ri
\left\{\frac{2(1y)_\mu}{[yz]^{1-\e}[1y]^{2-\e}}  - \frac{(yz)_{\mu}}{[yz]^{2-\e}[1y]^{1-\e}} \right\} + \no\\
+ \frac{1}{8(1-\e)^2}\frac{(31)_\nu}{[31]^2}\pd_\sigma^{(2)}\int~Dy~k\delta(2y)\int~Dz \Pi_{\rho\nu}(z3) \le \pd^{(z)}_\mu  \Pi_{\rho\sigma}(z2)\ri
\frac{(yz)_{\mu}}{[yz]^{1-\e}[1y]^{1-\e}}+   \no\\
+ \frac{k}{4(1-\e)}\frac{(31)_\nu}{[31]^2}\frac{(21)_\sigma}{[12]^{2-\e}}\int~Dz \Pi_{\rho\nu}(z3) \le \pd^{(z)}_\mu  \Pi_{\rho\sigma}(z2)\ri \frac{(1z)_{\mu}}{[1z]^{1-\e}}  = \no\\
 \frac{(31)_\nu}{[31]^2}\int~Dy~\frac{(2y)_\sigma}{[2y]^{2-\e}}\int~Dz \Pi_{\rho\nu}(z3) \le \pd^{(z)}_\mu  \Pi_{\rho\sigma}(z2)\ri
\left\{\frac{2(1y)_\mu}{[yz]^{1-\e}[1y]^{2-\e}}  - \frac{(yz)_{\mu}}{[yz]^{2-\e}[1y]^{1-\e}} \right\} + \no\\
+ \frac{k}{8(1-\e)^2}\frac{(31)_\nu}{[31]^2}\pd_\sigma^{(2)}\frac{1}{[12]^{1-\e}}\int~Dz \Pi_{\rho\nu}(z3) \le \pd^{(z)}_\mu  \Pi_{\rho\sigma}(z2)\ri \frac{(2z)_{\mu}}{[2z]^{1-\e}} +  \no \\
+ \frac{k}{4(1-\e)}\frac{(31)_\nu}{[31]^2}\frac{(21)_\sigma}{[12]^{2-\e}}\int~Dz \Pi_{\rho\nu}(z3) \le \pd^{(z)}_\mu  \Pi_{\rho\sigma}(z2)\ri \frac{(1z)_{\mu}}{[1z]^{1-\e}} = \no\\
 \frac{(31)_\nu}{[31]^2}\int~Dy~\frac{(2y)_\sigma}{[2y]^{2-\e}}\int~Dz \Pi_{\rho\nu}(z3) \le \pd^{(z)}_\mu  \Pi_{\rho\sigma}(z2)\ri
\left\{\frac{2(1y)_\mu}{[yz]^{1-\e}[1y]^{2-\e}}  - \frac{(yz)_{\mu}}{[yz]^{2-\e}[1y]^{1-\e}} \right\} + \no\\
- \frac{k}{8(1-\e)^2}\frac{(31)_\nu}{[31]^2}\pd_\sigma^{(2)}\frac{1}{[12]^{1-\e}}\int~Dz \le \pd^{(z)}_\mu  \Pi_{\rho\nu}(z3) \ri \Pi_{\rho\sigma}(z2) \frac{(2z)_{\mu}}{[2z]^{1-\e}} + \no  \\
+ \frac{k}{4(1-\e)^2}\frac{(31)_\nu}{[31]^2}\pd_\sigma^{(2)}\frac{1}{[12]^{1-\e}}\int~Dz  \Pi_{\rho\nu}(z3)  \Pi_{\rho\sigma}(z2) \frac{1}{[2z]^{1-\e}} + \no\\
+ \frac{k}{4}\frac{(31)_\nu}{[31]^2}\frac{(21)_\sigma}{[12]^{2-\e}}\int~Dz \Pi_{\rho\nu}(z3) \le \pd^{(z)}_\mu  \Pi_{\rho\sigma}(z2)\ri \frac{(1z)_{\mu}}{[1z]} = \no\\
\frac{(31)_\nu}{[31]^2}\int~Dy~\frac{(2y)_\sigma}{[2y]^{2-\e}}\int~Dz \Pi_{\rho\nu}(z3) \le \pd^{(z)}_\mu  \Pi_{\rho\sigma}(z2)\ri
\left\{\frac{2(1y)_\mu}{[yz]^{1-\e}[1y]^{2-\e}}  - \frac{(yz)_{\mu}}{[yz]^{2-\e}[1y]^{1-\e}} \right\} + \no\\
- \frac{k}{8}\frac{(31)_\nu}{[31]^2}\pd_\sigma^{(2)}\frac{1}{[12]}\int~Dz \le \pd^{(z)}_\mu  \Pi_{\rho\nu}(z3) \ri \Pi_{\rho\sigma}(z2) \frac{(2z)_{\mu}}{[2z]} +  \no \\
+ \frac{3 k}{8(1-\e)^2}\frac{(31)_\nu}{[31]^2}\pd_\rho^{(2)}\frac{1}{[12]^{1-\e}}\int~Dz  \Pi_{\rho\nu}(z3) \frac{1}{[2z]^{2-2\e}} + \no\\
+ \frac{k}{4}\frac{(31)_\nu}{[31]^2}\frac{(21)_\sigma}{[12]^{2}}\int~Dz \Pi_{\rho\nu}(z3) \le \pd^{(z)}_\mu  \Pi_{\rho\sigma}(z2)\ri \frac{(1z)_{\mu}}{[1z]} = 
\no \\
\frac{(31)_\nu}{[31]^2}\int~Dy~\frac{(2y)_\sigma}{[2y]^{2-\e}}\int~Dz \Pi_{\rho\nu}(z3) \le \pd^{(z)}_\mu  \Pi_{\rho\sigma}(z2)\ri
\left\{\frac{2(1y)_\mu}{[yz]^{1-\e}[1y]^{2-\e}}  - \frac{(yz)_{\mu}}{[yz]^{2-\e}[1y]^{1-\e}} \right\} + \no\\
- \frac{k}{8}\frac{(31)_\nu}{[31]^2}\pd_\sigma^{(2)}\frac{1}{[12]}\int~Dz \le \pd^{(z)}_\mu  \Pi_{\rho\nu}(z3) \ri \Pi_{\rho\sigma}(z2) \frac{(2z)_{\mu}}{[2z]} +  \no \\
+ \frac{3 k}{4 (1-\e)}\frac{(31)_\nu}{[31]^2} \frac{(12)_\rho}{[12]^{2-\e}}\int~Dz  \Pi_{\rho\nu}(z3) \frac{1}{[2z]^{2-2\e}}   \no\\
+ \frac{k}{4}\frac{(31)_\nu}{[31]^2}\frac{(21)_\sigma}{[12]^{2}}\int~Dz \Pi_{\rho\nu}(z3) \le \pd^{(z)}_\mu  \Pi_{\rho\sigma}(z2)\ri \frac{(1z)_{\mu}}{[1z]}  \equiv \no\\
\equiv T_{21} + T_{22} + T_{23} + T_{24} 
\ .
\label{SLT}
\end{eqnarray}
Integrals $T_{22}$ and $T_{24}$ are finite,  integral $T_{23}$ is singular. Since integral $T_{2}$ is finite, $T_{21}$ + $T_{23}$ should be finite.  
The sum $T_{21} + T_{23}$ is calculated in Appendix B.

\subsubsection{Calculation of $T_{22}$ and $T_{24}$}

In Eq. (\ref{SLT}) for $T_2$ there are two finite single  integrals,   
\begin{eqnarray}
T_{24} = \frac{k}{4}\frac{(31)_\nu}{[31]^2}\frac{(21)_\sigma}{[12]^{2}}\int~Dz \Pi_{\rho\nu}(z3) \le \pd^{(z)}_\mu  \Pi_{\rho\sigma}(z2)\ri \frac{(1z)_{\mu}}{[1z]} = \no\\
k\left[\frac{1/4}{[12]^2[31]^2} + \frac{1/4}{[12][23][31]^2} + \frac{-1/4}{[12]^2[23][31]} \right] \no\\
+ k\left[\frac{1/4}{[12][31]^2} +  \frac{-1/2[23]}{[12]^2[31]^2}    + \frac{1/2}{[12]^2[31]}   \right]J(1,1,1) \no\\
+ k\left[ \frac{-1/8}{[12]^2[23]^2} + \frac{1/2}{[12]^2[31]^2} + \frac{-1/8}{[23]^2[31]^2}  + \frac{-1/8}{[12][23][31]^2} \right. \no\\
+ \left. \frac{1/4}{[12][23]^2[31]} + \frac{-3/8}{[12]^2[23][31]} \right]\ln{[12]} \no\\
+ k\left[ \frac{1/8}{[12]^2[23]^2} + \frac{-1/2}{[12]^2[31]^2} + \frac{1/8}{[23]^2[31]^2} + \frac{1/8}{[12][23][31]^2} \right.\no\\ 
+ \left. \frac{-1/4}{[12][23]^2[31]} + \frac{3/8}{[12]^2[23][31]} \right]\ln{[31]},  \label{SLT3}
\end{eqnarray}
and 
\begin{eqnarray}
T_{22} = - \frac{k}{8}\frac{(31)_\nu}{[31]^2}\pd_\sigma^{(2)}\frac{1}{[12]}\int~Dz \le \pd^{(z)}_\mu  \Pi_{\rho\nu}(z3) \ri \Pi_{\rho\sigma}(z2) \frac{(2z)_{\mu}}{[2z]}  = \label{SLT4}\\
k\left[\frac{3/8}{[12]^2[23]^2} + \frac{-3/4}{[12]^2[31]^2} + \frac{3/8}{[12]^2[31]^2}   
+ \frac{3/8}{[12][23][31]^2}   +  \frac{-3/4}{[12][23]^2[31]}    +  \frac{3/8}{[12]^2[23][31]}            \right]  \no
\ .
\end{eqnarray}

\subsubsection{Total result for $T_2$}

Combinining $T_{22}, T_{24}$ with the result of Appendix B for $T_{21} + T_{23},$
\begin{eqnarray*}
- T_2 = T_{22} + T_{24} + K_3 + M_3 = \\
k\left[\frac{11/16}{[12]^2[23]^2} + \frac{-9/16}{[12]^2[31]^2} + \frac{7/16}{[23]^2[31]^2}  +  \frac{5/8}{[12][23][31]^2} 
+  \frac{-9/8}{[12][23]^2[31]} +  \frac{-1/8}{[12]^2[23][31]} \right] \\
+ k\left[\frac{1/4}{[12][31]^2} + \frac{1/4[12]}{[23]^2[31]^2} + \frac{-1/4}{[23][31]^2} + \frac{-1/2[23]}{[12]^2[31]^2}  
+  \frac{1/2}{[12]^2[31]} + \frac{-1/4}{[23]^2[31]}  \right]J(1,1,1) \\
+ k\left[\frac{-1/8}{[12]^2[23]^2} + \frac{1/2}{[12]^2[31]^2} + \frac{1/8}{[23]^2[31]^2} + \frac{-1/8}{[12][23][31]^2}  + \frac{1/4}{[12][23]^2[31]} 
+ \frac{-3/8}{[12]^2[23][31]} \right] \ln[12] \\
+ k\left[\frac{-3/16}{[12]^2[23]^2} + \frac{3/8}{[12]^2[31]^2} + \frac{1/16}{[23]^2[31]^2} + \frac{-11/16}{[12][23][31]^2} + \frac{1/8}{[12][23]^2[31]} 
+ \frac{-3/16}{[12]^2[23][31]} \right]\ln[23] \\
+ k\left[\frac{5/16}{[12]^2[23]^2} + \frac{-7/8}{[12]^2[31]^2} + \frac{-3/16}{[23]^2[31]^2} + \frac{13/16}{[12][23][31]^2} + \frac{-3/8}{[12][23]^2[31]} 
+ \frac{9/16}{[12]^2[23][31]} \right]\ln[31] \\
\\
+ \left[\frac{1/2}{[12]^2[23]^2} + \frac{-1}{[23]^2[31]^2} +  \frac{1/2}{[12][23]^2[31]} +  \frac{3/2}{[12]^2[23][31]}   \right] \\
+ \left[\frac{-3}{[12][23]^2} + \frac{1}{[12][31]^2} + \frac{1/2[12]}{[23]^2[31]^2}  +\frac{-1}{[23][31]^2}  + \frac{-2}{[12]^2[31]} +  \frac{3/2}{[23]^2[31]} 
+ \frac{[31]}{[12]^2[23]^2}  \right] J(1,1,1) \\
+ \left[\frac{-2}{[12]^2[23]^2} + \frac{-1/2}{[12][23][31]^2} + \frac{5/2}{[12][23]^2[31]} + \frac{1}{[12]^2[23][31]} \right]\ln[12] \\
+ \left[\frac{-1/2}{[12]^2[23]^2} +  \frac{1}{[23]^2[31]^2} + \frac{-1}{[12][23][31]^2} +  \frac{-1/2}{[12][23]^2[31]} +  \frac{-5/2}{[12]^2[23][31]}  \right]\ln[23] \\
+ \left[\frac{5/2}{[12]^2[23]^2} +   \frac{-1}{[23]^2[31]^2}   + \frac{3/2}{[12][23][31]^2}   + \frac{-2}{[12][23]^2[31]} + \frac{3/2}{[12]^2[23][31]}  \right]\ln[31]
\ .
\end{eqnarray*}

\subsection{The third line of (\ref{start})}

\begin{eqnarray}
T_3 \equiv \frac{(31)_\nu}{[31]^2}\int~Dy~\frac{(2y)_\sigma}{[2y]^{2-\e}}\frac{(1y)_\L}{[1y]^{2-\e}}
\int~Dz \le\pd^{(z)}_\mu\Pi_{\rho\sigma}(z2)\ri\Pi_{\rho\lambda}(zy) \Pi_{\mu\nu}(z3) 
\ .
\label{TL}
\end{eqnarray}
Integral $T_3$ contributes to 
the total result for diagram $(c)$ with factor $2.$ 
We use simple algebra to represent integral (\ref{TL}) as combination of terms 
with simpler Lorentz structure, by using for $\Pi_{\rho\L}(zy) (1y)_{\L}/[1y]^{2-\e}$ 
the form of Eq.~(\ref{IBP1}). It can be checked that expression (\ref{TL}) can be 
transformed identically as in the case of $T_1$, Eqs.~(\ref{line1})-(\ref{FLT}),
with the only substitutions $\Pi_{\rho \nu}(z3) \mapsto \Pi_{\rho \sigma}(z2)$
and $\Pi_{\mu \sigma}(z2) \mapsto \Pi_{\mu \nu}(z3)$. This leads to a form
completely analogous to Eq.~(\ref{FLT})
\begin{eqnarray}
T_3 =
\frac{(31)_\nu}{[31]^2}\int~Dy~\frac{(2y)_\sigma}{[2y]^{2-\e}}\int~Dz \le\pd^{(z)}_\mu\Pi_{\rho\sigma}(z2)\ri \Pi_{\mu\nu}(z3)
\left\{\frac{2(1y)_\rho}{[yz]^{1-\e}[1y]^{2-\e}}  - \frac{(yz)_{\rho}}{[yz]^{2-\e}[1y]^{1-\e}} \right\} + \no\\
+ \frac{k}{8(1-\e)^2}\frac{(31)_\nu}{[31]^2}\pd_\sigma^{(2)}\frac{1}{[12]^{1-\e}}\int~Dz \le\pd^{(z)}_\mu\Pi_{\rho\sigma}(z2)\ri \Pi_{\mu\nu}(z3) \frac{(2z)_{\rho}}{[2z]^{1-\e}} + \no\\
+ \frac{k}{4(1-\e)}\frac{(31)_\nu}{[31]^{2-\e}}\frac{(21)_\sigma}{[12]^{2-\e}}\int~Dz \le\pd^{(z)}_\mu\Pi_{\rho\sigma}(z2)\ri \Pi_{\mu\nu}(z3) \frac{(1z)_{\rho}}{[1z]^{1-\e}} \equiv \no\\
\equiv T_{31} + T_{32} + T_{33} \equiv T_{311} + T_{312} + T_{32} + T_{33} 
\ .
\label{TLT}
\end{eqnarray}
Single integrals $T_{32}$ and $T_{33}$ are finite. Since $T_3$ is finite, it means that $T_{31}$ is also finite. Moreover, the finiteness (UV and IR)  of $T_{31}$ 
can be checked by index counting directly. Now we consider $T_{31},$  $T_{31} \equiv T_{311} + T_{312}.$  Both $T_{311}$ and $T_{312}$ are finite.
Integrals $T_{311}$ and $T_{312}$ are calculated in Appendices C and D,
respectively.

\subsubsection{Calculation of $T_{32}$ and $T_{33}$}

Eq. (\ref{TLT}) contains two finite integrals,
\begin{eqnarray}
T_{33} \equiv \frac{k}{4}\frac{(31)_\nu}{[31]^{2}}\frac{(21)_\sigma}{[12]^{2}}\int~Dz \le\pd^{(z)}_\mu\Pi_{\rho\sigma}(z2)\ri \Pi_{\mu\nu}(z3) \frac{(1z)_{\rho}}{[1z]} = \no\\
k\left[\frac{1/4}{[12]^2[23]^2} + \frac{1/4}{[12]^2[31]^2}  + \frac{1/4}{[23]^2[31]^2}  +  \frac{-1/2}{[12][23]^2[31]}  \right] \no\\
+ k\left[\frac{1/4}{[12][31]^2} +  \frac{-1/2[23]}{[12]^2[31]^2} + \frac{1/4}{[12]^2[31]} \right]J(1,1,1)\no \\
+ k\left[\frac{1/8}{[12]^2[23]^2} + \frac{-1/2}{[12]^2[31]^2}  + \frac{-1/8}{[23]^2[31]^2}  +  \frac{-1/8}{[12][23][31]^2} +  \frac{1/8}{[12]^2[23][31]} \right]\ln[12] \no\\
+ k\left[\frac{1}{[12]^2[31]^2}   \right]\ln[23] \no\\
+ k\left[\frac{-1/8}{[12]^2[23]^2}  + \frac{-1/2}{[12]^2[31]^2}   + \frac{1/8}{[23]^2[31]^2}  +  \frac{1/8}{[12][23][31]^2} +  \frac{-1/8}{[12]^2[23][31]} \right]\ln[31], \label{TLT3}
\end{eqnarray}
and 
\begin{eqnarray}
T_{32}\equiv  \frac{k}{8}\frac{(31)_\nu}{[31]^2}\pd_\sigma^{(2)}\frac{1}{[12]}\int~Dz \le\pd^{(z)}_\mu\Pi_{\rho\sigma}(z2)\ri \Pi_{\mu\nu}(z3) \frac{(2z)_{\rho}}{[2z]} = \no\\
k\left[\frac{-1/8}{[12]^2[23]^2} + \frac{1/4}{[12]^2[31]^2}  + \frac{-1/8}{[23]^2[31]^2}  +  \frac{-1/8}{[12][23][31]^2}  +  \frac{1/4}{[12][23]^2[31]}   
+  \frac{-1/8}{[12]^2[23][31]}  \right] 
\ .
\label{TLT4}
\end{eqnarray}

\subsubsection{Total result for $T_3$}

Using the result of Appendices C and D, and of Eqs. (\ref{TLT3}) and (\ref{TLT4}), 
\begin{eqnarray*}
T_3 \equiv T_{32} + T_{33} + K_4 + M_4 + K_5 + M_5 = \\
k\left[\frac{5/16}{[12]^2[23]^2} + \frac{9/16}{[12]^2[31]^2}  + \frac{1/16}{[23]^2[31]^2}  +  \frac{3/8}{[12][23][31]^2} +  \frac{-3/8}{[12][23]^2[31]} 
  +  \frac{-3/8}{[12]^2[23][31]}  \right] \\
+ k\left[\frac{1/4[12]}{[23]^2[31]^2} +   \frac{-1/4}{[23][31]^2} +  \frac{-1/2[23]}{[12]^2[31]^2} +   \frac{1/4}{[12]^2[31]} +  \frac{-1/4}{[23]^2[31]} \right]J(1,1,1) \\
+ k\left[\frac{1/4}{[12]^2[23]^2} + \frac{-3/4}{[12]^2[31]^2} +  \frac{-1/4}{[23]^2[31]^2} + \frac{-1/4}{[12][23][31]^2} + \frac{-1/4}{[12][23]^2[31]}  
+ \frac{1/4}{[12]^2[23][31]} \right]\ln[12] \\
+ k\left[\frac{-1/16}{[12]^2[23]^2}    +   \frac{13/8}{[12]^2[31]^2} +    \frac{3/16}{[23]^2[31]^2} +  \frac{-1/16}{[12][23][31]^2}  + \frac{-1/8}{[12][23]^2[31]} 
+ \frac{-1/16}{[12]^2[23][31]}   \right]\ln[23] \\
+ k\left[\frac{-3/16}{[12]^2[23]^2}    +   \frac{-7/8}{[12]^2[31]^2} +   \frac{1/16}{[23]^2[31]^2} + \frac{5/16}{[12][23][31]^2}  + \frac{3/8}{[12][23]^2[31]} 
+ \frac{-3/16}{[12]^2[23][31]}\right]\ln[31] 
\\
+ \left[\frac{5/2}{[12]^2[23]^2} +  \frac{1}{[12][23][31]^2}  +  \frac{-5/2}{[12][23]^2[31]}   +  \frac{-1/2}{[12]^2[23][31]}     \right] \\
+ \left[\frac{1/2[12]}{[23]^2[31]^2} + \frac{-1}{[23][31]^2}  +    \frac{1/2}{[23]^2[31]}+    \frac{-[31]}{[12]^2[23]^2}  \right]J(1,1,1) \\
+ \left[\frac{1/2}{[12][23][31]^2} + \frac{-5/2}{[12][23]^2[31]} + \frac{-1}{[12]^2[23][31]} \right]\ln[12] \\
+ \left[\frac{-5/2}{[12]^2[23]^2} +   \frac{5/2}{[12][23]^2[31]}  + \frac{-1/2}{[12]^2[23][31]}  \right]\ln[23] \\
+ \left[\frac{5/2}{[12]^2[23]^2}  + \frac{-1/2}{[12][23][31]^2}   + \frac{3/2}{[12]^2[23][31]} \right]\ln[31]
\ .
\end{eqnarray*}

\section{The result for diagram $c$}

Summing up the contribution for all three lines of Eq. (\ref{start}), we obtain the result for diagram $(c).$

\begin{eqnarray*}
T \equiv -2T_1 +2T_2 + 2T_3 = \\ 
k\left[\frac{-11/8}{[12]^2[23]^2} + \frac{29/8}{[12]^2[31]^2}  + \frac{-15/8}{[23]^2[31]^2}  + \frac{-13/4}{[12][23][31]^2} +  \frac{13/4}{[12][23]^2[31]} 
+  \frac{-9/4}{[12]^2[23][31]}  \right] \\
+ k\left[\frac{1/2}{[12][31]^2} +  \frac{1/4[12]}{[23]^2[31]^2} +    \frac{-1}{[23][31]^2} +   \frac{-[23]}{[12]^2[31]^2}    + \frac{1}{[12]^2[31]} 
+    \frac{-1/4}{[23]^2[31]} \right]J(1,1,1) \\
+ k\left[\frac{1}{[12]^2[23]^2} + \frac{-3/2}{[12]^2[31]^2}  + \frac{-1/2}{[23]^2[31]^2} + \frac{3/4}{[12][23][31]^2} + \frac{-9/4}{[12][23]^2[31]} 
+ \frac{1/2}{[12]^2[23][31]} \right]\ln[12] \\
+ k\left[\frac{-1/8}{[12]^2[23]^2}    +   \frac{5/4}{[12]^2[31]^2} +   \frac{3/8}{[23]^2[31]^2}  + \frac{7/8}{[12][23][31]^2} + \frac{-1/4}{[12][23]^2[31]} 
+ \frac{-9/8}{[12]^2[23][31]} \right]\ln[23] \\
+ k\left[\frac{-7/8}{[12]^2[23]^2}    +   \frac{1/4}{[12]^2[31]^2} +   \frac{1/8}{[23]^2[31]^2} + \frac{-13/8}{[12][23][31]^2}   
+ \frac{5/2}{[12][23]^2[31]} + \frac{5/8}{[12]^2[23][31]}\right]\ln[31] 
\\
+ \left[\frac{1}{[12]^2[23]^2} + \frac{-2}{[12]^2[31]^2}  +   \frac{-2}{[12][23][31]^2}  +  \frac{-1}{[12][23]^2[31]}    
+  \frac{1}{[12]^2[23][31]}    \right] \\
+ \left[\frac{6}{[12][23]^2} + \frac{-4}{[12][31]^2}  + \frac{2[12]}{[23]^2[31]^2}   + \frac{-4}{[23][31]^2} + \frac{2}{[12]^2[31]} + \frac{-6}{[23]^2[31]}  
+ \frac{-2[31]}{[12]^2[23]^2}  \right]J(1,1,1) \\
+ \left[\frac{2}{[12]^2[23]^2} + \frac{2}{[23]^2[31]^2}  + \frac{4}{[12][23][31]^2}   + \frac{-10}{[12][23]^2[31]} + \frac{-2}{[12]^2[23][31]} \right]\ln[12] \\
+ \left[\frac{-1}{[12]^2[23]^2} +   \frac{1}{[12][23]^2[31]}   + \frac{-1}{[12]^2[23][31]}  \right]\ln[23] \\
+ \left[\frac{-1}{[12]^2[23]^2} + \frac{-2}{[23]^2[31]^2}    +     \frac{-4}{[12][23][31]^2} +   \frac{9}{[12][23]^2[31]}   
+ \frac{3}{[12]^2[23][31]}  \right]\ln[31] 
\ .
\end{eqnarray*}
Self-consistency requires the value $k=-4.$ This follows from formulas of Ref.~\cite{Cvetic:2006iu}.  Thus, the result for diagram $(c)$ is 
\begin{eqnarray*}
T = \left[\frac{13/2}{[12]^2[23]^2} + \frac{-33/2}{[12]^2[31]^2}  + \frac{15/2}{[23]^2[31]^2}  +  \frac{11}{[12][23][31]^2}  
+  \frac{-14}{[12][23]^2[31]} +  \frac{10}{[12]^2[23][31]}  \right] \\
+ \left[\frac{6}{[12][23]^2} + \frac{-6}{[12][31]^2}  +  \frac{[12]}{[23]^2[31]^2}  + \frac{4[23]}{[12]^2[31]^2} 
+ \frac{-2}{[12]^2[31]}  + \frac{-5}{[23]^2[31]} + \frac{-2[31]}{[12]^2[23]^2} \right]J(1,1,1) \\
+ \left[\frac{-2}{[12]^2[23]^2} +  \frac{6}{[12]^2[31]^2} +   \frac{4}{[23]^2[31]^2} +  \frac{1}{[12][23][31]^2} +  \frac{-1}{[12][23]^2[31]} 
+ \frac{-4}{[12]^2[23][31]} \right]\ln[12] \\
+ \left[\frac{-1/2}{[12]^2[23]^2} + \frac{-5}{[12]^2[31]^2}    +   \frac{-3/2}{[23]^2[31]^2}   +   \frac{-7/2}{[12][23][31]^2} +   \frac{2}{[12][23]^2[31]}   
+ \frac{7/2}{[12]^2[23][31]}  \right]\ln[23] \\
+ \left[\frac{5/2}{[12]^2[23]^2} + \frac{-1}{[12]^2[31]^2}    +   \frac{-5/2}{[23]^2[31]^2}   +   \frac{5/2}{[12][23][31]^2} +   \frac{-1}{[12][23]^2[31]}   
+ \frac{1/2}{[12]^2[23][31]}  \right]\ln[31] 
\ .
\end{eqnarray*}

\section{Conclusion}

We have treated in this paper the vertex of gluon self-interaction in the position space, significantly reducing the number of indices of the Lorentz group.
This reduction allowed us to obtain the result for the third contribution to the $Lcc$ diagram by using {\it Mathematica} software.  It is clear from the index counting arguments 
that diagram $(c)$ does not diverge in the limit $\e \rar 0$, due to transversality of the gauge propagator in the Landau gauge. The transversality makes 
the entire diagram convergent superficially, and makes also its subgraphs convergent.  $ {\cal N} = 4$ supersymmetry does not play any role in the scale independence of the
diagrams $(a),(b)$ and $(c).$ This is a pure effect of the Landau gauge, and this result will be true in any gauge theory, for example in pure QCD. 
$ {\cal N} = 4$ supersymmetry is important only for cancellation  of poles between diagrams $(d)$ and $(e)$  \footnote{The full result for the sum of all
the five two-loop planar diagrams can be found in Ref. \cite{Cvetic:2007fp}}. Scale independence of the kernels of dressed mean fields 
in the effective action of gauge theories and gravity,  and conformal invariance of the effective action of dressed mean fields for these theories can play an important role 
for different types of theories such as the ${\cal N}=8$ supergravity \cite{Bjerrum-Bohr:2006yw,Kang:2004cs,Green:2006gt,Bern:2006kd}, 
Chern-Simons theory near the RG fixed points \cite{Avdeev:1992jt}, 
massless gauge theory near fixed points in the coupling space, topological field theories in higher dimensions, finite ${\cal N}=1$ supersymmetric theories 
\cite{Jones:1986vp,Ermushev:1986cu,Kazakov:1991th,Kazakov:1995cy,Kondrashuk:1997uf}.  To introduce masses in the theory, we need to use softly broken 
supersymmetry, in which the couplings are spacetime-independent background superfields. The relation between the RG functions of softly broken and 
rigid theories was
discovered in Refs.~\cite{Yamada:1994id,Jack:1997pa,Avdeev:1997vx}.  The relation between the correlators of softly broken and rigid theories
can be found by a trick of general change of variables in superspace \cite{Kondrashuk:1999de}.

\subsection*{Acknowledgments}

The work of G.C. was supported by 
in part by Fondecyt (Chile) grant \#1050512. 
The work of I.K. was supported in part by Ministry of Education (Chile) under grant 
Mecesup FSM9901 and by DGIP UTFSM, by  Fondecyt (Chile) grant \#1040368, 
and by Departamento de Ciencias B\'asicas de la Universidad 
del B\'\i o-B\'\i o, Chill\'an (Chile).

\begin{appendix}

\section[]{}
\setcounter{equation}{0}
\label{App:A}

\begin{eqnarray*} 
T_{112} = - \frac{(31)_\nu}{[31]^2}\int~Dy~\frac{(2y)_\sigma}{[2y]^{2-\e}}\int~Dz \le\pd^{(z)}_\mu\Pi_{\rho\nu}(z3)\ri \Pi_{\mu\sigma}(z2)\frac{(yz)_{\rho}}{[yz]^{2-\e}[1y]^{1-\e}} = \no\\
+ \frac{1}{2(1-\e)}\frac{(31)_\nu}{[31]^2}\int~Dy~ \frac{(2y)_\sigma}{[2y]^{2-\e}[1y]^{1-\e}} \int~Dz \le \pd^{(z)}_\mu \Pi_{\rho\nu}(z3)\ri \Pi_{\mu\sigma}(z2)
\le \pd_\rho^{(y)} \frac{1}{[yz]^{1-\e}} \ri = \no\\
+ \frac{1}{2(1-\e)}\frac{(31)_\nu}{[31]^2}\int~Dy~ \le\pd_\rho^{(y)}\frac{(2y)_\sigma}{[2y]^{2-\e}[1y]^{1-\e}}\ri \int~Dz \Pi_{\rho\nu}(z3) \Pi_{\mu\sigma}(z2)
\le \pd^{(z)}_\mu \frac{1}{[yz]^{1-\e}} \ri = \no\\
+ \frac{1}{2(1-\e)}\frac{(31)_\nu}{[31]^2}\int~Dy~ \le\pd_\rho^{(y)}\frac{(2y)_\sigma}{[2y]^{2-\e}}\ri \frac{1}{[1y]^{1-\e}} \int~Dz \Pi_{\rho\nu}(z3) \Pi_{\mu\sigma}(z2)
\le \pd^{(z)}_\mu \frac{1}{[yz]^{1-\e}} \ri  \no\\
+ \frac{(31)_\nu}{[31]^2} \int~Dy~  \frac{(2y)_\sigma}{[2y]^{2-\e}}\frac{(1y)_\rho}{[1y]^{2-\e}} \int~Dz \Pi_{\rho\nu}(z3)\Pi_{\mu\sigma}(z2) \le \pd^{(z)}_\mu \frac{1}{[yz]^{1-\e}} \ri
\equiv \frac{I_1}{2} + \frac{I_2}{2} 
\end{eqnarray*}
Integral $I_2$ is $-T_{111}$ and is given by Eq. (\ref{FLT2.1}). We do several steps to calculate integral $I_1.$
Integral $I_1$ can be done by using the formula (\ref{trick2}). Integrals $I_1$ nad $I_2$ are finite double integrals. 
The first contribution to $I_1$ is (with factor  $2(1-\e)$)
\begin{eqnarray*}
A_1 \equiv \frac{(31)_\nu}{[31]^2}\int~Dy~ \le\pd_\rho^{(y)}\frac{(2y)_\sigma}{[2y]^{2-\e}}\ri \frac{1}{[1y]^{1-\e}} \int~Dz \Pi_{\rho\nu}(z3)\frac{2(yz)_\sigma}{[2z]^{1-\e}[yz]^{2-\e}} = \\
- \frac{1}{1-\e}\frac{(31)_\nu}{[31]^2}\int~Dy~ \le\pd_\sigma^{(y)}\frac{(2y)_\rho}{[2y]^{2-\e}}\ri \frac{1}{[1y]^{1-\e}}
\int~Dz \Pi_{\rho\nu}(z3) \le\pd^{(y)}_\sigma \frac{1}{[yz]^{1-\e}}\ri  \frac{1}{[2z]^{1-\e}} = \\
- \frac{k}{2(1-\e)}\frac{(31)_\nu}{[31]^2}\int~Dy~ \frac{(2y)_\rho}{[2y]^{2-\e}} \delta(1y) \int~Dz \Pi_{\rho\nu}(z3) \frac{1}{[yz]^{1-\e}[2z]^{1-\e}} \\
+ \frac{k}{4(1-\e)^2}\frac{(31)_\nu}{[31]^2}\int~Dy~ \le\pd_\rho^{(y)}\delta(2y)\ri\frac{1}{[1y]^{1-\e}}  \int~Dz \Pi_{\rho\nu}(z3) \frac{1}{[yz]^{1-\e}[2z]^{1-\e}} \\
+ \frac{1}{2(1-\e)}\frac{(31)_\nu}{[31]^2}\int~Dy~ \frac{(2y)_\rho}{[2y]^{2-\e}[1y]^{1-\e}} \int~Dz \Pi_{\rho\nu}(z3) k\delta(yz)  \frac{1}{[2z]^{1-\e}} = \\
- \frac{k}{2}\frac{(31)_\nu}{[31]^2} \frac{(21)_\rho}{[12]^{2}} \int~Dz \Pi_{\rho\nu}(z3) \frac{1}{[1z][2z]}
+ \frac{k}{2(1-\e)}\frac{(31)_\nu}{[31]^2} \frac{(21)_\rho}{[12]^{2-\e}}  \int~Dz \Pi_{\rho\nu}(z3) \frac{1}{[2z]^{2-2\e}} \\
+ \frac{k}{2(1-\e)}\frac{(31)_\nu}{[31]^2} \int~Dz \Pi_{\rho\nu}(z3) \frac{(2z)_\rho}{[2z]^{3-2\e} [1z]^{1-\e}} = \\
\\
- \frac{k}{2}\frac{(31)_\nu}{[31]^2} \frac{(21)_\rho}{[12]^{2}} \int~Dz \Pi_{\rho\nu}(z3) \frac{1}{[1z][2z]}
+ \frac{k}{2(1-\e)}\frac{(31)_\nu}{[31]^2} \frac{(21)_\rho}{[12]^{2-\e}}  \int~Dz \Pi_{\rho\nu}(z3) \frac{1}{[2z]^{2-2\e}} \\
- \frac{k}{4(1-\e)}\frac{(31)_\nu}{[31]^2} \int~Dz \Pi_{\rho\nu}(z3) \frac{(1z)_\rho}{[2z]^{2-2\e} [1z]^{2-\e}} = \\
- \frac{k}{2}A_{11} - \frac{k}{4(1-\e)}M_{10} - \frac{k}{8(1-\e)}M_{11}
\end{eqnarray*}
Integral $A_1$ is singular in UV region. Integral $A_{11}$ is finite.  Integral $M_{10}$ and and integral $M_{11}$ are defined in 
Section 3.1.1. The second contribution to $I_1$ is (with factor $1/2$)  
\begin{eqnarray*}
A_2 \equiv \frac{(31)_\nu}{[31]^2}\int~Dy~ \le\pd_\rho^{(y)}\frac{(2y)_\sigma}{[2y]^{2-\e}}\ri \frac{1}{[1y]^{1-\e}} \int~Dz \Pi_{\rho\nu}(z3)
\pd^2_{(z)}\le \frac{(2z)_{\sigma}}{[2z]^{1-\e}[yz]^{1-\e}} \ri  = \\
- 2(1-\e)\frac{(31)_\nu}{[31]^2}\int~Dy~\frac{(2y)_\sigma}{[2y]^{2-\e}}\frac{(1y)_\rho}{[1y]^{2-\e}}
\int~Dz~ \Pi_{\rho\nu}(z3) \pd^2_{(z)}\le \frac{(2z)_{\sigma}}{[2z]^{1-\e}[yz]^{1-\e}} \ri  \\
+ 2(1-\e)\frac{(31)_\nu}{[31]^2}\int~Dy~\frac{(2y)_\sigma}{[2y]^{2-\e}}\frac{1}{[1y]^{1-\e}}
\int~Dz~\le \frac{2g_{\rho\nu}}{[3z]^{1-\e}} - \pd_\nu^{(3)} \frac{(3z)_{\rho}}{[3z]^{1-\e}} \ri
\pd^2_{(z)}\le \frac{(2z)_{\sigma}(yz)_\rho}{[2z]^{1-\e}[yz]^{2-\e}} \ri  = \\
-I_3  - k\frac{(31)_\nu}{[31]^2}\frac{(23)_\sigma}{[23]} \pd_\nu^{(3)} \pd_\sigma^{(2)} J(1,1,1) \\
+ 2 \frac{(31)_\nu}{[31]^2} \pd_\nu^{(3)} \int~Dy~ \frac{(2y)_\sigma}{[2y]^{2-\e}}\frac{1}{[1y]^{1-\e}}
\int~Dz~ \le \pd_\rho^{(z)}  \frac{1}{[3z]^{1-\e}}\ri  \le \pd_\rho^{(z)} \frac{1}{[yz]^{1-\e}} \ri  \frac{(2z)_{\sigma}}{[2z]^{1-\e}}  = \\
\\
-I_3  - k\frac{(31)_\nu}{[31]^2}\frac{(23)_\sigma}{[23]} \pd_\nu^{(3)} \pd_\sigma^{(2)} J(1,1,1) \\
- 4(1-\e)\frac{(31)_\nu}{[31]^2} \pd_\nu^{(3)} \int~Dy~Dz \frac{(2y)_\sigma(2z)_{\sigma}}{[2y]^{2-\e}[1y]^{1-\e}[3z]^{1-\e}[yz]^{1-\e}[2z]^{2-\e}} \\
+ \frac{k}{2}\frac{(31)_\nu}{[31]^2} \pd_\nu^{(3)} \frac{(23)_{\sigma}}{[23]} \pd_\sigma^{(2)} J(1,1,1)
- k \frac{(31)_\nu}{[31]^2} \pd_\nu^{(3)} \int~Dy~ \frac{1}{[1y]^{1-\e}[3y]^{1-\e}[2y]^{2-2\e}}  = \\
\\
-I_3  - k\frac{(31)_\nu}{[31]^2}\frac{(23)_\sigma}{[23]} \pd_\nu^{(3)} \pd_\sigma^{(2)} J(1,1,1) \\
- \frac{1}{2}\frac{(31)_\nu}{[31]^2} \pd_\nu^{(3)}\pd_{(2)}^2 \int~Dy~Dz \frac{1}{[2y][1y][3z][yz][2z]} \\
+ \frac{k}{2(1-\e)}\frac{(31)_\nu}{[31]^2}\frac{1}{[12]^{1-\e}} \pd_\nu^{(3)} \int~Dz \frac{1}{[3z]^{1-\e}[2z]^{2-2\e}} \\
+ \frac{k}{2(1-\e)} \frac{(31)_\nu}{[31]^2} \pd_\nu^{(3)} \frac{1}{[23]^{1-\e}} \int~Dy~Dz \frac{1}{[2y]^{2-2\e}[1y]^{1-\e}} \\
+ \frac{k}{2}\frac{(31)_\nu}{[31]^2} \pd_\nu^{(3)} \frac{(23)_{\sigma}}{[23]} \pd_\sigma^{(2)} J(1,1,1)
- k \frac{(31)_\nu}{[31]^2} \pd_\nu^{(3)} \int~Dy~ \frac{1}{[1y]^{1-\e}[3y]^{1-\e}[2y]^{2-2\e}} = \\
\\
k\left[\frac{1}{[23]^2[31]^2} + \frac{2}{[12][23][31]^2}  + \frac{-1}{[12][23]^2[31]} \right] \no\\
+ k\left[\frac{-1/2[12]}{[23]^2[31]^2} +  \frac{1}{[23][31]^2}  + \frac{1/2}{[23]^2[31]} \right]J(1,1,1) \no\\
+ k\left[\frac{1}{[23]^2[31]^2} + \frac{1/2}{[12][23][31]^2}  + \frac{1/2}{[12][23]^2[31]} \right]\ln{[12]} \no\\
+ k\left[\frac{-1}{[12]^2[31]^2} + \frac{-1}{[23]^2[31]^2}   + \frac{1}{[12][23]^2[31]} + \frac{1}{[12]^2[23][31]} \right]\ln{[23]} \no\\
+ k\left[\frac{1}{[12]^2[31]^2} + \frac{-1/2}{[12][23][31]^2} + \frac{-3/2}{[12][23]^2[31]^2} + \frac{-1}{[12]^2[23][31]} \right]\ln{[31]} \no\\
\no\\
+ \left[\frac{2}{[12]^2[23]^2} + \frac{2}{[12]^2[31]^2} + \frac{2}{[23]^2[31]^2}  + \frac{4}{[12][23][31]^2} 
+ \frac{-4}{[12][23]^2[31]}  + \frac{-4}{[12]^2[23][31]} \right] \no\\
+ \left[\frac{-4}{[12][23]^2} + \frac{2}{[12][31]^2} + \frac{-2[12]}{[23]^2[31]^2} + \frac{4}{[23][31]^2} 
+ \frac{6}{[23]^2[31]} \right]J(1,1,1) \no\\
+ \left[ \frac{-2}{[23]^2[31]^2} + \frac{-2}{[12][23][31]^2}  + \frac{4}{[12][23]^2[31]} \right]\ln{[12]} \no\\
+ \left[\frac{-2}{[12]^2[23]^2} + \frac{-2}{[23]^2[31]^2} + \frac{2}{[12][23][31]^2} + \frac{4}{[12][23]^2[31]}  
+ \frac{2}{[12]^2[23][31]}         \right]\ln{[23]} \no\\
+ \left[\frac{2}{[12]^2[23]^2} + \frac{4}{[23]^2[31]^2} + \frac{-8}{[12][23]^2[31]} + \frac{-2}{[12]^2[23][31]} \right]\ln{[31]}
\end{eqnarray*}
Integral $A_2$ is finite in UV and IR. Integral $I_3 = J_2.$  The third contribution to $I_1$ is (with factor $-1/2$)
\begin{eqnarray*}
A_3 \equiv \frac{(31)_\nu}{[31]^2}\int~Dy~ \le\pd_\rho^{(y)}\frac{(2y)_\sigma}{[2y]^{2-\e}}\ri \frac{1}{[1y]^{1-\e}} \int~Dz \Pi_{\rho\nu}(z3)
\le \pd^2_{(z)}\frac{(2z)_{\sigma}}{[2z]^{1-\e}} \ri \frac{1}{[yz]^{1-\e}} = \\
-2\frac{(31)_\nu}{[31]^2}\int~Dy~ \le\pd_\sigma^{(2)}\frac{(2y)_\rho}{[2y]^{2-\e}}\ri \frac{1}{[1y]^{1-\e}} \int~Dz \Pi_{\rho\nu}(z3)
\le \pd^{(2)}_{\sigma}\frac{1}{[2z]^{1-\e}} \ri \frac{1}{[yz]^{1-\e}} = \\
-\frac{(31)_\nu}{[31]^2}\pd_{(2)}^2\int~Dy~Dz \Pi_{\rho\nu}(z3)  \frac{(2y)_\rho}{[2y]^{2}[1y][2z][yz]} \\
- \frac{k}{2(1-\e)}\frac{(31)_\nu}{[31]^2}\int~Dy~ \pd_\rho^{(2)} \delta(2y) \frac{1}{[1y]^{1-\e}} \int~Dz \Pi_{\rho\nu}(z3) \frac{1}{[2z]^{1-\e}[yz]^{1-\e}} \\
+ \frac{(31)_\nu}{[31]^2}\int~Dy~ \frac{(2y)_\rho}{[2y]^{2-\e}} \frac{1}{[1y]^{1-\e}} \int~Dz \Pi_{\rho\nu}(z3) k\delta(2z) \frac{1}{[yz]^{1-\e}} = \\
\\
-\frac{(31)_\nu}{[31]^2}\pd_{(2)}^2\int~Dy~Dz \Pi_{\rho\nu}(z3)  \frac{(2y)_\rho}{[2y]^{2}[1y][2z][yz]} \\
- k\frac{(31)_\nu}{[31]^2}\frac{(12)_\rho}{[12]^{2-\e}} \int~Dz \Pi_{\rho\nu}(z3) \frac{1}{[2z]^{2-2\e}}
+ k\frac{(31)_\nu}{[31]^2}\Pi_{\rho\nu}(23) \int~Dy~ \frac{(2y)_\rho}{[1y]^{1-\e}[2y]^{3-2\e}} = \\
-\frac{(31)_\nu}{[31]^2}\pd_{(2)}^2\int~Dy~Dz \Pi_{\rho\nu}(z3)  \frac{(2y)_\rho}{[2y]^{2}[1y][2z][yz]} \\
- k\frac{(31)_\nu}{[31]^2}\frac{(12)_\rho}{[12]^{2-\e}} \int~Dz \Pi_{\rho\nu}(z3) \frac{1}{[2z]^{2-2\e}}
- \frac{k}{2}\frac{(31)_\nu}{[31]^2}\Pi_{\rho\nu}(23) \int~Dy~ \frac{(1y)_\rho}{[1y]^{2-\e}[2y]^{2-2\e}} \equiv \\
\equiv -A_{31} - \frac{k}{2}M_{10} - \frac{k}{4}M_{11}
\end{eqnarray*}
Intergal $A_3$ is singular in UV. Integral $A_{31}$ is finite. The fourth contribution to $I_1$ is (with factor $-1/2$)
\begin{eqnarray*}
A_4 \equiv \frac{(31)_\nu}{[31]^2}\int~Dy~ \le\pd_\rho^{(y)}\frac{(2y)_\sigma}{[2y]^{2-\e}}\ri \frac{1}{[1y]^{1-\e}} \int~Dz \Pi_{\rho\nu}(z3)
\frac{(2z)_{\sigma}}{[2z]^{1-\e}} \le \pd^2_{(z)} \frac{1}{[yz]^{1-\e}} \ri  = \\
k\frac{(31)_\nu}{[31]^2}\int~Dy~ \le\pd_\rho^{(y)}\frac{(2y)_\sigma}{[2y]^{2-\e}}\ri \frac{1}{[1y]^{1-\e}} \Pi_{\rho\nu}(y3)\frac{(2y)_{\sigma}}{[2y]^{1-\e}}   = \\
k\frac{(31)_\nu}{[31]^2}\int~Dy~ \le - \frac{g_{\sigma\rho}}{[2y]^{2-\e}} + 2(2-\e)\frac{(2y)_\rho(2y)_\sigma}{[2y]^{3-\e}} \ri
\frac{1}{[1y]^{1-\e}} \Pi_{\rho\nu}(y3)\frac{(2y)_{\sigma}}{[2y]^{1-\e}}   = \\
k(3 - 2\e)\frac{(31)_\nu}{[31]^2}\int~Dy~ \Pi_{\rho\nu}(y3)\frac{(2y)_{\rho}}{[2y]^{3-2\e}[1y]^{1-\e}} = \\
-k\frac{3 - 2\e}{2}\frac{(31)_\nu}{[31]^2}\int~Dy~ \Pi_{\rho\nu}(y3)\frac{(1y)_{\rho}}{[2y]^{2-2\e}[1y]^{2-\e}} = 
-k\frac{3 - 2\e}{4}M_{11}
\end{eqnarray*}
This integral is singular. We collect all singular integrals from $A_1,$ $A_3$ and $A_4.$ 
\begin{eqnarray*}
M_2 \equiv 
- \frac{k}{2}M_{10}  - \frac{k}{4}M_{11} + \frac{k}{4}M_{10}  + \frac{k}{8}M_{12} + k\frac{3 - 2\e}{8}M_{11} =  \\
- \frac{k}{4}M_{10}  + \frac{k}{8}M_{12} + k\frac{1 - 2\e}{8}M_{11} =  \\
k\left[\frac{1/8}{[12]^2[23]^2} + \frac{-5/8}{[12]^2[31]^2} + \frac{1/8}{[12]^2[23]^2}  + \frac{1/2}{[12][23][31]^2}  + \frac{-1/4}{[12][23]^2[31]}  
+  \frac{1/2}{[12]^2[23][31]}      \right]  \\
+ k\left[\frac{-1/4}{[12]^2[23]^2} + \frac{1/2}{[12]^2[31]^2} + \frac{-1/4}{[23]^2[31]^2} + \frac{-1/4}{[12][23][31]^2} + \frac{1/2}{[12][23]^2[31]} 
+ \frac{-1/4}{[12]^2[23][31]} \right]\ln[12] \\
+ k\left[\frac{1/8}{[12]^2[23]^2} + \frac{-1/4}{[12]^2[31]^2} + \frac{1/8}{[23]^2[31]^2} + \frac{1/8}{[12][23][31]^2} + \frac{-1/4}{[12][23]^2[31]} 
+  \frac{1/8}{[12]^2[23][31]} \right]\ln[23] \\
+ k\left[\frac{1/8}{[12]^2[23]^2} + \frac{-1/4}{[12]^2[31]^2} + \frac{1/8}{[23]^2[31]^2} + \frac{1/8}{[12][23][31]^2} + \frac{-1/4}{[12][23]^2[31]} 
+  \frac{1/8}{[12]^2[23][31]} \right]\ln[31] 
 \end{eqnarray*}
Combining two finite integrals from $A_1$ and $A_3,$ we obtain 
\begin{eqnarray*}
K_2 \equiv - k A_{11} + \frac{1}{2}A_{31} = k\left[\frac{-1/2}{[12]^2[31]^2}  + \frac{-1/2}{[12][23][31]^2}   +  \frac{1/2}{[12]^2[23][31]} \right] \no \\
+ k\left[\frac{-1/2}{[12][31]^2} + \frac{[23]}{[12]^2[31]^2}  +\frac{-1}{[12]^2[31]}  \right]J(1,1,1) \no\\
+ k\left[\frac{-1/2}{[12]^2[31]^2} +  \frac{1/2}{[12]^2[23][31]} \right]\ln[12] + k \left[\frac{1/2}{[12]^2[31]^2} + \frac{-1/2}{[12]^2[23][31]} \right] \ln[31] \no
\\
+ \left[\frac{1}{[12]^2[23]^2} + \frac{-1}{[12][23]^2[31]}  + \frac{-1}{[12]^2[23][31]}     \right] \no\\
+ \left[\frac{1}{[12][23]^2} + \frac{1}{[12]^2[31]} + \frac{-[31]}{[12]^2[23]^2}  \right]J(1,1,1) \no\\
+ \left[\frac{1}{[12]^2[23]^2} + \frac{-2}{[12][23]^2[31]} + \frac{-1}{[12]^2[23][31]} \right]\ln[12] \no\\
+ \left[\frac{-1}{[12]^2[23]^2} + \frac{1}{[12][23]^2[31]} +  \frac{1}{[12]^2[23][31]} \right]\ln[23] + \frac{1}{[12][23]^2[31]} \ln[31]   
\end{eqnarray*}
Collecting all the parts together, 
\begin{eqnarray*}
I_{1} \equiv K_2 + \frac{1}{2}A_2 + M_2 
\end{eqnarray*}
The result is written in Eq. (\ref{FLT2.2}).

\section[]{}
\setcounter{equation}{0}
\label{App:B}

In this Appendix we provide the calculation of $T_{21}.$ We will reduce the number of indices further by means of simple algebra, 
\begin{eqnarray}
\Pi_{\rho\sigma}(2z) \frac{(2y)_\sigma}{[2y]^{2-\e}} =  \le \frac{g_{\rho\sigma}}{[2z]^{1-\e}} + 2(1-\e)\frac{(2z)_\rho (2z)_\sigma}{[2z]^{2-\e}}\ri \frac{(2y)_\sigma}{[2y]^{2-\e}} = \no\\
\le \frac{2g_{\rho\sigma}}{[2z]^{1-\e}} - \pd_\sigma^{(2)} \frac{(2z)_{\rho}}{[2z]^{1-\e}} \ri \frac{(2y)_\sigma}{[2y]^{2-\e}} =
\frac{2(2y)_\rho}{[2z]^{1-\e}[2y]^{2-\e}}  - \le\pd_\sigma^{(2)} \frac{(2z)_{\rho}}{[2z]^{1-\e}}\ri \frac{(2y)_\sigma}{[2y]^{2-\e}}  = \no\\
\frac{2(2y)_\rho}{[2z]^{1-\e}[2y]^{2-\e}}  +  \frac{1}{2(1-\e)}\le\pd_\sigma^{(2)} \frac{(2z)_{\rho}}{[2z]^{1-\e}}\ri\le \pd_\sigma^{(2)} \frac{1}{[2y]^{1-\e}}\ri  = \no\\
\frac{2(2y)_\rho}{[2z]^{1-\e}[2y]^{2-\e}} +  \frac{1}{4(1-\e)}\left[ \pd^2_{(2)}\le \frac{(2z)_{\rho}}{[2z]^{1-\e}[2y]^{1-\e}} \ri  
- \le \pd^2_{(2)}\frac{(2z)_{\rho}}{[2z]^{1-\e}} \ri\frac{1}{[2y]^{1-\e}} \right. \no\\
\left. - \frac{(2z)_{\rho}}{[2z]^{1-\e}} \le \pd^2_{(2)} \frac{1}{[2y]^{1-\e}} \ri\right] \label{trick3}
\end{eqnarray}
We need to calculate the following integral
\begin{eqnarray*}
T_{21} = \frac{(31)_\nu}{[31]^2}\int~Dy~\frac{(2y)_\sigma}{[2y]^{2-\e}}\int~Dz \Pi_{\rho\nu}(z3) \le \pd^{(z)}_\mu  \Pi_{\rho\sigma}(z2)\ri \times \\
\times \left\{\frac{2(1y)_\mu}{[yz]^{1-\e}[1y]^{2-\e}}  - \frac{(yz)_{\mu}}{[yz]^{2-\e}[1y]^{1-\e}} \right\} = \no\\
 \frac{(31)_\nu}{[31]^2}\int~Dy~Dz \Pi_{\rho\nu}(z3) \le \pd^{(z)}_\mu  \Pi_{\rho\sigma}(z2) \frac{(2y)_\sigma}{[2y]^{2-\e}} \ri 
\left\{\frac{2(1y)_\mu}{[yz]^{1-\e}[1y]^{2-\e}}  - \frac{(yz)_{\mu}}{[yz]^{2-\e}[1y]^{1-\e}} \right\}  
\end{eqnarray*}
Now we take into account the following equality,
\begin{eqnarray}
\frac{2(1y)_\mu}{[yz]^{1-\e}[1y]^{2-\e}}  - \frac{(yz)_{\mu}}{[yz]^{2-\e}[1y]^{1-\e}}  =  \no\\
\frac{1}{1-\e}\pd^{(y)}_\mu \le\frac{1}{[yz]^{1-\e}[1y]^{1-\e}}\ri + \frac{(yz)_{\mu}}{[yz]^{2-\e}[1y]^{1-\e}} \label{trick4}
\end{eqnarray}
The first term on the l.h.s. of Eq. (\ref{trick4}) correponds to the finite part of $T_{21},$ while the second term on the l.h.s. of Eq. (\ref{trick4}) contains 
a singular contribution.  On the r.h.s. we produce a mixture of singular and finite contributions. However, it is more easy to work with such a representation.  
We do several steps to calculate integral $T_{21},$ taking into account Eq. (\ref{trick3}). The first contribution (with factor $1/(1-\e)$) to $T_{21}$ is 
\begin{eqnarray*}
B_1 \equiv \frac{(31)_\nu}{[31]^2}\int~Dy~Dz \Pi_{\rho\nu}(z3) \le \pd^{(z)}_\mu  \frac{2(2y)_\rho}{[2z]^{1-\e}[2y]^{2-\e}}  \ri \pd^{(y)}_\mu \le\frac{1}{[yz]^{1-\e}[1y]^{1-\e}}\ri = \\
\frac{(31)_\nu}{[31]^2}\int~Dy~Dz \Pi_{\rho\nu}(z3) \le \pd^{(z)}_\mu  \frac{2(2y)_\mu}{[2z]^{1-\e}[2y]^{2-\e}}  \ri \pd^{(y)}_\rho \le\frac{1}{[yz]^{1-\e}[1y]^{1-\e}}\ri = \\
\frac{1}{1-\e}\frac{(31)_\nu}{[31]^2}\int~Dy~Dz \Pi_{\rho\nu}(z3) \le \pd^{(z)}_\mu  \frac{1}{[2z]^{1-\e}} \ri \le\pd^{(y)}_\mu \frac{1}{[2y]^{1-\e}} \ri
\pd^{(y)}_\rho \le\frac{1}{[yz]^{1-\e}[1y]^{1-\e}}\ri = \\
\frac{1}{2(1-\e)}\frac{(31)_\nu}{[31]^2}\int~Dy~Dz \Pi_{\rho\nu}(z3)\left[ \pd^2_{(2)}\le \frac{1}{[2y]^{1-\e}[2z]^{1-\e}} \ri  \right.\\
\left. - \le \pd^2_{(2)} \frac{1}{[2y]^{1-\e}} \ri \frac{1}{[2z]^{1-\e}}    -  \frac{1}{[2y]^{1-\e}}  \le \pd^2_{(2)} \frac{1}{[2z]^{1-\e}} \ri \right]
\pd^{(y)}_\rho \le\frac{1}{[yz]^{1-\e}[1y]^{1-\e}}\ri = \\
\\
\frac{1}{2(1-\e)}\frac{(31)_\nu}{[31]^2}\pd^2_{(2)} \int~Dy~Dz \Pi_{\rho\nu}(z3) \frac{1}{[2y]^{1-\e}[2z]^{1-\e}} \pd^{(y)}_\rho \le\frac{1}{[yz]^{1-\e}[1y]^{1-\e}}\ri \\
- \frac{1}{2(1-\e)}\frac{(31)_\nu}{[31]^2}\int~Dy~Dz \Pi_{\rho\nu}(z3)k\delta(2y) \frac{1}{[2z]^{1-\e}} \pd^{(y)}_\rho \le\frac{1}{[yz]^{1-\e}[1y]^{1-\e}}\ri \\
- \frac{1}{2(1-\e)}\frac{(31)_\nu}{[31]^2}\int~Dy~Dz \Pi_{\rho\nu}(z3)k\delta(2z) \frac{1}{[2y]^{1-\e}} \pd^{(y)}_\rho \le\frac{1}{[yz]^{1-\e}[1y]^{1-\e}}\ri = \\
\frac{(31)_\nu}{[31]^2}\pd^2_{(2)} \int~Dy~Dz \Pi_{\rho\nu}(z3) \frac{1}{[2y]^{1-\e}[2z]^{1-\e}} \frac{(zy)_\rho}{[yz]^{2-\e}[1y]^{1-\e}} \\
+ \frac{(31)_\nu}{[31]^2}\pd^2_{(2)} \int~Dy~Dz \Pi_{\rho\nu}(z3) \frac{1}{[2y]^{1-\e}[2z]^{1-\e}} \frac{(1y)_\rho}{[yz]^{1-\e}[1y]^{2-\e}} \\
- \frac{k}{2(1-\e)}\frac{(31)_\nu}{[31]^2}\int~Dz \Pi_{\rho\nu}(z3)\frac{1}{[2z]^{1-\e}} \pd^{(2)}_\rho \le\frac{1}{[2z]^{1-\e}[12]^{1-\e}}\ri \\
- \frac{k}{2(1-\e)}\frac{(31)_\nu}{[31]^2} \Pi_{\rho\nu}(23) \int~Dy  \frac{1}{[2y]^{1-\e}} \pd^{(y)}_\rho \le\frac{1}{[y2]^{1-\e}[1y]^{1-\e}}\ri = \\
\\
- \frac{(31)_\nu}{[31]^2}\pd^2_{(2)} \int~Dy~Dz \Pi_{\rho\nu}(z3) \frac{(yz)_\rho}{[yz]^{2}[1y][2y][2z]} \\
+ \frac{(31)_\nu}{[31]^2}\pd^2_{(2)} \int~Dy~Dz \Pi_{\rho\nu}(z3) \frac{(1y)_\rho}{[yz][1y]^{2}[2y][2z]} \\
+  k\frac{(31)_\nu}{[31]^2}\frac{(21)_\rho}{[12]^{2-\e}} \int~Dz \Pi_{\rho\nu}(z3)\frac{1}{[2z]^{2-2\e}}  
+  k\frac{(31)_\nu}{[31]^2} \Pi_{\rho\nu}(23) \int~Dy  \frac{(2y)_\rho}{[2y]^{3-2\e}[1y]^{1-\e}} = \\
\\
- \frac{(31)_\nu}{[31]^2}\pd^2_{(2)} \int~Dy~Dz \Pi_{\rho\nu}(z3) \frac{(yz)_\rho}{[yz]^{2}[1y][2y][2z]} \\
+ \frac{(31)_\nu}{[31]^2}\pd^2_{(2)} \int~Dy~Dz \Pi_{\rho\nu}(z3) \frac{(1y)_\rho}{[yz][1y]^{2}[2y][2z]} \\
+  k\frac{(31)_\nu}{[31]^2}\frac{(21)_\rho}{[12]^{2-\e}} \int~Dz \Pi_{\rho\nu}(z3)\frac{1}{[2z]^{2-2\e}}  
-  \frac{k}{2}\frac{(31)_\nu}{[31]^2} \Pi_{\rho\nu}(23) \int~Dy  \frac{(1y)_\rho}{[2y]^{2-2\e}[1y]^{2-\e}} \equiv  \\
\equiv - B_{11} + J_{31}/2 -\frac{k}{2}M_{10} -\frac{k}{4}M_{12} 
\end{eqnarray*}
Integral $B_1$ is singular in UV.  Integral $B_{11}$ is finite. We realize the strategy to present all the integrals in terms 
of basic integrals used to calculate $T_1.$ The second contribution to $T_{21}$ (with factor 1) is 
\begin{eqnarray*}
B_2 \equiv \frac{(31)_\nu}{[31]^2}\int~Dy~Dz \Pi_{\rho\nu}(z3) \le \pd^{(z)}_\mu  \frac{2(2y)_\rho}{[2z]^{1-\e}[2y]^{2-\e}} \ri \frac{(yz)_{\mu}}{[yz]^{2-\e}[1y]^{1-\e}}  = \\
\frac{1}{2(1-\e)}\frac{(31)_\nu}{[31]^2}\int~Dy~Dz \Pi_{\rho\nu}(z3) \frac{2(2y)_\rho}{[2y]^{2-\e}} \le \pd^{(z)}_\mu  \frac{1}{[2z]^{1-\e}}  \ri
 \le \pd^{(z)}_\mu  \frac{1}{[yz]^{1-\e}} \ri  \frac{1}{[1y]^{1-\e}}  = \\
\frac{1}{4(1-\e)}\frac{(31)_\nu}{[31]^2}\int~Dy~Dz \Pi_{\rho\nu}(z3) \frac{2(2y)_\rho}{[2y]^{2-\e}} \left[ \pd^2_{(z)}\le \frac{1}{[2z]^{1-\e}[yz]^{1-\e}} \ri  \right.\\
\left. - \le \pd^2_{(z)} \frac{1}{[yz]^{1-\e}} \ri \frac{1}{[2z]^{1-\e}}    -  \frac{1}{[yz]^{1-\e}}  \le \pd^2_{(z)} \frac{1}{[2z]^{1-\e}} \ri \right] \frac{1}{[1y]^{1-\e}}  = \\
\frac{1}{4(1-\e)}\frac{(31)_\nu}{[31]^2}\pd^2_{(3)}\int~Dy~Dz \Pi_{\rho\nu}(z3) \frac{2(2y)_\rho}{[2y]^{2-\e}[2z]^{1-\e}[yz]^{1-\e}[1y]^{1-\e}}  \\
- \frac{k}{4(1-\e)}\frac{(31)_\nu}{[31]^2}\int~Dy~Dz \Pi_{\rho\nu}(z3) \frac{2(2y)_\rho}{[2y]^{2-\e}}\delta(yz)\frac{1}{[2z]^{1-\e}[1y]^{1-\e}}  \\
- \frac{k}{4(1-\e)}\frac{(31)_\nu}{[31]^2}\int~Dy~Dz \Pi_{\rho\nu}(z3) \frac{2(2y)_\rho}{[2y]^{2-\e}[yz]^{1-\e}} \delta(2z)  \frac{1}{[1y]^{1-\e}}  = \\
\frac{1}{4(1-\e)}\frac{(31)_\nu}{[31]^2}\pd^2_{(3)}\int~Dy~Dz \Pi_{\rho\nu}(z3) \frac{2(2y)_\rho}{[2y]^{2-\e}[2z]^{1-\e}[yz]^{1-\e}[1y]^{1-\e}}  \\
- \frac{k}{4(1-\e)}\frac{(31)_\nu}{[31]^2}\int~Dz \Pi_{\rho\nu}(z3) \frac{2(2z)_\rho}{[2z]^{3-2\e}[1z]^{1-\e}}
- \frac{k}{4(1-\e)}\frac{(31)_\nu}{[31]^2}\int~Dy \Pi_{\rho\nu}(23) \frac{2(2y)_\rho}{[2y]^{3-2\e}[1y]^{1-\e}} = \\
\\
\frac{1}{2}\frac{(31)_\nu}{[31]^2}\pd^2_{(3)}\int~Dy~Dz \Pi_{\rho\nu}(z3) \frac{(2y)_\rho}{[2y]^{2}[2z][yz][1y]}  \\
+ \frac{k}{4(1-\e)}\frac{(31)_\nu}{[31]^2}\int~Dz \Pi_{\rho\nu}(z3) \frac{(1z)_\rho}{[2z]^{2-2\e}[1z]^{2-\e}}
+ \frac{k}{4(1-\e)}\frac{(31)_\nu}{[31]^2}\Pi_{\rho\nu}(23)  \int~Dy \frac{(1y)_\rho}{[2y]^{2-2\e}[1y]^{2-\e}} \equiv \\
\equiv B_{21}/2 + \frac{k}{8(1-\e)}M_{11} + \frac{k}{8(1-\e)}M_{12}
\end{eqnarray*}
Integral $B_2$ is singular in UV. Integral $B_{21}$ is finite. $M$-integrals are singular. The third contribution to $T_{21}$ correponds to the second term  in Eq. (\ref{trick3}) 
(with factor $ (1/4)/(1-\e)^2 $) is 
\begin{eqnarray*}
B_3 \equiv \frac{(31)_\nu}{[31]^2}\int~Dy~Dz \Pi_{\rho\nu}(z3) \pd^2_{(2)}\le \pd^{(z)}_\mu\frac{(2z)_{\rho}}{[2z]^{1-\e}[2y]^{1-\e}} \ri 
\pd^{(y)}_\mu \le\frac{1}{[yz]^{1-\e}[1y]^{1-\e}}\ri = \\
\frac{(31)_\nu}{[31]^2} \pd^2_{(2)} \int~Dy~Dz \Pi_{\rho\nu}(z3) \le \pd^{(z)}_\rho \frac{(2z)_{\mu}}{[2z]^{1-\e}[2y]^{1-\e}} \ri \pd^{(y)}_\mu \le\frac{1}{[yz]^{1-\e}[1y]^{1-\e}}\ri = \\
\frac{(31)_\nu}{[31]^2} \pd^2_{(2)} \int~Dy~Dz \Pi_{\rho\nu}(z3) \le \pd^{(y)}_\mu \frac{(2z)_{\mu}}{[2z]^{1-\e}[2y]^{1-\e}} \ri \pd^{(z)}_\rho \le\frac{1}{[yz]^{1-\e}[1y]^{1-\e}}\ri = \\
2(1-\e)\frac{(31)_\nu}{[31]^2} \pd^2_{(2)} \int~Dy~Dz \Pi_{\rho\nu}(z3) \frac{(2z)_{\mu}(2y)_\mu}{[2z]^{1-\e}[2y]^{2-\e}}\pd^{(z)}_\rho \le\frac{1}{[yz]^{1-\e}[1y]^{1-\e}}\ri = \\
2\frac{(31)_\nu}{[31]^2} \pd^2_{(2)} \int~Dy~Dz \Pi_{\rho\nu}(z3) \frac{(2z)_{\mu}(2y)_\mu}{[2z][2y]^{2}}\pd^{(z)}_\rho \le\frac{1}{[yz][1y]}\ri
\end{eqnarray*}
Integral $B_3$ is finite integral in UV and IR. Another contribution to  $T_{21}$  that corresponds to the second term in Eq. (\ref{trick3}) is  (with factor $ (1/4)/(1-\e) $  )
\begin{eqnarray*}
B_4 \equiv \frac{(31)_\nu}{[31]^2}\int~Dy~Dz \Pi_{\rho\nu}(z3) \pd^2_{(2)}\le \pd^{(z)}_\mu\frac{(2z)_{\rho}}{[2z]^{1-\e}[2y]^{1-\e}} \ri \frac{(yz)_{\mu}}{[yz]^{2-\e}[1y]^{1-\e}} = \\
\frac{1}{2(1-\e)}\frac{(31)_\nu}{[31]^2}\pd^2_{(2)} \int~Dy~Dz \Pi_{\rho\nu}(z3) \frac{1}{[2y]^{1-\e}} \le \pd^{(z)}_\mu \frac{(2z)_{\rho}}{[2z]^{1-\e}}\ri
\le \pd^{(z)}_\mu  \frac{1}{[yz]^{1-\e}} \ri  \frac{1}{[1y]^{1-\e}} =  \\
\frac{1}{4(1-\e)}\frac{(31)_\nu}{[31]^2}\pd^2_{(2)} \int~Dy~Dz \Pi_{\rho\nu}(z3) \frac{1}{[2y]^{1-\e}} \left[\pd^2_{(z)}\le \frac{(2z)_{\rho}}{[2z]^{1-\e}}\frac{1}{[yz]^{1-\e}} \ri \right.\\
- \left.  \le \pd^2_{(z)} \frac{(2z)_{\rho}}{[2z]^{1-\e}} \ri \frac{1}{[yz]^{1-\e}}  -  \le\pd^2_{(z)}\frac{1}{[yz]^{1-\e}}\ri \frac{(2z)_{\rho}}{[2z]^{1-\e}} \right]\frac{1}{[1y]^{1-\e}}
=  \\
\frac{1}{4(1-\e)}\frac{(31)_\nu}{[31]^2}\pd^2_{(2)} \pd^2_{(3)} \int~Dy~Dz \Pi_{\rho\nu}(z3) \frac{(2z)_{\rho}}{[2z]^{1-\e}[yz]^{1-\e}[1y]^{1-\e}[2y]^{1-\e}} \\
- \frac{(31)_\nu}{[31]^2}\pd^2_{(2)} \int~Dy~Dz \Pi_{\rho\nu}(z3) \frac{1}{[2y]^{1-\e}} \frac{1}{[2z]^{1-\e}}  \frac{(yz)_\rho}{[yz]^{2-\e}} \frac{1}{[1y]^{1-\e}} \\
- \frac{k}{4(1-\e)}\frac{(31)_\nu}{[31]^2}\pd^2_{(2)} \int~Dy~Dz \Pi_{\rho\nu}(z3) \frac{1}{[2y]^{1-\e}} \delta(yz)\frac{(2z)_{\rho}}{[2z]^{1-\e}} \frac{1}{[1y]^{1-\e}} = \\
\frac{1}{4(1-\e)}\frac{(31)_\nu}{[31]^2}\pd^2_{(2)} \pd^2_{(3)} \int~Dy~Dz \Pi_{\rho\nu}(z3) \frac{(2z)_{\rho}}{[2z]^{1-\e}[yz]^{1-\e}[1y]^{1-\e}[2y]^{1-\e}} \\
- \frac{(31)_\nu}{[31]^2}\pd^2_{(2)} \int~Dy~Dz \Pi_{\rho\nu}(z3) \frac{(yz)_\rho}{[yz]^{2-\e}[1y]^{1-\e}[2y]^{1-\e}[2z]^{1-\e}} \\
- \frac{k}{4(1-\e)}\frac{(31)_\nu}{[31]^2}\pd^2_{(2)} \int~Dz \Pi_{\rho\nu}(z3)  \frac{(2z)_{\rho}}{[2z]^{2-2\e}[1z]^{1-\e}} = \\
\\
\frac{1}{4}\frac{(31)_\nu}{[31]^2}\pd^2_{(2)} \pd^2_{(3)} \int~Dy~Dz \Pi_{\rho\nu}(z3) \frac{(2z)_{\rho}}{[2z][yz][1y][2y]} \\
- \frac{(31)_\nu}{[31]^2}\pd^2_{(2)} \int~Dy~Dz \Pi_{\rho\nu}(z3) \frac{(yz)_\rho}{[yz]^{2}[1y][2y][2z]} 
- \frac{k}{4}\frac{(31)_\nu}{[31]^2}\pd^2_{(2)} \int~Dz \Pi_{\rho\nu}(z3)  \frac{(2z)_{\rho}}{[2z]^{2}}\frac{1}{[1z]} \equiv \\
\equiv  \frac{1}{4}B_{41} - B_{11} - \frac{k}{4}B_{42}.
\end{eqnarray*}
Integral $B_4$ is finite in UV and IR. Integrals $B_{41},$ $B_{11}$ and $B_{42}$  are finite. A contribution to $T_{21}$ that corresponds to the third term in Eq. (\ref{trick3}) 
is (with factor $ -(1/4)/(1-\e)^2 $)
\begin{eqnarray*}
B_5 \equiv \frac{(31)_\nu}{[31]^2}\int~Dy~Dz \Pi_{\rho\nu}(z3) \le \pd^2_{(2)}\pd^{(z)}_\mu \frac{(2z)_{\rho}}{[2z]^{1-\e}} \ri
\frac{1}{[2y]^{1-\e}} \pd^{(y)}_\mu \le\frac{1}{[yz]^{1-\e}[1y]^{1-\e}}\ri = \\
\\
-4(1-\e)\frac{(31)_\nu}{[31]^2}\int~Dy~Dz \Pi_{\rho\nu}(z3) \le \pd^{(z)}_\rho \frac{(2z)_{\mu}}{[2z]^{2-\e}} \ri
\frac{1}{[2y]^{1-\e}} \pd^{(y)}_\mu \le\frac{1}{[yz]^{1-\e}[1y]^{1-\e}}\ri = \\
- 8(1-\e)^2\frac{(31)_\nu}{[31]^2} \int~Dy~Dz \Pi_{\rho\nu}(z3)   \frac{(2z)_{\mu}}{[2z]^{2-\e}} \frac{(2y)_{\mu}}{[2y]^{2-\e}} \pd^{(z)}_\rho \le\frac{1}{[yz]^{1-\e}[1y]^{1-\e}}\ri = \\
- \frac{(31)_\nu}{[31]^2} \int~Dy~Dz \Pi_{\rho\nu}(z3) \left[ \pd^2_{(2)}\le \frac{1}{[2z]^{1-\e}[2y]^{1-\e}} \ri  \right.\\
\left. - \le \pd^2_{(2)} \frac{1}{[2y]^{1-\e}} \ri \frac{1}{[2z]^{1-\e}}    -  \frac{1}{[2y]^{1-\e}}  \le \pd^2_{(2)} \frac{1}{[2z]^{1-\e}} \ri \right]
\pd^{(z)}_\rho \le\frac{1}{[yz]^{1-\e}[1y]^{1-\e}}\ri = \\
- \frac{(31)_\nu}{[31]^2}  \pd^2_{(2)} \int~Dy~Dz \Pi_{\rho\nu}(z3) \frac{1}{[2z]^{1-\e}[2y]^{1-\e}} \pd^{(z)}_\rho \le\frac{1}{[yz]^{1-\e}[1y]^{1-\e}}\ri  \\
+ \frac{(31)_\nu}{[31]^2} \int~Dy~Dz \Pi_{\rho\nu}(z3) k\delta(2y)  \frac{1}{[2z]^{1-\e}} \pd^{(z)}_\rho \le\frac{1}{[yz]^{1-\e}[1y]^{1-\e}}\ri \\
+\frac{(31)_\nu}{[31]^2} \int~Dy~Dz \Pi_{\rho\nu}(z3) \frac{1}{[2y]^{1-\e}} k\delta(2z) \pd^{(z)}_\rho \le\frac{1}{[yz]^{1-\e}[1y]^{1-\e}}\ri = \\
+ 2(1-\e)\frac{(31)_\nu}{[31]^2}  \pd^2_{(2)} \int~Dy~Dz \Pi_{\rho\nu}(z3) \frac{(2z)_\rho}{[2z]^{2-\e}[2y]^{1-\e}[yz]^{1-\e}[1y]^{1-\e}}  \\
+ k\frac{(31)_\nu}{[31]^2} \int~Dz \Pi_{\rho\nu}(z3) \frac{1}{[2z]^{1-\e}} \pd^{(z)}_\rho \le\frac{1}{[2z]^{1-\e}[12]^{1-\e}}\ri \\
+  k\frac{(31)_\nu}{[31]^2} \int~Dy \Pi_{\rho\nu}(23) \frac{1}{[2y]^{1-\e}} \pd^{(2)}_\rho \le\frac{1}{[y2]^{1-\e}[1y]^{1-\e}}\ri = \\
+ 2\frac{(31)_\nu}{[31]^2}  \pd^2_{(2)} \int~Dy~Dz \Pi_{\rho\nu}(z3) \frac{(2z)_\rho}{[2z]^{2}[2y][yz][1y]}  \\
-  2k(1-\e)\frac{(31)_\nu}{[31]^2} \int~Dy \Pi_{\rho\nu}(23) \frac{(2y)_\rho}{[2y]^{3-2\e}[1y]^{1-\e}} = \\
\\
+ 2\frac{(31)_\nu}{[31]^2}  \pd^2_{(2)} \int~Dy~Dz \Pi_{\rho\nu}(z3) \frac{(2z)_\rho}{[2z]^{2}[2y][yz][1y]}  \\
+ k(1-\e)\frac{(31)_\nu}{[31]^2} \Pi_{\rho\nu}(23) \int~Dy  \frac{(1y)_\rho}{[2y]^{2-2\e}[1y]^{2-\e}} \equiv 2B_{51} + \frac{k(1-\e)}{2}M_{12}.
\end{eqnarray*}
Integral $B_{51}$ is finite. Another contribution to $T_{21}$ that corresponds to the third term in Eq. (\ref{trick3}) is (with factor $ -(1/4)/(1-\e)$)
\begin{eqnarray*}
B_6 \equiv \frac{(31)_\nu}{[31]^2}\int~Dy~Dz \Pi_{\rho\nu}(z3) \le \pd^2_{(2)}\pd^{(z)}_\mu \frac{(2z)_{\rho}}{[2z]^{1-\e}} \ri  
\frac{1}{[2y]^{1-\e}} \frac{(yz)_{\mu}}{[yz]^{2-\e}[1y]^{1-\e}} = \\
- 4(1-\e)\frac{(31)_\nu}{[31]^2}\int~Dy~Dz \Pi_{\rho\nu}(z3) \le \pd^{(z)}_\rho \frac{(2z)_{\mu}}{[2z]^{2-\e}} \ri  \frac{1}{[2y]^{1-\e}} \frac{(yz)_{\mu}}{[yz]^{2-\e}[1y]^{1-\e}} = \\
- \frac{(31)_\nu}{[31]^2}\int~Dy~Dz \Pi_{\rho\nu}(z3) \frac{1}{[2y]^{1-\e}[1y]^{1-\e}}
\left[\pd^2_{(z)}\le \frac{(2z)_{\rho}}{[2z]^{2-\e}}\frac{1}{[yz]^{1-\e}} \ri \right.\\
- \left.  \le \pd^2_{(z)} \frac{(2z)_{\rho}}{[2z]^{2-\e}} \ri \frac{1}{[yz]^{1-\e}}  -  \le\pd^2_{(z)}\frac{1}{[yz]^{1-\e}}\ri \frac{(2z)_{\rho}}{[2z]^{2-\e}} \right] = \\
\\
- \frac{(31)_\nu}{[31]^2}\pd^2_{(3)}\int~Dy~Dz \Pi_{\rho\nu}(z3)\frac{(2z)_{\rho}}{[2z]^{2-\e}[yz]^{1-\e}[2y]^{1-\e}[1y]^{1-\e}} \\
- \frac{1}{2(1-\e)}\frac{(31)_\nu}{[31]^2}\int~Dy~Dz \Pi_{\rho\nu}(z3) \frac{1}{[2y]^{1-\e}[1y]^{1-\e}} k\pd_\rho^{(2)}\delta(2z) \frac{1}{[yz]^{1-\e}}   \\
+ \frac{(31)_\nu}{[31]^2}\int~Dy~Dz \Pi_{\rho\nu}(z3) \frac{1}{[2y]^{1-\e}[1y]^{1-\e}} k\delta(yz) \frac{(2z)_{\rho}}{[2z]^{2-\e}} = \\
\\
-  \frac{(31)_\nu}{[31]^2}\pd^2_{(3)}\int~Dy~Dz \Pi_{\rho\nu}(z3)\frac{(2z)_{\rho}}{[2z]^{2-\e}[yz]^{1-\e}[2y]^{1-\e}[1y]^{1-\e}} \\
-  k\frac{(31)_\nu}{[31]^2}\int~Dy~Dz \Pi_{\rho\nu}(z3) \frac{1}{[2y]^{1-\e}[1y]^{1-\e}} \delta(2z) \frac{(yz)_\rho}{[yz]^{2-\e}}   
+  k\frac{(31)_\nu}{[31]^2}\int~Dz \Pi_{\rho\nu}(z3) \frac{(2z)_{\rho}}{[2z]^{3-2\e}[1z]^{1-\e}} = \\
- \frac{(31)_\nu}{[31]^2}\pd^2_{(3)}\int~Dy~Dz \Pi_{\rho\nu}(z3)\frac{(2z)_{\rho}}{[2z]^{2}[yz][2y][1y]} \\
- k\frac{(31)_\nu}{[31]^2}\Pi_{\rho\nu}(23)\int~Dz \frac{(y2)_\rho}{[2y]^{3-2\e}[1y]^{1-\e}}  
+ k\frac{(31)_\nu}{[31]^2}\int~Dz \Pi_{\rho\nu}(z3) \frac{(2z)_{\rho}}{[2z]^{3-2\e}[1z]^{1-\e}} = \\
- \frac{(31)_\nu}{[31]^2}\pd^2_{(3)}\int~Dy~Dz \Pi_{\rho\nu}(z3)\frac{(2z)_{\rho}}{[2z]^{2}[yz][2y][1y]} \\
- \frac{k}{2}\frac{(31)_\nu}{[31]^2}\Pi_{\rho\nu}(23)\int~Dz \frac{(1y)_\rho}{[2y]^{2-2\e}[1y]^{2-\e}}  
- \frac{k}{2}\frac{(31)_\nu}{[31]^2}\int~Dz \Pi_{\rho\nu}(z3) \frac{(1z)_{\rho}}{[2z]^{2-2\e}[1z]^{2-\e}}  \equiv \\
\equiv -B_{61} - \frac{k}{4}M_{12}   - \frac{k}{4}M_{11}
\end{eqnarray*}
Integral $B_{61}$ is finite. Another  contribution to $T_{21}$ that corresponds to the fourth term in Eq. (\ref{trick3}) is (with factor $ -(1/4)/((1-\e)^2) $ )
\begin{eqnarray*}
B_7 \equiv \frac{(31)_\nu}{[31]^2}\int~Dy~Dz \Pi_{\rho\nu}(z3) \le \pd^{(z)}_\mu\frac{(2z)_{\rho}}{[2z]^{1-\e}}\ri
\le \pd^2_{(2)} \frac{1}{[2y]^{1-\e}} \ri \pd^{(y)}_\mu \le\frac{1}{[yz]^{1-\e}[1y]^{1-\e}}\ri = \\
\frac{(31)_\nu}{[31]^2}\int~Dy~Dz \Pi_{\rho\nu}(z3) \le \pd^{(z)}_\mu \frac{(2z)_{\rho}}{[2z]^{1-\e}}\ri 
k\delta(2y)\pd^{(y)}_\mu \le\frac{1}{[yz]^{1-\e}[1y]^{1-\e}}\ri = \\
\\
k\frac{(31)_\nu}{[31]^2}\int~Dz \Pi_{\rho\nu}(z3) \le \pd^{(z)}_\mu \frac{(2z)_{\rho}}{[2z]^{1-\e}}\ri \pd^{(2)}_\mu \le\frac{1}{[2z]^{1-\e}[12]^{1-\e}}\ri = \\
-k(1-\e)\frac{(31)_\nu}{[31]^2}\frac{(12)_\rho}{[12]^{2-\e}}\int~Dz \Pi_{\rho\nu}(z3) \frac{1}{[2z]^{2-2\e}} = -\frac{k(1-\e)}{2}M_{10}
\end{eqnarray*}
Another contribution to $T_{21}$ that corresponds to the fourth term in Eq. (\ref{trick3}) is (with factor $ -(1/4)/(1-\e)$)
\begin{eqnarray*}
\frac{(31)_\nu}{[31]^2}\int~Dy~Dz \Pi_{\rho\nu}(z3) \le \pd^{(z)}_\mu\frac{(2z)_{\rho}}{[2z]^{1-\e}}\ri \le \pd^2_{(2)} \frac{1}{[2y]^{1-\e}} \ri\frac{(yz)_{\mu}}{[yz]^{2-\e}[1y]^{1-\e}} = \\
\frac{(31)_\nu}{[31]^2}\int~Dy~Dz \Pi_{\rho\nu}(z3) \le \pd^{(z)}_\mu \frac{(2z)_{\rho}}{[2z]^{1-\e}}\ri k\delta(2y) \frac{(yz)_{\mu}}{[yz]^{2-\e}[1y]^{1-\e}}  = \\
k\frac{(31)_\nu}{[31]^2}\int~Dy~Dz \Pi_{\rho\nu}(z3) \le \pd^{(z)}_\mu \frac{(2z)_{\rho}}{[2z]^{1-\e}}\ri \frac{(2z)_{\mu}}{[2z]^{2-\e}[12]^{1-\e}} = 0  
\end{eqnarray*}
This integral is zero due to transversality of the gluon propagator. Combining the singular contribution $T_{23}$ 
from Eq. (\ref{SLT}) with the singular single integrals from $B_1,$  $B_2,$ $B_5,$  $B_6$ and $B_7$ we obtain 
\begin{eqnarray*}
M_ 3 \equiv \frac{3 k}{8 (1-\e)}M_{10} -  \frac{k}{2(1-\e)}M_{10}  - \frac{k}{4(1-\e)}M_{12} + \frac{k}{8(1-\e)}M_{11} + \frac{k}{8(1-\e)}M_{12} \\
 - \frac{k}{8(1-\e)}M_{12} + \frac{k}{16(1-\e)}M_{12} + \frac{k}{16(1-\e)}M_{11} + \frac{k}{8(1-\e)}M_{10} = 
\\
k\left[\frac{3/16}{[12]^2[23]^2} + \frac{3/16}{[12]^2[31]^2}  +\frac{3/16}{[23]^2[31^2}  + \frac{-3/8}{[12][23][31]^2}   
+  \frac{-3/8}{[12][23]^2[31]} +  \frac{-3/8}{[12]^2[23][31]} \right] \\
+ k\left[\frac{-3/16}{[12]^2[23]^2} + \frac{3/8}{[12]^2[31]^2} + \frac{-3/16}{[23]^2[31]^2} + \frac{-3/16}{[12][23][31]^2}  
+ \frac{3/8}{[12][23]^2[31]} +  \frac{-3/16}{[12]^2[23][31]} \right]\ln[23] \\
+ k\left[\frac{3/16}{[12]^2[23]^2} + \frac{-3/8}{[12]^2[31]^2} + \frac{3/16}{[23]^2[31]^2} + \frac{3/16}{[12][23][31]^2} 
  + \frac{-3/8}{[12][23]^2[31]} +  \frac{3/16}{[12]^2[23][31]} \right]\ln[31] 
\end{eqnarray*}
Combining finite contributions from finite double integrals and single finite integral of $B_1,..., B_7$ we obtain  
\begin{eqnarray*}
K_3 \equiv -B_{11} + J_{31}/2 + B_{21}/2 + B_3/4 + B_{41}/16 -B_{11}/4 - kB_{42}/16 - B_{51}/2 + B_{61}/4 = \\
k\left[\frac{1/8}{[12]^2[23]^2} + \frac{-1/4}{[12]^2[31]^2} + \frac{-1/8}{[23]^2[31]^2} +  \frac{3/8}{[12][23][31]^2} +  \frac{1/8}{[12]^2[23][31]} \right] \\
+ k\left[\frac{1/4[12]}{[23]^2[31]^2}  + \frac{-1/4}{[23][31]^2} + \frac{-1/4}{[23]^2[31]}  \right]J(1,1,1) + k \frac{1/4}{[23]^2[31]^2} \ln[12] \\
+ k\left[\frac{1/4}{[23]^2[31]^2} + \frac{-1/2}{[12][23][31]^2} + \frac{-1/4}{[12][23]^2[31]} \right]\ln[23] \\
+ k\left[\frac{-1/2}{[23]^2[31]^2} + \frac{1/2}{[12][23][31]^2} + \frac{1/4}{[12][23]^2[31]} \right]\ln[31] \\
\\
+ \left[\frac{1/2}{[12]^2[23]^2} + \frac{-1}{[23]^2[31]^2} +  \frac{1/2}{[12][23]^2[31]} +  \frac{3/2}{[12]^2[23][31]}   \right] \\
+ \left[\frac{-3}{[12][23]^2} + \frac{1}{[12][31]^2} + \frac{1/2[12]}{[23]^2[31]^2}  +\frac{-1}{[23][31]^2}  + \frac{-2}{[12]^2[31]} +  \frac{3/2}{[23]^2[31]} 
+  \frac{[31]}{[12]^2[23]^2}  \right] J(1,1,1) \\
+ \left[\frac{-2}{[12]^2[23]^2} + \frac{-1/2}{[12][23][31]^2} + \frac{5/2}{[12][23]^2[31]} + \frac{1}{[12]^2[23][31]} \right]\ln[12] \\
+ \left[\frac{-1/2}{[12]^2[23]^2} +  \frac{1}{[23]^2[31]^2} + \frac{-1}{[12][23][31]^2} +  \frac{-1/2}{[12][23]^2[31]} +  \frac{-5/2}{[12]^2[23][31]}  \right]\ln[23] \\
+ \left[\frac{5/2}{[12]^2[23]^2} +   \frac{-1}{[23]^2[31]^2}   + \frac{3/2}{[12][23][31]^2}   + \frac{-2}{[12][23]^2[31]} + \frac{3/2}{[12]^2[23][31]}  \right]\ln[31] 
\end{eqnarray*}
Thus, 
\begin{eqnarray*}
T_{21} + T_{23} = K_3 + M_3 
\end{eqnarray*}

\section[]{}
\setcounter{equation}{0}
\label{App:C}

In this Appendix integral $T_{311}$ is calculated,
\begin{eqnarray*}
T_{311} = \frac{(31)_\nu}{[31]^2}\int~Dy~\frac{(2y)_\sigma}{[2y]^{2-\e}}\int~Dz \le\pd^{(z)}_\mu\Pi_{\rho\sigma}(z2)\ri \Pi_{\mu\nu}(z3)\frac{2(1y)_\rho}{[yz]^{1-\e}[1y]^{2-\e}} = \\
\frac{(31)_\nu}{[31]^2}\int~Dy~~Dz \Pi_{\mu\nu}(z3)\le\pd^{(z)}_\mu\Pi_{\rho\sigma}(z2) \frac{(2y)_\sigma}{[2y]^{2-\e}} \ri \frac{2(1y)_\rho}{[yz]^{1-\e}[1y]^{2-\e}} 
\end{eqnarray*}
According to  Eq. (\ref{trick3}) the first contribution to  $T_{311}$ is (with factor 1)
\begin{eqnarray*}
C_1 \equiv \frac{(31)_\nu}{[31]^2}\int~Dy~Dz \Pi_{\mu\nu}(z3) \le \pd^{(z)}_\mu  \frac{2(2y)_\rho}{[2z]^{1-\e}[2y]^{2-\e}}  \ri \frac{2(1y)_\rho}{[yz]^{1-\e}[1y]^{2-\e}} = \\
\frac{(31)_\nu}{[31]^2}\int~Dy~Dz \Pi_{\mu\nu}(z3) \le \pd^{(z)}_\mu  \frac{1}{[2z]^{1-\e}} \ri 
\frac{2(2y)_\rho}{[2y]^{2-\e}}  \frac{2(1y)_\rho}{[1y]^{2-\e}}\frac{1}{[yz]^{1-\e}}   = \\
\frac{1}{(1-\e)^2}\frac{(31)_\nu}{[31]^2}\int~Dy~Dz \Pi_{\mu\nu}(z3) \le \pd^{(z)}_\mu  \frac{1}{[2z]^{1-\e}} \ri 
 \le \pd^{(y)}_\rho \frac{1}{[2y]^{1-\e}} \ri  \le \pd^{(y)}_\rho \frac{1}{[1y]^{1-\e}} \ri \frac{1}{[yz]^{1-\e}}   = \\
\frac{1}{2(1-\e)^2}\frac{(31)_\nu}{[31]^2}\int~Dy~Dz \Pi_{\mu\nu}(z3) \le \pd^{(z)}_\mu  \frac{1}{[2z]^{1-\e}} \ri 
\left[ \pd^2_{(y)}\le \frac{1}{[2y]^{1-\e}[1y]^{1-\e}} \ri  \right.\\
\left. - \le \pd^2_{(y)} \frac{1}{[2y]^{1-\e}} \ri \frac{1}{[1y]^{1-\e}}    -  \frac{1}{[2y]^{1-\e}}  \le \pd^2_{(y)} \frac{1}{[1y]^{1-\e}} \ri \right] \frac{1}{[yz]^{1-\e}}   = \\
\frac{1}{2(1-\e)^2}\frac{(31)_\nu}{[31]^2}\int~Dy~Dz \Pi_{\mu\nu}(z3) \le \pd^{(z)}_\mu  \frac{1}{[2z]^{1-\e}} \ri 
\frac{1}{[2y]^{1-\e}[1y]^{1-\e}}   k\delta(yz)    \\
- \frac{1}{2(1-\e)^2}\frac{(31)_\nu}{[31]^2}\int~Dy~Dz \Pi_{\mu\nu}(z3) \le \pd^{(z)}_\mu  \frac{1}{[2z]^{1-\e}} \ri 
\frac{1}{[2y]^{1-\e}[yz]^{1-\e}}   k\delta(1y)    \\
- \frac{1}{2(1-\e)^2}\frac{(31)_\nu}{[31]^2}\int~Dy~Dz \Pi_{\mu\nu}(z3) \le \pd^{(z)}_\mu  \frac{1}{[2z]^{1-\e}} \ri 
\frac{1}{[1y]^{1-\e}[yz]^{1-\e}}   k\delta(2y)  =   \\
\frac{k}{2(1-\e)^2}\frac{(31)_\nu}{[31]^2}\int~Dz \Pi_{\mu\nu}(z3) \le \pd^{(z)}_\mu  \frac{1}{[2z]^{1-\e}} \ri 
\frac{1}{[2z]^{1-\e}[1z]^{1-\e}}      \\
- \frac{k}{2(1-\e)^2}\frac{(31)_\nu}{[31]^2}\int~Dz \Pi_{\mu\nu}(z3) \le \pd^{(z)}_\mu  \frac{1}{[2z]^{1-\e}} \ri 
\frac{1}{[12]^{1-\e}[1z]^{1-\e}}      \\
- \frac{k}{2(1-\e)^2}\frac{(31)_\nu}{[31]^2}\int~Dz \Pi_{\mu\nu}(z3) \le \pd^{(z)}_\mu  \frac{1}{[2z]^{1-\e}} \ri 
\frac{1}{[12]^{1-\e}[2z]^{1-\e}}    =   \\
\frac{k}{1-\e}\frac{(31)_\nu}{[31]^2}\int~Dz \Pi_{\mu\nu}(z3) \frac{(2z)_\mu}{[2z]^{3-2\e}[1z]^{1-\e}}   
- k\frac{(31)_\nu}{[31]^2[12]}\int~Dz \Pi_{\mu\nu}(z3) \frac{(2z)_\mu}{[2z]^{2}[1z]}  =    \\
- \frac{k}{2(1-\e)}\frac{(31)_\nu}{[31]^2}\int~Dz \Pi_{\mu\nu}(z3) \frac{(1z)_\mu}{[2z]^{2-2\e}[1z]^{2-\e}}    
- k\frac{(31)_\nu}{[31]^2[12]}\int~Dz \Pi_{\mu\nu}(z3) \frac{(2z)_\mu}{[2z]^{2}[1z]}  \equiv  \\
\equiv - \frac{k}{4(1-\e)}M_{11}   - k C_{11}  
\end{eqnarray*}
Integral $C_{11}$ is finite. The second contribution to $T_{311}$ (with factor $1/(4(1-\e))$) is  
\begin{eqnarray*}
C_2 \equiv \frac{(31)_\nu}{[31]^2}\int~Dy~Dz \Pi_{\mu\nu}(z3) \le \pd^{(z)}_\mu \pd^2_{(2)}\le \frac{(2z)_{\rho}}{[2z]^{1-\e}[2y]^{1-\e}} \ri \ri \frac{2(1y)_\rho}{[yz]^{1-\e}[1y]^{2-\e}} = \\
\frac{1}{1-\e}\pd^2_{(2)} \frac{(31)_\nu}{[31]^2}\int~Dy~Dz \Pi_{\mu\nu}(z3) \le\pd^{(z)}_\rho \frac{(2z)_{\mu}}{[2z]^{1-\e}} \ri \frac{1}{[2y]^{1-\e}} \frac{1}{[yz]^{1-\e}}  
\le \pd^{(y)}_\rho \frac{1}{[1y]^{1-\e}} \ri = \\
- \frac{1}{1-\e}\pd^2_{(2)} \frac{(31)_\nu}{[31]^2}\int~Dy~Dz \Pi_{\mu\nu}(z3) \le\pd^{(z)}_\rho \frac{(2z)_{\mu}}{[2z]^{1-\e}} \ri \le \pd^{(y)}_\rho \frac{1}{[2y]^{1-\e}} \ri 
\frac{1}{[yz]^{1-\e}}  \frac{1}{[1y]^{1-\e}} \\
- \frac{1}{1-\e}\pd^2_{(2)} \frac{(31)_\nu}{[31]^2}\int~Dy~Dz \Pi_{\mu\nu}(z3) \le\pd^{(z)}_\rho \frac{(2z)_{\mu}}{[2z]^{1-\e}} \ri \frac{1}{[2y]^{1-\e}} 
\le \pd^{(y)}_\rho \frac{1}{[yz]^{1-\e}} \ri  \frac{1}{[1y]^{1-\e}} = \\
2 \pd^2_{(2)} \frac{(31)_\nu}{[31]^2}\int~Dy~Dz \Pi_{\mu\nu}(z3) \frac{(2z)_{\rho}}{[2z]^{1-\e}}  \le \pd^{(y)}_\rho \frac{1}{[2y]^{1-\e}} \ri 
\frac{(yz)_\mu}{[yz]^{2-\e}}  \frac{1}{[1y]^{1-\e}} \\
- \frac{1}{1-\e}\pd^2_{(2)} \frac{(31)_\nu}{[31]^2}\int~Dy~Dz \Pi_{\mu\nu}(z3) \le\pd^{(z)}_\rho \frac{(2z)_{\mu}}{[2z]^{1-\e}} \ri \frac{1}{[2y]^{1-\e}} 
\le \pd^{(y)}_\rho \frac{1}{[yz]^{1-\e}} \ri  \frac{1}{[1y]^{1-\e}} = \\
4(1-\e)\pd^2_{(2)} \frac{(31)_\nu}{[31]^2}\int~Dy~Dz \Pi_{\mu\nu}(z3) \frac{(2z)_{\rho}}{[2z]^{1-\e}} \frac{(2y)_\rho}{[2y]^{2-\e}}  
\frac{(yz)_\mu}{[yz]^{2-\e}}  \frac{1}{[1y]^{1-\e}} \\
+ \frac{1}{1-\e}\pd^2_{(2)} \frac{(31)_\nu}{[31]^2}\int~Dy~Dz \Pi_{\mu\nu}(z3) \le\pd^{(z)}_\rho \frac{(2z)_{\mu}}{[2z]^{1-\e}} \ri 
\le \pd^{(z)}_\rho \frac{1}{[yz]^{1-\e}} \ri  \frac{1}{[2y]^{1-\e}[1y]^{1-\e}} = \\
2\pd^2_{(2)} \frac{(31)_\nu}{[31]^2}\int~Dy~Dz \Pi_{\mu\nu}(z3) \frac{[2z]+[2y]-[yz]}{[2z]^{1-2\e}[2y]^{2-2\e}}  \frac{(yz)_\mu}{[yz]^{2}[1y]} \\
+ \frac{1}{2(1-\e)}\pd^2_{(2)} \frac{(31)_\nu}{[31]^2}\int~Dy~Dz \Pi_{\mu\nu}(z3) 
\left[\pd^2_{(z)}\le\frac{(2z)_{\mu}}{[2z]^{1-\e}[yz]^{1-\e}}\ri  +  4(1-\e)\frac{(2z)_{\mu}}{[2z]^{2-\e}[yz]^{1-\e}} \right. \\ 
\left.  - \frac{(2z)_{\mu}}{[2z]^{1-\e}} k\delta(yz) \right]  \frac{1}{[2y]^{1-\e}[1y]^{1-\e}} = \\
2\pd^2_{(2)} \frac{(31)_\nu}{[31]^2}\int~Dy~Dz \Pi_{\mu\nu}(z3) \frac{(yz)_\mu}{[2z]^{-2\e}[2y]^{2-2\e}[yz]^2[1y]} \\
+ 2\pd^2_{(2)} \frac{(31)_\nu}{[31]^2}\int~Dy~Dz \Pi_{\mu\nu}(z3) \frac{(yz)_\mu}{[2z][2y][yz]^2[1y]} \\
- 2\pd^2_{(2)} \frac{(31)_\nu}{[31]^2}\int~Dy~Dz \Pi_{\mu\nu}(z3) \frac{(yz)_\mu}{[2z]^{1-2\e}[2y]^{2-2\e}[yz][1y]} \\
+ \frac{1}{2}\frac{(31)_\nu}{[31]^2} \pd^2_{(2)}\pd^2_{(3)} \int~Dy~Dz \Pi_{\mu\nu}(z3) \frac{(2z)_{\mu}}{[2z][yz][2y][1y]}  \\
+ 2\frac{(31)_\nu}{[31]^2} \pd^2_{(2)}  \int~Dy~Dz \Pi_{\mu\nu}(z3) \frac{(2z)_{\mu}}{[2z]^{2}[yz][2y][1y]} 
- \frac{k}{2}\frac{(31)_\nu}{[31]^2} \pd^2_{(2)}\int~Dz \Pi_{\mu\nu}(z3) \frac{(2z)_{\mu}}{[2z]^{2}[1z]} = \\
4\e\pd^2_{(2)} \frac{(31)_\nu}{[31]^2}\int~Dy~Dz \Pi_{\mu\nu}(z3) \frac{(2z)_\mu}{[2z]^{1-2\e}[2y]^{2-2\e}[yz][1y]} \\
+ 2\pd^2_{(2)} \frac{(31)_\nu}{[31]^2}\int~Dy~Dz \Pi_{\mu\nu}(z3) \frac{(yz)_\mu}{[2z][2y][yz]^2[1y]} \\
+ 2\pd^2_{(2)} \frac{(31)_\nu}{[31]^2}\int~Dy~Dz \Pi_{\mu\nu}(z3) \frac{(2y)_\mu}{[2z]^{1-2\e}[2y]^{2-2\e}[yz][1y]} \\
- 2\pd^2_{(2)} \frac{(31)_\nu}{[31]^2}\int~Dy~Dz \Pi_{\mu\nu}(z3) \frac{(2z)_\mu}{[2z]^{1-2\e}[2y]^{2-2\e}[yz][1y]} \\
+ \frac{1}{2}\frac{(31)_\nu}{[31]^2} \pd^2_{(2)}\pd^2_{(3)} \int~Dy~Dz \Pi_{\mu\nu}(z3) \frac{(2z)_{\mu}}{[2z][yz][2y][1y]}  \\
+ 2\frac{(31)_\nu}{[31]^2} \pd^2_{(2)}  \int~Dy~Dz \Pi_{\mu\nu}(z3) \frac{(2z)_{\mu}}{[2z]^{2}[yz][2y][1y]} 
- \frac{k}{2}\frac{(31)_\nu}{[31]^2} \pd^2_{(2)}\int~Dz \Pi_{\mu\nu}(z3) \frac{(2z)_{\mu}}{[2z]^{2}[1z]}  = \\
- \frac{(2 - 4\e)A(1-2\e,1,1)}{\e(1-2\e)}\pd^2_{(2)} \frac{(31)_\nu}{[31]^2[12]^{1-\e}}\int~Dz \Pi_{\mu\nu}(z3)
 \frac{(2z)_\mu}{[2z]^{2-3\e}[1z]^{\e}} \\
+ 2\pd^2_{(2)} \frac{(31)_\nu}{[31]^2}\int~Dy~Dz \Pi_{\mu\nu}(z3) \frac{(yz)_\mu}{[2z][2y][yz]^2[1y]} \\
+ 2\pd^2_{(2)} \frac{(31)_\nu}{[31]^2}\int~Dy~Dz \Pi_{\mu\nu}(z3) \frac{(2y)_\mu}{[2z]^{1-2\e}[2y]^{2-2\e}[yz][1y]} \\
+ \frac{1}{2}\frac{(31)_\nu}{[31]^2} \pd^2_{(2)}\pd^2_{(3)} \int~Dy~Dz \Pi_{\mu\nu}(z3) \frac{(2z)_{\mu}}{[2z][yz][2y][1y]}  \\
+ 2\frac{(31)_\nu}{[31]^2} \pd^2_{(2)}  \int~Dy~Dz \Pi_{\mu\nu}(z3) \frac{(2z)_{\mu}}{[2z]^{2}[yz][2y][1y]} 
- \frac{k}{2}\frac{(31)_\nu}{[31]^2} \pd^2_{(2)}\int~Dz \Pi_{\mu\nu}(z3) \frac{(2z)_{\mu}}{[2z]^{2}[1z]} = \\
\\
2\pd^2_{(2)} \frac{(31)_\nu}{[31]^2[12]}\int~Dz \Pi_{\mu\nu}(z3) \frac{(1z)_\mu}{[2z][1z]} \\
+ 2\pd^2_{(2)} \frac{(31)_\nu}{[31]^2}\int~Dy~Dz \Pi_{\mu\nu}(z3) \frac{(yz)_\mu}{[2z][2y][yz]^2[1y]} \\
+ 2\pd^2_{(2)} \frac{(31)_\nu}{[31]^2}\int~Dy~Dz \Pi_{\mu\nu}(z3) \frac{(2y)_\mu}{[2z][2y]^{2}[yz][1y]} \\
+ \frac{1}{2}\frac{(31)_\nu}{[31]^2} \pd^2_{(2)}\pd^2_{(3)} \int~Dy~Dz \Pi_{\mu\nu}(z3) \frac{(2z)_{\mu}}{[2z][yz][2y][1y]}  \\
+ 2\frac{(31)_\nu}{[31]^2} \pd^2_{(2)}  \int~Dy~Dz \Pi_{\mu\nu}(z3) \frac{(2z)_{\mu}}{[2z]^{2}[yz][2y][1y]} 
- \frac{k}{2}\frac{(31)_\nu}{[31]^2} \pd^2_{(2)}\int~Dz \Pi_{\mu\nu}(z3) \frac{(2z)_{\mu}}{[2z]^{2}[1z]} \equiv \\
\equiv 2C_{21} + 2B_{11} + 2A_{31} + \frac{1}{2}B_{41} + 2B_{51} - \frac{k}{2}B_{42}
\end{eqnarray*}
All $A$-integrals and $B$-integrals  are finite and were calculated in Appendices A and B, respectively. Integral $C_{21}$ is finite. 
The third contribution to $T_{311}$ is (with factor $-1/(4(1-\e))$) 
\begin{eqnarray*}
C_3 \equiv 
\frac{(31)_\nu}{[31]^2}\int~Dy~Dz \Pi_{\mu\nu}(z3) \le \pd^{(z)}_\mu  \le \pd^2_{(2)}\frac{(2z)_{\rho}}{[2z]^{1-\e}}\ri \frac{1}{[2y]^{1-\e}} \ri \frac{2(1y)_\rho}{[yz]^{1-\e}[1y]^{2-\e}} = \\
- 4 \frac{(31)_\nu}{[31]^2}\int~Dy~Dz \Pi_{\mu\nu}(z3) \le\pd^{(z)}_\rho \frac{(2z)_{\mu}}{[2z]^{2-\e}} \ri \frac{1}{[2y]^{1-\e}} \frac{1}{[yz]^{1-\e}}  
\le \pd^{(y)}_\rho \frac{1}{[1y]^{1-\e}} \ri = \\
\\
4 \frac{(31)_\nu}{[31]^2}\int~Dy~Dz \Pi_{\mu\nu}(z3) \le\pd^{(z)}_\rho \frac{(2z)_{\mu}}{[2z]^{2-\e}} \ri \le \pd^{(y)}_\rho \frac{1}{[2y]^{1-\e}} \ri 
\frac{1}{[yz]^{1-\e}}  \frac{1}{[1y]^{1-\e}} \\
+ 4 \frac{(31)_\nu}{[31]^2}\int~Dy~Dz \Pi_{\mu\nu}(z3) \le\pd^{(z)}_\rho \frac{(2z)_{\mu}}{[2z]^{2-\e}} \ri \frac{1}{[2y]^{1-\e}} 
\le \pd^{(y)}_\rho \frac{1}{[yz]^{1-\e}} \ri  \frac{1}{[1y]^{1-\e}} = \\
-8(1-\e)\frac{(31)_\nu}{[31]^2}\int~Dy~Dz \Pi_{\mu\nu}(z3) \frac{(2z)_{\rho}}{[2z]^{2-\e}}\le \pd^{(y)}_\rho \frac{1}{[2y]^{1-\e}} \ri 
\frac{(yz)_\mu}{[yz]^{2-\e}}  \frac{1}{[1y]^{1-\e}} \\
- 4 \frac{(31)_\nu}{[31]^2}\int~Dy~Dz \Pi_{\mu\nu}(z3) \le\pd^{(z)}_\rho \frac{(2z)_{\mu}}{[2z]^{2-\e}} \ri \frac{1}{[2y]^{1-\e}} 
\le \pd^{(z)}_\rho \frac{1}{[yz]^{1-\e}} \ri  \frac{1}{[1y]^{1-\e}} = \\
\\
-4\frac{(31)_\nu}{[31]^2}\int~Dy~Dz \Pi_{\mu\nu}(z3)  \le \pd^{(2)}_\rho \frac{1}{[2z]^{1-\e}} \ri     \le \pd^{(2)}_\rho \frac{1}{[2y]^{1-\e}} \ri 
\frac{(yz)_\mu}{[yz]^{2-\e}}  \frac{1}{[1y]^{1-\e}} \\
- 4 \frac{(31)_\nu}{[31]^2}\int~Dy~Dz \Pi_{\mu\nu}(z3) \le\pd^{(z)}_\rho \frac{(2z)_{\mu}}{[2z]^{2-\e}} \ri \frac{1}{[2y]^{1-\e}} 
\le \pd^{(z)}_\rho \frac{1}{[yz]^{1-\e}} \ri  \frac{1}{[1y]^{1-\e}} = \\
- 2\pd^2_{(2)} \frac{(31)_\nu}{[31]^2}\int~Dy~Dz \Pi_{\mu\nu}(z3)  \frac{(yz)_\mu}{[2z]^{1-\e}[2y]^{1-\e}[yz]^{2-\e}[1y]^{1-\e}} \\
+ 2\frac{(31)_\nu}{[31]^2}\int~Dy~Dz \Pi_{\mu\nu}(z3)  \frac{(yz)_\mu}{[2y]^{1-\e}[yz]^{2-\e}[1y]^{1-\e}}k\delta(2z) \\
+ 2\frac{(31)_\nu}{[31]^2}\int~Dy~Dz \Pi_{\mu\nu}(z3)  \frac{(yz)_\mu}{[2z]^{1-\e}[yz]^{2-\e}[1y]^{1-\e}}k\delta(2y) \\
- 2 \frac{(31)_\nu}{[31]^2}\int~Dy~Dz \Pi_{\mu\nu}(z3) 
\left[\pd^2_{(z)}\le\frac{(2z)_{\mu}}{[2z]^{2-\e}[yz]^{1-\e}}\ri  -  \frac{k}{2(1-\e)}\pd_\mu^{(z)}\delta(2z)\frac{1}{[yz]^{1-\e}} \right. \\ 
\left.  - \frac{(2z)_{\mu}}{[2z]^{2-\e}} k\delta(yz) \right]  \frac{1}{[2y]^{1-\e}[1y]^{1-\e}} = \\
\\
- 2\pd^2_{(2)} \frac{(31)_\nu}{[31]^2}\int~Dy~Dz \Pi_{\mu\nu}(z3)  \frac{(yz)_\mu}{[2z][2y][yz]^{2}[1y]} \\
+ 2k \frac{(31)_\nu}{[31]^2}\int~Dy~ \Pi_{\mu\nu}(23)  \frac{(y2)_\mu}{[2y]^{1-\e}[2y]^{2-\e}[1y]^{1-\e}} \\
+ 2k \frac{(31)_\nu}{[31]^2[12]^{1-\e}}\int~Dz \Pi_{\mu\nu}(z3)  \frac{(2z)_\mu}{[2z]^{3-2\e}} \\
- 2 \frac{(31)_\nu}{[31]^2}\pd^2_{(3)}\int~Dy~Dz \Pi_{\mu\nu}(z3) \frac{(2z)_{\mu}}{[2z]^{2-\e}[yz]^{1-\e}[2y]^{1-\e}[1y]^{1-\e}} \\
- 2 k\frac{(31)_\nu}{[31]^2}\int~Dy~Dz \Pi_{\mu\nu}(z3) \frac{(yz)_\mu}{[yz]^{2-\e}[2y]^{1-\e}[1y]^{1-\e}}\delta(2z)  \\
+ 2 k \frac{(31)_\nu}{[31]^2}\int~Dy~Dz \Pi_{\mu\nu}(z3) \frac{(2z)_{\mu}}{[2z]^{2-\e}} \delta(yz) \frac{1}{[2y]^{1-\e}[1y]^{1-\e}} = \\
\\
- 2\pd^2_{(2)} \frac{(31)_\nu}{[31]^2}\int~Dy~Dz \Pi_{\mu\nu}(z3)  \frac{(yz)_\mu}{[2z][2y][yz]^{2}[1y]} 
- 2k \frac{(31)_\nu}{[31]^2}\int~Dy~ \Pi_{\mu\nu}(23)  \frac{(2y)_\mu}{[2y]^{3-2\e}[1y]^{1-\e}} \\
- 2 \frac{(31)_\nu}{[31]^2}\pd^2_{(3)}\int~Dy~Dz \Pi_{\mu\nu}(z3) \frac{(2z)_{\mu}}{[2z]^{2}[yz][2y][1y]} 
+ 2 k\frac{(31)_\nu}{[31]^2}\int~Dy~ \Pi_{\mu\nu}(23) \frac{(2y)_\mu}{[2y]^{3-2\e}[1y]^{1-\e}}  \\
+ 2 k \frac{(31)_\nu}{[31]^2}\int~Dz \Pi_{\mu\nu}(z3) \frac{(2z)_{\mu}}{[2z]^{3-2\e}[1z]^{1-\e}} = \\
\\
- 2\pd^2_{(2)} \frac{(31)_\nu}{[31]^2}\int~Dy~Dz \Pi_{\mu\nu}(z3)  \frac{(yz)_\mu}{[2z][2y][yz]^{2}[1y]} \\
- 2 \frac{(31)_\nu}{[31]^2}\pd^2_{(3)}\int~Dy~Dz \Pi_{\mu\nu}(z3) \frac{(2z)_{\mu}}{[2z]^{2}[yz][2y][1y]} 
+ 2 k \frac{(31)_\nu}{[31]^2}\int~Dy~Dz \Pi_{\mu\nu}(z3) \frac{(2z)_{\mu}}{[2z]^{3-2\e}[1z]^{1-\e}} = \\
\\
- 2\pd^2_{(2)} \frac{(31)_\nu}{[31]^2}\int~Dy~Dz \Pi_{\mu\nu}(z3)  \frac{(yz)_\mu}{[2z][2y][yz]^{2}[1y]} \\
- 2 \frac{(31)_\nu}{[31]^2}\pd^2_{(3)}\int~Dy~Dz \Pi_{\mu\nu}(z3) \frac{(2z)_{\mu}}{[2z]^{2}[yz][2y][1y]} 
- k \frac{(31)_\nu}{[31]^2}\int~Dy~Dz \Pi_{\mu\nu}(z3) \frac{(1z)_{\mu}}{[2z]^{2-2\e}[1z]^{2-\e}} \equiv \\
\equiv  -2B_{11} - 2B_{61} - \frac{k}{2}M_{11}.
\end{eqnarray*}
The fourth contribution to $T_{311}$ is (with factor $-1/(4(1-\e))$). 
\begin{eqnarray*}
C_4 \equiv \frac{(31)_\nu}{[31]^2}\int~Dy~Dz \Pi_{\mu\nu}(z3) \le \pd^{(z)}_\mu  \frac{(2z)_{\rho}}{[2z]^{1-\e}} \ri 
\le \pd^2_{(2)} \frac{1}{[2y]^{1-\e}} \ri \frac{2(1y)_\rho}{[yz]^{1-\e}[1y]^{2-\e}} = \\
k\frac{(31)_\nu}{[31]^2}\int~Dy~Dz \Pi_{\mu\nu}(z3) \le \pd^{(z)}_\mu  \frac{(2z)_{\rho}}{[2z]^{1-\e}} \ri \delta(2y)\frac{2(1y)_\rho}{[yz]^{1-\e}[1y]^{2-\e}} = \\
k\frac{(31)_\nu}{[31]^2}\int~Dy~Dz \Pi_{\mu\nu}(z3) \le \pd^{(z)}_\mu  \frac{(2z)_{\rho}}{[2z]^{1-\e}} \ri\frac{2(12)_\rho}{[2z]^{1-\e}[12]^{2-\e}} = \\
k\frac{(31)_\nu}{[31]^2}\frac{2(12)_\rho}{[12]^{2-\e}}\int~Dz \Pi_{\mu\nu}(z3) \le \pd^{(z)}_\mu  \frac{(2z)_{\rho}}{[2z]^{1-\e}} \ri\frac{1}{[2z]^{1-\e}} = \\
- 2(1-\e)k\frac{(31)_\nu}{[31]^2}\frac{2(12)_\rho}{[12]^{2-\e}}\int~Dz \Pi_{\mu\nu}(z3) \frac{(2z)_{\rho} (2z)_\mu}{[2z]^{3-2\e}} = \\
- \frac{k}{2}\frac{(31)_\nu}{[31]^2}\frac{2(12)_\rho}{[12]^{2-\e}}\int~Dz \Pi_{\rho\nu}(z3) \frac{1}{[2z]^{2-2\e}} = -\frac{k}{2}M_{10}
\end{eqnarray*}
The sum of the singular integrals from contribution $C_1,$ $C_3,$ and $C_4$ is  
\begin{eqnarray*}
M_4 \equiv - \frac{k}{4(1-\e)}M_{11}  + \frac{k}{8(1-\e)}M_{11} + \frac{k}{8(1-\e)}M_{10} = \\
k\left[\frac{1/8}{[12]^2[23]^2} + \frac{-1/4}{[12]^2[31]^2} + \frac{1/8}{[23]^2[31]^2} + \frac{1/8}{[12][23][31]^2} + \frac{-1/4}{[12][23]^2[31]} + \frac{1/8}{[12]^2[23][31]} \right]\ln[12] \\
+ k\left[\frac{-1/8}{[12]^2[23]^2} + \frac{1/4}{[12]^2[31]^2} + \frac{-1/8}{[23]^2[31]^2} + \frac{-1/8}{[12][23][31]^2} + \frac{1/4}{[12][23]^2[31]} + \frac{-1/8}{[12]^2[23][31]} \right]\ln[31] 
\end{eqnarray*}
The sum of the finite integrals from $C_1,$ $C_2,$ $C_3,$ and $C_4$ is 
\begin{eqnarray*}  
K_4 \equiv - kC_{11} + \frac{1}{2}C_{21} + \frac{1}{2}B_{11} + \frac{1}{2}A_{31} + \frac{1}{8}B_{41} + \frac{1}{2}B_{51} - \frac{k}{8}B_{42} + \frac{1}{2}B_{11} + \frac{1}{2}B_{61} = \\
k\left[\frac{1/4}{[12]^2[23]^2} + \frac{-1/4}{[23]^2[31]^2} +  \frac{5/4}{[12][23][31]^2}  +   \frac{-1/4}{[12]^2[23][31]} \right] \\
+ k\left[\frac{-1/2}{[12][31]^2} + \frac{1/2[12]}{[23]^2[31]^2} + \frac{-1/2}{[23][31]^2}  + \frac{-1/2}{[23]^2[31]}         \right]J(1,1,1) \\
+ k\left[\frac{-1/2}{[23]^2[31]^2} + \frac{-1/2}{[12][23][31]^2} \right]\ln[12] 
+ k\left[\frac{1}{[12]^2[31]^2}    +   \frac{1/2}{[23]^2[31]^2}  + \frac{-1/2}{[12][23]^2[31]} \right]\ln[23] \\
+ k\left[\frac{-1}{[12]^2[31]^2} + \frac{1/2}{[12][23][31]^2}   + \frac{1/2}{[12][23]^2[31]} \right]\ln[31] 
\\
+ \left[\frac{2}{[12]^2[23]^2} +  \frac{2}{[12][23][31]^2} +  \frac{-2}{[12][23]^2[31]}   \right] \\
+ \left[\frac{1}{[12][23]^2} + \frac{[12]}{[23]^2[31]^2}  + \frac{-2}{[23][31]^2}  + \frac{-1}{[23]^2[31]} + \frac{-[31]}{[12]^2[23]^2}  \right]J(1,1,1) \\
+ \left[\frac{1}{[12][23][31]^2} + \frac{-2}{[12][23]^2[31]} + \frac{-1}{[12]^2[23][31]} \right]\ln[12] 
+ \left[\frac{-2}{[12]^2[23]^2} + \frac{2}{[12][23]^2[31]} \right]\ln[23] \\
+ \left[\frac{2}{[12]^2[23]^2} + \frac{-1}{[12][23][31]^2}  + \frac{1}{[12]^2[23][31]} \right]\ln[31] 
\end{eqnarray*}
The result for $T_{311}$ is  
\begin{eqnarray*}
T_{311} = M_4 + K_4
\end{eqnarray*}

\section[]{}
\setcounter{equation}{0}
\label{App:D}

In this Appendix we calculate $T_{312}$
\begin{eqnarray*}
T_{312} = - \frac{(31)_\nu}{[31]^2}\int~Dy~\frac{(2y)_\sigma}{[2y]^{2-\e}}\int~Dz \le\pd^{(z)}_\mu\Pi_{\rho\sigma}(z2)\ri \Pi_{\mu\nu}(z3)\frac{(yz)_\rho}{[yz]^{2-\e}[1y]^{1-\e}} = \\
- \frac{(31)_\nu}{[31]^2}\int~Dy~~Dz \Pi_{\mu\nu}(z3)\le\pd^{(z)}_\mu\Pi_{\rho\sigma}(z2) \frac{(2y)_\sigma}{[2y]^{2-\e}} \ri \frac{(yz)_\rho}{[yz]^{2-\e}[1y]^{1-\e}} 
\end{eqnarray*}
We use Eq. (\ref{trick3}) to present as a sum of four contributions, and the first contribution to $T_{312}$ (with factor $-1$) is 
\begin{eqnarray*}
D_1 \equiv  \frac{(31)_\nu}{[31]^2}\int~Dy~Dz \Pi_{\mu\nu}(z3) \le \pd^{(z)}_\mu  \frac{2(2y)_\rho}{[2z]^{1-\e}[2y]^{2-\e}}  \ri \frac{(yz)_\rho}{[yz]^{2-\e}[1y]^{1-\e}} = \\
\frac{(31)_\nu}{[31]^2}\int~Dy~Dz \Pi_{\mu\nu}(z3) \le \pd^{(z)}_\mu  \frac{1}{[2z]^{1-\e}} \ri 
\frac{2(2y)_\rho}{[2y]^{2-\e}}   \frac{(yz)_\rho}{[yz]^{2-\e}}\frac{1}{[1y]^{1-\e}}   = \\
- \frac{1}{2(1-\e)^2}\frac{(31)_\nu}{[31]^2}\int~Dy~Dz \Pi_{\mu\nu}(z3) \le \pd^{(z)}_\mu  \frac{1}{[2z]^{1-\e}} \ri 
 \le \pd^{(y)}_\rho \frac{1}{[2y]^{1-\e}} \ri  \le \pd^{(y)}_\rho \frac{1}{[yz]^{1-\e}} \ri \frac{1}{[1y]^{1-\e}}   = \\
- \frac{1}{4(1-\e)^2}\frac{(31)_\nu}{[31]^2}\int~Dy~Dz \Pi_{\mu\nu}(z3) \le \pd^{(z)}_\mu  \frac{1}{[2z]^{1-\e}} \ri 
\left[ \pd^2_{(y)}\le \frac{1}{[2y]^{1-\e}[yz]^{1-\e}} \ri  \right.\\
\left. - \le \pd^2_{(y)} \frac{1}{[2y]^{1-\e}} \ri \frac{1}{[yz]^{1-\e}}    -  \frac{1}{[2y]^{1-\e}}  \le \pd^2_{(y)} \frac{1}{[yz]^{1-\e}} \ri \right] \frac{1}{[1y]^{1-\e}}   = \\
-\frac{1}{4(1-\e)^2}\frac{(31)_\nu}{[31]^2}\int~Dy~Dz \Pi_{\mu\nu}(z3) \le \pd^{(z)}_\mu  \frac{1}{[2z]^{1-\e}} \ri 
\frac{1}{[2y]^{1-\e}[yz]^{1-\e}}   k\delta(1y)    \\
+ \frac{1}{4(1-\e)^2}\frac{(31)_\nu}{[31]^2}\int~Dy~Dz \Pi_{\mu\nu}(z3) \le \pd^{(z)}_\mu  \frac{1}{[2z]^{1-\e}} \ri 
\frac{1}{[1y]^{1-\e}[yz]^{1-\e}}   k\delta(2y)    \\
+ \frac{1}{4(1-\e)^2}\frac{(31)_\nu}{[31]^2}\int~Dy~Dz \Pi_{\mu\nu}(z3) \le \pd^{(z)}_\mu  \frac{1}{[2z]^{1-\e}} \ri 
\frac{1}{[2y]^{1-\e}[1y]^{1-\e}}   k\delta(yz)  =   \\
-\frac{k}{4(1-\e)^2}\frac{(31)_\nu}{[31]^2}\int~Dz \Pi_{\mu\nu}(z3) \le \pd^{(z)}_\mu  \frac{1}{[2z]^{1-\e}} \ri 
\frac{1}{[12]^{1-\e}[1z]^{1-\e}}      \\
+ \frac{k}{4(1-\e)^2}\frac{(31)_\nu}{[31]^2}\int~Dz \Pi_{\mu\nu}(z3) \le \pd^{(z)}_\mu  \frac{1}{[2z]^{1-\e}} \ri 
\frac{1}{[12]^{1-\e}[2z]^{1-\e}}      \\
+ \frac{k}{4(1-\e)^2}\frac{(31)_\nu}{[31]^2}\int~Dz \Pi_{\mu\nu}(z3) \le \pd^{(z)}_\mu  \frac{1}{[2z]^{1-\e}} \ri 
\frac{1}{[1z]^{1-\e}[2z]^{1-\e}}    =   \\
\\
- \frac{k}{2(1-\e)}\frac{(31)_\nu}{[31]^2}\frac{1}{[12]^{1-\e}}\int~Dz \Pi_{\mu\nu}(z3) \frac{(2z)_\mu}{[2z]^{2-2\e}[1z]^{1-\e}} \\   
+ \frac{k}{2(1-\e)}\frac{(31)_\nu}{[31]^2}\int~Dz \Pi_{\mu\nu}(z3) \frac{(2z)_\mu}{[2z]^{3-2\e}[1z]^{1-\e}}  =    \\
- \frac{k}{4(1-\e)}\frac{(31)_\nu}{[31]^2}\int~Dz \Pi_{\mu\nu}(z3) \frac{(1z)_\mu}{[2z]^{2-2\e}[1z]^{2-\e}}   
- \frac{k}{2}\frac{(31)_\nu}{[31]^2[12]}\int~Dz \Pi_{\mu\nu}(z3) \frac{(2z)_\mu}{[2z]^{2}[1z]} \equiv \\
\equiv - \frac{k}{8(1-\e)}M_{11} - \frac{k}{2}C_{11}
\end{eqnarray*}
The second contribution to $T_{312}$ is (with factor $-1/(4(1-\e))$) 
\begin{eqnarray*}
D_2 \equiv \frac{(31)_\nu}{[31]^2}\int~Dy~Dz \Pi_{\mu\nu}(z3) \le \pd^{(z)}_\mu \pd^2_{(2)}\le \frac{(2z)_{\rho}}{[2z]^{1-\e}[2y]^{1-\e}} \ri \ri \frac{(yz)_\rho}{[yz]^{2-\e}[1y]^{1-\e}} = \\
\pd^2_{(2)} \frac{(31)_\nu}{[31]^2}\int~Dy~Dz \Pi_{\mu\nu}(z3) \le\pd^{(z)}_\mu \frac{(2z)_{\rho}}{[2z]^{1-\e}[2y]^{1-\e}}\ri  \frac{(yz)_\rho}{[yz]^{2-\e}[1y]^{1-\e}} = \\
\frac{1}{2(1-\e)}\pd^2_{(2)} \frac{(31)_\nu}{[31]^2}\int~Dy~Dz \Pi_{\mu\nu}(z3) \frac{1}{[2y]^{1-\e}[1y]^{1-\e}} \le\pd^{(z)}_\rho \frac{(2z)_{\mu}}{[2z]^{1-\e}}\ri \le \pd^{(z)}_\rho
\frac{1}{[yz]^{1-\e}}\ri = \\
\frac{1}{4(1-\e)}\pd^2_{(2)} \frac{(31)_\nu}{[31]^2}\int~Dy~Dz \Pi_{\mu\nu}(z3) \frac{1}{[2y]^{1-\e}[1y]^{1-\e}} 
\left[ \pd^2_{(z)}\le \frac{(2z)_{\mu}}{[2z]^{1-\e}[yz]^{1-\e}} \ri  \right.\\
\left. - \le \pd^2_{(z)} \frac{(2z)_{\mu}}{[2z]^{1-\e}} \ri \frac{1}{[yz]^{1-\e}}    -  \frac{(2z)_{\mu}}{[2z]^{1-\e}}  \le \pd^2_{(z)} \frac{1}{[yz]^{1-\e}} \ri \right] = \\
\frac{1}{4}\frac{(31)_\nu}{[31]^2}\pd^2_{(2)}\pd^2_{(3)}\int~Dy~Dz \Pi_{\mu\nu}(z3) \frac{(2z)_\mu}{[2y][1y][2z][yz]} \\
+ \pd^2_{(2)} \frac{(31)_\nu}{[31]^2}\int~Dy~Dz \Pi_{\mu\nu}(z3) \frac{(2z)_\mu}{[2y][1y][2z]^2[yz]}   \\
- \frac{k}{4(1-\e)}\pd^2_{(2)} \frac{(31)_\nu}{[31]^2}\int~Dy~Dz \Pi_{\mu\nu}(z3) \frac{1}{[2y]^{1-\e}[1y]^{1-\e}} \frac{(2z)_{\mu}}{[2z]^{1-\e}}  \delta(yz) = \\
\frac{1}{4}\frac{(31)_\nu}{[31]^2}\pd^2_{(2)}\pd^2_{(3)}\int~Dy~Dz \Pi_{\mu\nu}(z3) \frac{(2z)_\mu}{[2y][1y][2z][yz]} \\
+ \frac{(31)_\nu}{[31]^2}\pd^2_{(2)}\int~Dy~Dz \Pi_{\mu\nu}(z3) \frac{(2z)_\mu}{[2y][1y][2z]^2[yz]}  
- \frac{k}{4}\frac{(31)_\nu}{[31]^2}\pd^2_{(2)}\int~~Dz \Pi_{\mu\nu}(z3) \frac{(2z)_\mu}{[2z]^{2}[1z]} \equiv  \\
\equiv \frac{1}{4}B_{41} + B_{51} - \frac{k}{4}B_{42}
\end{eqnarray*}
The third contribution to $T_{312}$ is (with factor  $1/(4(1-\e))$) 
\begin{eqnarray*}
D_3 \equiv \frac{(31)_\nu}{[31]^2}\int~Dy~Dz \Pi_{\mu\nu}(z3) \le \pd^{(z)}_\mu  \le \pd^2_{(2)}\frac{(2z)_{\rho}}{[2z]^{1-\e}}\ri \frac{1}{[2y]^{1-\e}} \ri 
\frac{(yz)_\rho}{[yz]^{2-\e}[1y]^{1-\e}} = \\
-4(1-\e)\frac{(31)_\nu}{[31]^2}\int~Dy~Dz \Pi_{\mu\nu}(z3) \le \pd^{(z)}_\mu \frac{(2z)_{\rho}}{[2z]^{2-\e}}\ri \frac{1}{[2y]^{1-\e}}  \frac{(yz)_\rho}{[yz]^{2-\e}[1y]^{1-\e}} = \\
-4(1-\e)\frac{(31)_\nu}{[31]^2}\int~Dy~Dz \Pi_{\mu\nu}(z3) \le \pd^{(z)}_\rho \frac{(2z)_{\mu}}{[2z]^{2-\e}}\ri \frac{1}{[2y]^{1-\e}}  \frac{(yz)_\rho}{[yz]^{2-\e}[1y]^{1-\e}} = \\
-2\frac{(31)_\nu}{[31]^2}\int~Dy~Dz \Pi_{\mu\nu}(z3) \frac{1}{[2y]^{1-\e}[1y]^{1-\e}} \le\pd^{(z)}_\rho \frac{(2z)_{\mu}}{[2z]^{2-\e}}\ri \le \pd^{(z)}_\rho\frac{1}{[yz]^{1-\e}}\ri = \\
-\frac{(31)_\nu}{[31]^2}\int~Dy~Dz \Pi_{\mu\nu}(z3) \frac{1}{[2y]^{1-\e}[1y]^{1-\e}} \left[ \pd^2_{(z)}\le \frac{(2z)_{\mu}}{[2z]^{2-\e}[yz]^{1-\e}} \ri  \right.\\
\left. - \le \pd^2_{(z)} \frac{(2z)_{\mu}}{[2z]^{2-\e}} \ri \frac{1}{[yz]^{1-\e}}    -  \frac{(2z)_{\mu}}{[2z]^{2-\e}}  \le \pd^2_{(z)} \frac{1}{[yz]^{1-\e}} \ri \right] = \\
- \frac{(31)_\nu}{[31]^2}\pd^2_{(3)}\int~Dy~Dz \Pi_{\mu\nu}(z3) \frac{(2z)_\mu}{[2y][1y][2z]^2[yz]} \\
- \frac{k}{2(1-\e)}\frac{(31)_\nu}{[31]^2}\int~Dy~Dz \Pi_{\mu\nu}(z3) \frac{1}{[2y]^{1-\e}[1y]^{1-\e}} \pd^{(2)}_\mu \delta(2z) \frac{1}{[yz]^{1-\e}}  \\
+ k\frac{(31)_\nu}{[31]^2}\int~Dy~Dz \Pi_{\mu\nu}(z3) \frac{1}{[2y]^{1-\e}[1y]^{1-\e}} \frac{(2z)_{\mu}}{[2z]^{2-\e}}  \delta(yz) = \\
\\
- \frac{(31)_\nu}{[31]^2}\pd^2_{(3)}\int~Dy~Dz \Pi_{\mu\nu}(z3) \frac{(2z)_\mu}{[2y][1y][2z]^2[yz]} \\
- k\frac{(31)_\nu}{[31]^2}\int~Dy~Dz \Pi_{\mu\nu}(z3) \frac{1}{[2y]^{1-\e}[1y]^{1-\e}} \delta(2z) \frac{(yz)_\mu}{[yz]^{2-\e}}  \\
+ k\frac{(31)_\nu}{[31]^2}\int~Dz \Pi_{\mu\nu}(z3) \frac{(2z)_{\mu}}{[2z]^{3-2\e} [1z]^{1-\e} }  = \\
\\
- \frac{(31)_\nu}{[31]^2}\pd^2_{(3)}\int~Dy~Dz \Pi_{\mu\nu}(z3) \frac{(2z)_\mu}{[2y][1y][2z]^2[yz]} 
- \frac{k}{2}\frac{(31)_\nu}{[31]^2} \Pi_{\mu\nu}(23) \int~Dy\frac{(1y)_\mu}{[2y]^{2-2\e}[1y]^{2-\e}}   \\
- \frac{k}{2}\frac{(31)_\nu}{[31]^2}\int~Dz \Pi_{\mu\nu}(z3) \frac{(1z)_{\mu}}{[2z]^{2-2\e} [1z]^{2-\e} } = - B_{61} - \frac{k}{4}M_{10}  - \frac{k}{4}M_{11}  
\end{eqnarray*}
The fourth contribution to $T_{312}$ is zero. Indeed, 
\begin{eqnarray*}
D_4 \equiv \frac{(31)_\nu}{[31]^2}\int~Dy~Dz \Pi_{\mu\nu}(z3) \le \pd^{(z)}_\mu  \frac{(2z)_{\rho}}{[2z]^{1-\e}} \ri 
\le \pd^2_{(2)} \frac{1}{[2y]^{1-\e}} \ri \frac{(yz)_\rho}{[yz]^{2-\e}[1y]^{1-\e}} = \\
k\frac{(31)_\nu}{[31]^2}\int~Dy~Dz \Pi_{\mu\nu}(z3) \le \pd^{(z)}_\mu  \frac{(2z)_{\rho}}{[2z]^{1-\e}} \ri \delta(2y)\frac{(yz)_\rho}{[yz]^{2-\e}[1y]^{1-\e}} = \\
k\frac{(31)_\nu}{[31]^2}\int~Dy~Dz \Pi_{\mu\nu}(z3) \le \pd^{(z)}_\mu  \frac{(2z)_{\rho}}{[2z]^{1-\e}} \ri\frac{(2z)_\rho}{[2z]^{2-\e}[12]^{1-\e}} = 0
\end{eqnarray*}
The sum of singular integral from $D_1$ and $D_3$ contribution is 
\begin{eqnarray*}
M_5 \equiv \frac{k}{8(1-\e)}M_{11 } - \frac{k}{16(1-\e)}M_{12} - \frac{k}{16(1-\e)}M_{11} = \\  
k\left[\frac{1/16}{[12]^2[23]^2} + \frac{1/16}{[12]^2[31]^2}  + \frac{1/16}{[23]^2[31]^2} +  \frac{-1/8}{[12][23][31]^2}   +  \frac{-1/8}{[12][23]^2[31]} 
+  \frac{-1/8}{[12]^2[23][31]} \right] \\
+ k\left[\frac{-1/16}{[12]^2[23]^2}  + \frac{1/8}{[12]^2[31]^2}  +  \frac{-1/16}{[23]^2[31]^2}  + \frac{-1/16}{[12][23][31]^2}  + \frac{1/8}{[12][23]^2[31]}  
\frac{-1/16}{[12][23]^2[31]}  \right]\ln[23] \\
+ k\left[\frac{1/16}{[12]^2[23]^2}  + \frac{-1/8}{[12]^2[31]^2}  +  \frac{1/16}{[23]^2[31]^2}  + \frac{1/16}{[12][23][31]^2}  + \frac{-1/8}{[12][23]^2[31]} +   
\frac{1/16}{[12][23]^2[31]} \right]\ln[31] 
\end{eqnarray*}
The sum of finite integral from $D_1,$ $D_2$ and $D_3$
\begin{eqnarray*}  
K_5 \equiv \frac{k}{2}C_{11} - \frac{1}{16}B_{41} - \frac{1}{4}B_{51} + \frac{k}{16}B_{42} - \frac{1}{4}B_{61} = \\
k\left[\frac{-1/8}{[12]^2[23]^2} + \frac{1/8}{[23]^2[31]^2}  +  \frac{-5/8}{[12][23][31]^2} +  \frac{1/8}{[12]^2[23][31]}  \right] \\
+ k\left[\frac{1/4}{[12][31]^2} +  \frac{-1/4[12]}{[23]^2[31]^2} +   \frac{1/4}{[23][31]^2} +  \frac{1/4}{[23]^2[31]} \right]J(1,1,1) \\
+ k\left[\frac{1/4}{[23]^2[31]^2} + \frac{1/4}{[12][23][31]^2} \right]\ln[12] 
+ k\left[\frac{-1/2}{[12]^2[31]^2}    +   \frac{-1/4}{[23]^2[31]^2}  + \frac{1/4}{[12][23]^2[31]} \right]\ln[23] \\
+ k\left[\frac{1/2}{[12]^2[31]^2}   + \frac{-1/4}{[12][23][31]^2}   + \frac{-1/4}{[12][23]^2[31]} \right]\ln[31] 
\\
+ \left[\frac{1/2}{[12]^2[23]^2} +  \frac{-1}{[12][23][31]^2}  +  \frac{-1/2}{[12][23]^2[31]}  +  \frac{-1/2}{[12]^2[23][31]}            \right] \\
+ \left[\frac{-1}{[12][23]^2} + \frac{-1/2[12]}{[23]^2[31]^2} + \frac{1}{[23][31]^2}  + \frac{3/2}{[23]^2[31]}  \right]J(1,1,1) \\
+ \left[\frac{-1/2}{[12][23][31]^2} + \frac{-1/2}{[12][23]^2[31]} \right]\ln[12] 
+ \left[\frac{-1/2}{[12]^2[23]^2} +   \frac{1/2}{[12][23]^2[31]}  +  \frac{-1/2}{[12]^2[23][31]}  \right]\ln[23] \\
+ \left[\frac{1/2}{[12]^2[23]^2} + \frac{1/2}{[12][23][31]^2}  + \frac{1/2}{[12]^2[23][31]} \right]\ln[31] 
\end{eqnarray*}
The result for $T_{312}$ is  
\begin{eqnarray*}
T_{312} = M_5 + K_5
\end{eqnarray*}

\end{appendix}


\begin{thebibliography}{99}


\bibitem{Cvetic:2004kx}
G.~Cveti\v{c}, I.~Kondrashuk and I.~Schmidt, ``Effective action of dressed 
mean fields for N = 4 super-Yang-Mills theory,''
Mod.\ Phys.\ Lett.\ A {\bf 21} (2006) 1127 
[hep-th/0407251].


\bibitem{Cvetic:2006kk}
G.~Cveti\v{c}, I.~Kondrashuk and I.~Schmidt,
``On the effective action of dressed mean fields for N = 4 super-Yang-Mills theory,'' 
in {\it Symmetry, Integrability and Geometry: Methods and Applications,} SIGMA (2006) 002
[math-ph/0601002].


\bibitem{Kondrashuk:2004pu}
I.~Kondrashuk and I.~Schmidt, ``Finiteness of n = 4 super-Yang-Mills 
effective action in terms of dressed n = 1 superfields,'' 
hep-th/0411150



\bibitem{Cvetic:2006iu}
G.~Cveti\v{c}, I.~Kondrashuk, A.~Kotikov and I.~Schmidt, ``Towards the two-loop Lcc vertex in Landau gauge,'' Int.\ J.\ Mod.\ Phys.\  A {\bf 22} (2007) 1905
[arXiv:hep-th/0604112].


\bibitem{Cvetic:2007fp}
G.~Cveti\v{c} and I.~Kondrashuk, ``Further results for the two-loop Lcc vertex in the Landau gauge,'' J. High Energy Phys. {\bf 2} (2008) 023,  arXiv:hep-th/0703138.





\bibitem{Cvetic:2002dx}
I.~Kondrashuk, G.~Cveti\v{c},  and I.~Schmidt, 
``Approach to solve Slavnov-Taylor identities in nonsupersymmetric non-Abelian gauge theories,''
Phys.\ Rev.\ D {\bf 67} (2003) 065006
[hep-ph/0203014].


\bibitem{Cvetic:2002in}
G.~Cveti\v{c}, I.~Kondrashuk and I.~Schmidt, 
``QCD effective action with 
dressing functions: Consistency checks inperturbative regime,'' 
Phys.\ Rev.\ D {\bf 67} (2003) 065007
[hep-ph/0210185].


\bibitem{Kondrashuk:2000br} 
I.~Kondrashuk, 
``The solution to Slavnov-Taylor identities in D4 N = 1 SYM,'' 
JHEP {\bf 0011}, 034 (2000)
[hep-th/0007136].


\bibitem{Kondrashuk:2000qb}
I.~Kondrashuk, 
``Renormalizations in softly broken N = 1 theories: Slavnov-Taylor identities,'' 
J.\ Phys.\ A {\bf 33} (2000) 6399
[hep-th/0002096].


\bibitem{Kondrashuk:2003tw}
I.~Kondrashuk,
``An approach to solve Slavnov-Taylor identity in D4 N = 1 supergravity,''
Mod.\ Phys.\ Lett.\ A {\bf 19} (2004) 1291
[gr-qc/0309075].



\bibitem{Blasi:1990xz} A.~Blasi, O.~Piguet and S.~P.~Sorella, 
``Landau gauge and finiteness,'' 
Nucl.\ Phys.\ B {\bf 356} (1991) 154.


\bibitem{Dudal:2003pe}
D.~Dudal {\it et al.},
``The anomalous dimension of the gluon-ghost mass operator in Yang-Mills theory,''
Phys.\ Lett.\ B {\bf 569} (2003) 57, 
[hep-th/0306116].


\bibitem{Dudal:2003np} D.~Dudal, H.~Verschelde, V.~E.~R.~Lemes, M.~S.~Sarandy, 
R.~F.~Sobreiro, S.~P.~Sorella and J.~A.~Gracey, 
``Renormalizability of the local composite operator A(mu)**2 in linear covariant gauges,'' 
Phys.\ Lett.\ B {\bf 574} (2003) 325, 
[hep-th/0308181].



\bibitem{Mathematica} {\it MATHEMATICA 6.0}, Wolfram Research, Inc.



\bibitem{BoSh} N.N. Bogolyubov, D.V. Shirkov, ``Introduction to the theory of 
              quantized fields,'' Moscow, Nauka, 1984 (English translation: 
              Intersci. ~Monogr. ~Phys. ~Astron. ~{\bf 3} ~(1959) ~1). 


\bibitem{Vasil} A.N. Vasiliev, ``Field theoric renormalization group in critical behaviour theory and stochastic dynamics'', Edition of St. Petersburg 
              Institute of Nuclear Physics, 1998. 



\bibitem{Slavnov:1972fg}
A.~A.~Slavnov, 
``Ward Identities In Gauge Theories,'' 
Theor.\ Math.\ Phys.\  {\bf 10} (1972) 99
[Teor.\ Mat.\ Fiz.\  {\bf 10} (1972) 153].



\bibitem{Taylor:1971ff}
J.~C.~Taylor, 
``Ward Identities And Charge Renormalization Of The Yang-Mills Field,''
Nucl.\ Phys.\ B {\bf 33} (1971) 436.



\bibitem{Slavnov:1974dg}
A.~A.~Slavnov,
``Renormalization Of Supersymmetric Gauge Theories. 2. Nonabelian Case,''
Nucl.\ Phys.\ B {\bf 97} (1975) 155.



\bibitem{Faddeev:1980be}
L.~D.~Faddeev and A.~A.~Slavnov,
``Gauge Fields. Introduction To Quantum Theory,''
Front.\ Phys.\  {\bf 50}, 1 (1980)
[Front.\ Phys.\  {\bf 83}, 1 (1990)];
``Introduction to quantum theory of gauge fields'', Moscow, Nauka, (1988).


\bibitem{Lee:1973hb}
B.~W.~Lee,
``Transformation Properties Of Proper Vertices In Gauge Theories,''
Phys.\ Lett.\ B {\bf 46} (1973) 214.



\bibitem{Zinn-Justin:1974mc}
J.~Zinn-Justin,
``Renormalization Of Gauge Theories,''
SACLAY-D.PH-T-74-88,
{\it Lectures given at Int. Summer Inst. for Theoretical Physics, 
Jul 29 - Aug 9, 1974, Bonn, West Germany}.


\bibitem{Becchi:1974md}
C.~Becchi, A.~Rouet and R.~Stora,
``Renormalization Of The Abelian Higgs-Kibble Model,''
Commun.\ Math.\ Phys.\  {\bf 42} (1975) 127.



\bibitem{Tyutin:1975qk}
I.~V.~Tyutin,
``Gauge Invariance In Field Theory And Statistical Physics In Operator Formalism,''
LEBEDEV-75-39 (in Russian), 1975. 



\bibitem{Bern:2005iz}
Z.~Bern, L.~J.~Dixon and V.~A.~Smirnov, 
``Iteration of planar amplitudes in maximally supersymmetric Yang-Mills theory at three loops and beyond,''
  Phys.\ Rev.\  D {\bf 72} (2005) 085001
  [hep-th/0505205].


\bibitem{Usyukina:1992jd}
  N.~I.~Usyukina and A.~I.~Davydychev,
  ``An Approach to the evaluation of three and four point ladder diagrams,''
  Phys.\ Lett.\  B {\bf 298} (1993) 363.


\bibitem{Usyukina:1993ch}
  N.~I.~Usyukina and A.~I.~Davydychev,
  ``Exact results for three and four point ladder diagrams with an arbitrary
  number of rungs,''
  Phys.\ Lett.\  B {\bf 305} (1993) 136.



\bibitem{Broadhurst:1993ib}
  D.~J.~Broadhurst,
  ``Summation of an infinite series of ladder diagrams,''
  Phys.\ Lett.\  B {\bf 307} (1993) 132.


\bibitem{Bern:2006ew}
  Z.~Bern, M.~Czakon, L.~J.~Dixon, D.~A.~Kosower and V.~A.~Smirnov,
  ``The Four-Loop Planar Amplitude and Cusp Anomalous Dimension in Maximally
  Supersymmetric Yang-Mills Theory,''
  Phys.\ Rev.\  D {\bf 75} (2007) 085010
  [arXiv:hep-th/0610248].



\bibitem{Nguyen:2007ya}
  D.~Nguyen, M.~Spradlin and A.~Volovich,
  ``New Dual Conformally Invariant Off-Shell Integrals,''
  Phys.\ Rev.\  D {\bf 77}, 025018 (2008)
  [arXiv:0709.4665 [hep-th]].




\bibitem{Drummond:2006rz}
  J.~M.~Drummond, J.~Henn, V.~A.~Smirnov and E.~Sokatchev,
  ``Magic identities for conformal four-point integrals,''
  JHEP {\bf 0701} (2007) 064
  [arXiv:hep-th/0607160].


\bibitem{Alday:2007hr}
  L.~F.~Alday and J.~M.~Maldacena,
  ``Gluon scattering amplitudes at strong coupling,''
  JHEP {\bf 0706} (2007) 064
  [arXiv:0705.0303 [hep-th]].


\bibitem{Kotikov:2000pm}
  A.~V.~Kotikov and L.~N.~Lipatov,
  ``NLO corrections to the BFKL equation in QCD and in supersymmetric gauge theories,''
  Nucl.\ Phys.\ B {\bf 582} (2000) 19
  [hep-ph/0004008].


\bibitem{Kotikov:2002ab}
  A.~V.~Kotikov and L.~N.~Lipatov,
  ``DGLAP and BFKL equations in the N = 4 supersymmetric gauge theory,''
  Nucl.\ Phys.\ B {\bf 661} (2003) 19
  [Erratum-ibid.\ B {\bf 685} (2004) 405]
  [hep-ph/0208220].


\bibitem{Kotikov:2003fb}
  A.~V.~Kotikov, L.~N.~Lipatov and V.~N.~Velizhanin,
  ``Anomalous dimensions of Wilson operators in N = 4 SYM theory,''
  Phys.\ Lett.\ B {\bf 557} (2003) 114
  [hep-ph/0301021].


\bibitem{Kotikov:2004er}
  A.~V.~Kotikov, L.~N.~Lipatov, A.~I.~Onishchenko and V.~N.~Velizhanin,
  ``Three-loop universal anomalous dimension of the Wilson operators in N =  4 SUSY Yang-Mills model,''
  Phys.\ Lett.\ B {\bf 595} (2004) 521
  [Erratum-ibid.\ B {\bf 632} (2006) 754]
  [hep-th/0404092].


\bibitem{Kotikov:2006ts}
  A.~V.~Kotikov and L.~N.~Lipatov,
  ``On the highest transcendentality in N = 4 SUSY,''
  Nucl.\ Phys.\  B {\bf 769}, 217 (2007)
  [arXiv:hep-th/0611204].


\bibitem{Isaev:2003tk}
  A.~P.~Isaev,
  ``Multi-loop Feynman integrals and conformal quantum mechanics,''
  Nucl.\ Phys.\  B {\bf 662} (2003) 461
  [arXiv:hep-th/0303056].


\bibitem{Vasiliev:1981dg}
  A.~N.~Vasiliev, Y.~M.~Pismak and Y.~R.~Khonkonen,
  ``1/N Expansion: Calculation Of The Exponents Eta And Nu In The Order 1/N**2
  For Arbitrary Number Of Dimensions,''
  Theor.\ Math.\ Phys.\  {\bf 47} (1981) 465
  [Teor.\ Mat.\ Fiz.\  {\bf 47} (1981) 291].


\bibitem{Kazakov:1984bw}
 D.~I.~Kazakov, 
``Analytical Methods For Multiloop Calculations: Two Lectures On The Method Of Uniqueness,''
JINR-E2-84-410.

\bibitem{Kazakov:1984km}
  D.~I.~Kazakov,
  ``The Method Of Uniqueness, A New Powerful Technique For Multiloop Calculations,''
  Phys.\ Lett.\ B {\bf 133} (1983) 406




\bibitem{Tkachov:1981wb}
  F.~V.~Tkachov,
  ``A Theorem On Analytical Calculability Of Four Loop Renormalization Group Functions,''
  Phys.\ Lett.\ B {\bf 100} (1981) 65.



\bibitem{Chetyrkin:1981qh}
  K.~G.~Chetyrkin and F.~V.~Tkachov,
  ``Integration By Parts: The Algorithm To Calculate Beta Functions In Four Loops,''
  Nucl.\ Phys.\ B {\bf 192} (1981) 159.


\bibitem{Chetyrkin:1980pr}
  K.~G.~Chetyrkin, A.~L.~Kataev and F.~V.~Tkachov,
  ``New Approach To Evaluation Of Multiloop Feynman Integrals: The Gegenbauer Polynomial X Space Technique,''
  Nucl.\ Phys.\ B {\bf 174} (1980) 345.


\bibitem{Celmaster:1980ji}
  W.~Celmaster and R.~J.~Gonsalves,
  ``Fourth Order QCD Contributions To The E+ E- Annihilation Cross-Section,''
  Phys.\ Rev.\ D {\bf 21} (1980) 3112.


\bibitem{Terrano:1980af}
  A.~E.~Terrano,
  ``A Method For Feynman Diagram Evaluation,''
  Phys.\ Lett.\ B {\bf 93} (1980) 424.


\bibitem{Lampe:1982av}
  B.~Lampe and G.~Kramer,
  ``Application Of Gegenbauer Integration Method To E+ E- Annihilation Process,''
  Phys.\ Scripta {\bf 28} (1983) 585.


\bibitem{Kotikov:1995cw}
A.~V.~Kotikov, 
``The Gegenbauer Polynomial Technique: the evaluation of a class of Feynman diagrams,'' 
Phys.\ Lett.\ B {\bf 375} (1996) 240 
[hep-ph/9512270].



\bibitem{Davydychev:1992xr}
  A.~I.~Davydychev,
  ``Recursive algorithm of evaluating vertex type Feynman integrals,''
  J.\ Phys.\ A {\bf 25}, 5587 (1992).



\bibitem{Bjerrum-Bohr:2006yw}
  N.~E.~J.~Bjerrum-Bohr, D.~C.~Dunbar, H.~Ita, W.~B.~Perkins and K.~Risager, ``The no-triangle hypothesis for N = 8 supergravity,''
  JHEP {\bf 0612}, 072 (2006)
  [arXiv:hep-th/0610043].




\bibitem{Kang:2004cs}
 K.~Kang and I.~Kondrashuk, 
``Semiclassical scattering amplitudes of dressed gravitons,''
hep-ph/0408168.


\bibitem{Green:2006gt}
  M.~B.~Green, J.~G.~Russo and P.~Vanhove,
  ``Non-renormalisation conditions in type II string theory and maximal supergravity,''
  JHEP {\bf 0702} (2007) 099
  [arXiv:hep-th/0610299].


\bibitem{Bern:2006kd}
  Z.~Bern, L.~J.~Dixon and R.~Roiban, 
``Is N = 8 supergravity ultraviolet finite?,''
Phys.\ Lett.\  B {\bf 644} (2007) 265
[hep-th/0611086].


\bibitem{Avdeev:1992jt}
L.~V.~Avdeev, D.~I.~Kazakov and I.~N.~Kondrashuk,
``Renormalizations in supersymmetric and nonsupersymmetric nonAbelian Chern-Simons field theories with matter,''
Nucl.\ Phys.\  B {\bf 391} (1993) 333.

\bibitem{Jones:1986vp}
  D.~R.~T.~Jones, 
`` Coupling constant renormalization and finite field theory,''
  Nucl.\ Phys.\  B {\bf 277} (1986) 153.


\bibitem{Ermushev:1986cu}
  A.~V.~Ermushev, D.~I.~Kazakov and O.~V.~Tarasov, 
``Finite N=1 Supersymmetric Grand Unified Theories,''
  Nucl.\ Phys.\  B {\bf 281} (1987) 72.


\bibitem{Kazakov:1991th}
  D.~I.~Kazakov and I.~N.~Kondrashuk, 
``Low-Energy Predictions Of Susy Guts: Minimal Versus Finite Model,''
  Int.\ J.\ Mod.\ Phys.\  A {\bf 7} (1992) 3869.


\bibitem{Kazakov:1995cy}
  D.~I.~Kazakov, M.~Y.~Kalmykov, I.~N.~Kondrashuk and A.~V.~Gladyshev,
``Softly Broken Finite Supersymmetric Grand Unified Theory,''
  Nucl.\ Phys.\  B {\bf 471} (1996) 389
  [hep-ph/9511419].

\bibitem{Kondrashuk:1997uf}
  I.~N.~Kondrashuk, 
``Reduction of the finite grand unification theory to the minimal supersymmetric standard model,''
  J.\ Exp.\ Theor.\ Phys.\  {\bf 84} (1997) 432
  [Zh.\ Eksp.\ Teor.\ Fiz.\  {\bf 111} (1997) 787].

\bibitem{Yamada:1994id}
  Y.~Yamada,
  ``Two loop renormalization group equations for soft SUSY breaking scalar interactions: Supergraph method,''
Phys.\ Rev.\ D {\bf 50} (1994) 3537
[hep-ph/9401241].

\bibitem{Jack:1997pa}
  I.~Jack and D.~R.~T.~Jones,
  ``The gaugino beta-function,''
  Phys.\ Lett.\ B {\bf 415} (1997) 383
  [hep-ph/9709364].

\bibitem{Avdeev:1997vx}
  L.~V.~Avdeev, D.~I.~Kazakov and I.~N.~Kondrashuk,
  ``Renormalizations in softly broken SUSY gauge theories,''
  Nucl.\ Phys.\ B {\bf 510} (1998) 289
  [hep-ph/9709397].

\bibitem{Kondrashuk:1999de}
  I.~Kondrashuk,
  ``On the relation between Green functions of the SUSY theory with and without soft terms,''
  Phys.\ Lett.\  B {\bf 470} (1999) 129
  [hep-th/9903167].

\end{thebibliography}
\end{document}